\RequirePackage{fix-cm}
\documentclass[twocolumn,epjc3]{svjour3}  
\smartqed  

\RequirePackage{graphicx}
\RequirePackage{times}
\RequirePackage{flushend}
\RequirePackage[numbers,sort&compress]{natbib}
\RequirePackage[colorlinks,citecolor=blue,urlcolor=blue,linkcolor=blue]{hyperref}
\RequirePackage{orcidlink}
\RequirePackage{amsmath,amsfonts,amssymb}
\RequirePackage{float}
 \RequirePackage{soul}

\newcommand{\orcid}[1]{\href{https://orcid.org/#1}{\textcolor[HTML]{A6CE39}{\aiOrcid}}}
\journalname{Eur. Phys. J. C}
\begin{document}


\title{Probing displaced top quark signature at the LHC Run~3
}



\author{Jeremy Andrea \orcidlink{0000-0002-8298-7560} \thanksref{addr1}
        \and
        Daniel Bloch \orcidlink{0000-0002-4535-5273} \thanksref{addr1} 
        \and
        \'Eric Conte \orcidlink{0000-0003-3154-3123} \thanksref{addr1,addr2} 
        \and
        Douja Darej\thanksref{addr1} 
        \and
        Robin~Ducrocq~\orcidlink{0000-0002-1245-7257}~\thanksref{e1,addr1} 
        \and
        Emery Nibigira \orcidlink{0000-0001-5821-291X} \thanksref{addr3} 
}

\thankstext{e1}{robin.ducrocq@iphc.cnrs.fr}


\institute{Universit\'e de Strasbourg, CNRS, IPHC UMR7178, F-67000 Strasbourg, France \label{addr1}
           \and
           Universit\'e de Haute-Alsace, 68100 Mulhouse, France \label{addr2}
           \and
           University of Tennessee, Knoxville, Tennessee 37996, USA \label{addr3}
}

\date{Received: date / Accepted: date}

\maketitle

\begin{abstract}
In the context of prospective studies for searches of new physics at the LHC Run~3, this paper investigates the relevance of using top quarks produced from new long-lived particles, and detected in the tracker volume of the ATLAS and CMS experiments. Such a signature, referred to as \sloppy\textsl{displaced top quarks}, leads to final states containing displaced vertices and a high multiplicity of displaced jets and tracks, thanks to the top quark decays. Therefore, it is a possible powerful tool for searching for new long-lived particles. Three simplified models based on supersymmetry are explicitly designed for the study of this signature. They differ according to the nature of the long-lived heavy particle which produces at least one top quark: electrically neutral or charged, coloured or non-coloured long-lived particle. For each model, a wide region of parameter space, consistent with a reasonable number of displaced top quarks decaying in a typical tracker volume has been probed. 
From this study, promising benchmarks are defined and experimental guidelines are suggested.
\end{abstract}



\section{Introduction}
After a long shutdown for technical maintenance and improvement, the LHC~\cite{Evans:2008zzb} Run~3 officially started the 5th of July 2022 and currently provides proton-proton collisions with an energy in the center-of-mass of 13.6~TeV (14~TeV in the future). Experimentalists explore the TeV physics through data acquired by the ATLAS~\cite{ATLAS:2008xda}, CMS~\cite{CMS:2008xjf} and LHCb~\cite{LHCb:2008vvz} detectors. The research program is extremely rich and two particular topics, relevant to this paper, can be highlighted: 
\begin{itemize}
    \item[$\bullet$] \sloppy The top quark physics relies on the measurement of the production, the decays, and the properties of the top quark~\cite{Wuchterl:2022iwh}. Its high mass, and the fact that its decay occurs before its hadronisation, confers to the top quark a unique place in the Standard Model (SM) of particle physics. Thus, it is experimentally relevant to use the top quark to explore several BSM (Beyond Standard Model) physics scenarios involving new heavy particles that couple preferentially to the top quark.
    
    \item[$\bullet$] Among the potential  BSM experimental signatures, several recent investigations focus on exotic particles with a lifetime long enough to allow them to decay beyond the first layer of detection~\cite{Alimena:2019zri}. While these event topologies benefit from less background sources, the reconstruction is usually difficult with traditional algorithms. It can lead to different kinds of reconstructed signatures such as displaced leptons~\cite{ATLAS:2020wjh,CMS:2021kdm}, displaced jets~\cite{CMS:2020iwv,ATLAS:2022zhj}, displaced vertices~\cite{ATLAS:2019fwx,Bondarenko:2019tss} or disappearing tracks~\cite{CMS:2020atg,ATLAS:2022rme}. Analyses devoted to the search of Heavy Stable Charged Particles (HSCP) can also be considered if the long-lived particle decays outside the entire volume of the detector~\cite{Khachatryan_2016,https://doi.org/10.48550/arxiv.2205.06013}.
\end{itemize}

This paper combines these two topics (top quark physics and long-lived particles) and proposes the investigation of scenarios at LHC Run~3 with $\sqrt{s}=14\ \text{TeV}$ where an exotic heavy long-lived particle decays into at least one top quark. Such a scenario benefits from the possibility to use trajectories of charged particles (tracks), to reconstruct and identify displaced vertices. In our study, the displacement of the top quark decay vertex with respect to the primary vertex must be consistent with the (ATLAS or CMS) tracker volume. Therefore, we take under consideration scenarios for which the long-lived particles decay within the tracker volume. In this case, the tracking algorithm must reconstruct tracks with missing hits in the first layers of the tracker detector. If the displacement in the transverse plane to the beam axis is lower than about $4~\textrm{cm}$\footnote{The first tracker detection layer is located at 3.3~cm for the ATLAS experiment~\cite{ATL-INDET-PUB-2016-001} (2.9~cm for the CMS experiment~\cite{Adam_2021}) away from the center of the detector.}, the good quality tracks coming from the top decay (electrically charged) final states should have been reconstructed by considering all the hits in the first tracker detection
layer. If it is even smaller, the SM top quark production becomes an important and irreducible background for a potential analysis.
At the opposite, if the transverse displacement is above $1~\textrm{m}$, the particle decays outside the tracker. The present signature will be referred to as \textsl{displaced top quark} in the following. As top quarks are produced only at the primary vertex in the SM, the observation of \textsl{displaced top quarks} would constitute a clear evidence of new physics.  \\

Non-prompt top quark studies have already been considered by CMS and ATLAS collaborations in an inclusive approach~\cite{RPVllp1}\cite{ATLAS:2022fag}\cite{PhysRevD.104.052011} through objects like displaced jets or displaced vertices. Exclusive research based on \textsl{displaced top quarks} still needs to be done. If the new physics is top-philic, exclusive research could offer a better sensitivity than inclusive research. Moreover, a \textsl{displaced top quark} is also a challenging signature on which the performance of traditional identification and reconstruction algorithms can be evaluated, such as the track reconstruction or the identification of b quark jets (although the latter algorithm is not required for building an experimental analysis dedicated to the search of \textsl{displaced top quarks}). 
Naturally, the detection of this signature involves a specific tune or improvement of the reconstruction of tracks which are initiating inside the tracker volume. \\

In the first stage, the study of the \textsl{displaced top quarks} signature requires simplified models, benchmark points, phy\-sics processes, and distributions of typical observables that this paper will provide. The different models are classified according to the charges of the long-lived particle, in particular, the electric charge and the colour. Three cases will be considered: mediation by an electrically neutral and uncoloured long-lived particle, an electrically charged and co\-loured long-lived particle (R-hadron based on a squark), and an electrically neutral and coloured long-lived particle (R-hadron based on a gluino).\\

This paper is organised as follows. The first section is devoted to the common methodology followed in the presentation of each long-lived particle scenario. Then, models and benchmarks will be described in a second section. For each scenario, the relevant region of the parameters space and key experimental observables will be discussed. Afterwards, a discussion will be given on several points of attention related to the online selection (achieved by the trigger system to reduce the data flow to acquire) 
and offline selections (built by the experimentalists on top of the reconstructed information). 
 Finally, the main results will be summarised in the last section.

\section{Methodology} 

\subsection{Targeted strategy}

The strategy developed for this paper consists in finding topologies of events in accordance with the definition of \textsl{displaced top quarks}. One additional requirement is also imposed: the set of parameters handling the production cross section of the long-lived mediator must differ from the set of parameters handling its decay. Such a configuration allows us to decouple the cross section from the flight distance of the long-lived particles, in order to get a large variety of top quark displacement values with a production cross section not too low. These particular topologies lead to the design of specific simplified models which will be classified according to the charges of the long-lived particle. Proposing benchmarks and experimental guidelines require to follow a common and explicit methodology:
\begin{itemize}
    \item [$\bullet$]A simplified model is defined for describing the \textsl{displaced top quark} signatures. 
    \item[$\bullet$] The production of the long-lived particle is studied by choosing one or several physical processes and by calculating the production cross section at LHC with an energy in the center-of-mass of 14 TeV.
    \item[$\bullet$] The average flight distance of the long-lived particle, which will be denoted $\langle d^\textrm{flight} \rangle$ subsequently, is determined by computing the proper time $\tau$ from the decay width of the particle, and the average kinetic factor $\langle \beta\gamma \rangle$, required to move from the long-lived particle rest frame to the laboratory frame according to $\langle d^\textrm{flight} \rangle=\langle \beta\gamma \rangle c \tau$.
    \item[$\bullet$] The most interesting parameter region allowing a \textsl{displaced top quark} is highlighted, considering that it decays within the tracker volume.
    \item[$\bullet$] Benchmarks are chosen by selecting points in the parameter space of the model for which the production cross section gives enough statistics with LHC Run~3 (assuming an integrated luminosity of $300~\textrm{fb}^{-1}$). A flight distance for the long-lived particle consistent with the tracker \sloppy  volume is also required, as well as particle masses leading to sufficiently high production cross section and compatible with the range of mass scrutinised by LHC Runs 1 and 2 analyses. These benchmarks could be the foundation stone of potential future experimental analyses dedicated to displaced top quarks.
    \item[$\bullet$] Distributions of key observables are shown and discussed, in order to highlight the properties of the benchmark points and their differences.
\end{itemize}

\subsection{Event generation} 

Cross section or width calculations and event generation require the application of several high-energy physics packages. For implementing the theoretical simplified models, the framework \textsc{FeynRules}~\cite{feynrules} is used in order to extract in an automated way the Feynman rules and to store them in terms of couplings and parameters into the UFO (Universal FeynRules Output) format~\cite{ufo}. This model is then interfaced to the Monte Carlo (MC) generator \sloppy \textsc{Madgraph\_aMC@NLO}~v.2.7.0~\cite{Alwall_2014} in order to generate the hard process at the leading-order (LO) of the quantum chromodynamics (QCD). No kinematics or geometrical cut is applied at this stage. The NNPDF30~\cite{Ball_2015} parton distribution functions (PDF) at LO QCD is used for the generation, and is embedded in the LHAPDF package~\cite{lhapdf}. At this point, the produced events contain only leptons and partons. The next step of the simulation is to apply a shower program, in our case \textsc{Pythia~8.306}~\cite{Pythia}, to handle initial and final states radiations, hadronisation, hadron decays and underlying events. All the allowed decay channels of the top quark (hadronic and leptonic) are accounted for. In order to describe best the LHC data, the \textsc{Pythia 8} process must be tuned with experimental measurements gathered into a MC tune including beam remnants and multiple parton interactions. As no MC tune is available for LHC at 14~TeV, the MC tune chosen for our analysis is CUETP8M2T4~\cite{mctunes} which has been prepared with LHC data taken at a $\sqrt{s}=13\ \text{TeV}$ and is consistent with the PDF choice made in this article. For coloured long-lived particles, the R-hadronisation is also achieved by \textsc{Pythia~8} with the default parameter values. In particular, the probability of producing a gluinoball $R^0(\tilde{g}g)$ is set to 10\%~\cite{Mackeprang:2007bqa}. Online and offline effects due to additional proton-proton interactions within the same or nearby bunch crossings have not been considered in this simulation.\\

Furthermore, the \textsc{MadAnalysis~5} (MA5) framework~\cite{Conte_2013,Conte_2014,Conte_2018} has been used not only to analyse the event samples but also to take into account a very simplified detector simulation, thanks to the SFS module~\cite{Araz_2021} embedded in the MA5 package. In this detector simulation, the fiducial acceptance of the tracker has been implemented by modelling the tracker volume by a hollow cylinder with an inner radius of 4~cm and an outer radius of 100~cm, extending up to $\pm 300\ \text{cm}$ along the beam axis. This implementation describes approximately the geometry of a tracker such as the ATLAS or CMS ones. The magnetic field has also been taken into consideration by taking a uniform value of B=3.8~T aligned with the beam axis (z-axis), which corresponds to the one of the CMS detector\footnote{Note that even though the magnetic field of the ATLAS detector is different, the tracking specifications between both experiments can be considered similar at the accuracy level of this phenomenological work.}. The simulation also benefits from one of the last developments~\cite{Araz_2022} of \textsc{MadAnalysis~5} which allows us to apply the magnetic field to all electrically charged particles, including charged long-lived particles. Then, the jets are reconstructed using the \textsc{FastJet} package~\cite{Cacciari_2012} with an anti-$k_T$ algorithm~\cite{Cacciari_2008} according to the following parameters: $p_{T}>30\ \text{GeV}$ with $p_T$ the transverse momentum, $\big|\eta\big| < 2.5$ and $\Delta R=\sqrt{\Delta\eta^2+\Delta\phi^2}<0.4$ (where $\Delta\eta$ is the difference in pseudorapidity and $\Delta\phi$ is the difference in azimuthal angle between two particles). 
 
 \subsection{Reconstruction of physical observables}\label{subsec:obs}

Different observables are used to characterise the benchmark points. Consistently with the \textsc{MA5}~\cite{Conte_2013} implementations, the MET (\textit{Missing Transverse Energy}) is defined as the modulus of the opposite of the vector sum of the transverse momenta of the isolated reconstructed objects. Similarly, the TET (\textit{Total Transverse Energy}) observable is defined as the sum of the magnitude of the transverse momenta of the isolated reconstructed objects. From these two observables, are excluded the invisible particles which are the neutrinos but also the neutral long-lived mediator if it decays beyond the volume of the calorimeter. This last statement is based on the assumption that the long-lived mediator, especially in the R-hadron case, do not interact with the calorimeter matter or its potential interaction is negligible. 
In particular, some expected properties 
of the R-hadrons interaction with matter~\cite{Kraan_2004} such as R-mesons changing into R-baryons, R-hadrons electric charge changing or R-hadrons \textsl{stopped} in the calorimeter and decaying at some later time, are not considered in this study.\\

The THT (\textit{Total Transverse Hadronic Energy}) or MHT (\textit{Missing Transverse Hadronic Energy}) observables are reconstructed on the same basis as the TET and MET observables, but they are restricted to hadronic particles.\\

Some specific transverse observables are also investigated such as $\alpha_T$ \cite{alphaT}. This observable is defined in the case of a dijet-event topology as: 
\begin{equation}
\alpha_T = \frac{{E_{T_2}}}{m_{T_{jj}}} \label{eq:alphat}
\end{equation}
with {$E_{T_2}$ the transverse energy} of the second hardest jet and $m_{T_{jj}}$ the transverse mass of the two hardest jets. Supersymmetric (and more generically BSM) events may be identified through high values of this observable. In the case of multijet events, all possible configurations of dijets are tested by combining all reconstructed jets into two pseudo-jets $pj_1$ and $pj_2$. The configuration which minimises the quantity $\Delta H_T=E_T^{pj_1}-E_T^{pj_2}$ is chosen, with $E_T$ the transverse energy. The $\alpha_T$-observable in Eq.~\ref{eq:alphat} can then be rewritten as a function of $\Delta H_T$ like:
\begin{equation}
\alpha_T = \frac12 \frac{1-\Delta H_T/\text{THT}}{\sqrt{1-(\text{MHT}/\text{THT})^2}}\ . \nonumber
\end{equation}\medskip

Finally, the transverse and longitudinal impact parameters $d_0$ and $d_z$ of the top quarks are also calculated for each benchmark, taking into account the magnitude of the magnetic field. The \textsc{MA5} definitions~\cite{Araz_2022} are reminded:
    \begin{equation}
    \begin{aligned}
     d_0 = &\  \textrm{sgn}\left(r\right)\cdot\left(\sqrt{x^2+y^2}-\left|r\right|\right)\ ,\\
    d_z = &\  z_v + \frac{p_z}{qB}\arctan \left( \frac{x p_x +yp_y}{yp_x - xp_y}\right)
    \end{aligned}\nonumber
    \end{equation}
(with $r=p_T/(qB)$, $x= x_v+p_y/qB$ and $y=y_v-p_x/qB$) where $q$ is the electric charge of the particle and $(p_x,p_y,p_z)$ is the momentum at the position of the displaced vertex $(x_v,y_v,z_v)$.\medskip

In this paper, \textsl{displaced tracks} are defined as charged particles with a production vertex located inside the tracker volume. Such tracks must have a transverse momentum $p_T$ greater than $1~\text{GeV}$.


\section{Models and benchmarks devoted to displaced top quarks}
\subsection{Mediation by an electrically neutral and uncoloured long-lived particle} 

In this section, we define a simplified model based on the R-parity violated MSSM (Minimal Supersymmetric Standard Model)~\cite{RPV_MSSM}. A long-lived neutralino is postulated, which is an electrically neutral and uncoloured particle. Two different processes at LHC are investigated in this context: the pair production of neutralinos and the pair production of smuons.
 
 \subsubsection{Model description}

The simplified model is based on the MSSM density Lagrangian, where the superpotential $W$ also contains R-parity violated couplings encapsulated in a term noted $W_{RPV}$:
\begin{equation}
W=W_{MSSM}+W_{RPV}\nonumber
\end{equation}
with:
\begin{equation}
\begin{aligned}
W_{RPV}  = &\  \epsilon_i \left( H_2 \cdot L_i \right)\  +\  \frac{1}{2} \lambda_{ijk} \left( L_i \cdot L_j \right) E_k^c \ \\
& + \lambda'_{ijk}  \left( L_i \cdot Q_j \right) D_k^c+\ \frac{1}{2} \lambda''_{ijk} U_i^cD_f^cD^c_k    \nonumber
\end{aligned}
\end{equation}

\noindent where $L_i$, $Q_i$ and $H_2$ are respectively the lepton, the quark and the up-Higgs SU(2) doublets superfields. The ‘‘ $\cdot$ ’’ binary operation is the SU(2) inner product. $E_i$, $U_i$ and $D_i$ are respectively the lepton, the up and down quark SU(2) singlets. The upper index $^c$ indicates the charge conjugate. The lower indices $i$, $j$ and $k$ can be equal to 1, 2 and 3 and match the lepton/quark generation. Finally, the trilinear couplings $\lambda_{ijk}$, $\lambda'_{ijk}$ and $\lambda''_{ijk}$  are dimensionless R-parity violated couplings, whereas the bilinear couplings $\epsilon_i$ have the dimension of energy. These couplings are taken real for simplification. Concerning the trilinear couplings:
\begin{itemize}
\item[$\bullet$] $\lambda_{ijk}$ leads to leptonic number violation. Due to the gauge symmetry, the coupling is antisymmetric according to the two first indices: $\lambda_{ijk}=-\lambda_{jik}$;
\item[$\bullet$] $\lambda'_{ijk}$ leads to leptonic and baryonic number violation;
\item[$\bullet$] $\lambda''_{ijk}$ leads to baryonic number violation. Due to the gauge symmetry, the coupling is antisymmetric according to the two first indices: $\lambda''_{ijk}=-\lambda''_{jik}$.
\end{itemize}

\begin{figure}[htbp]
    \centering
    \includegraphics[scale=0.18]{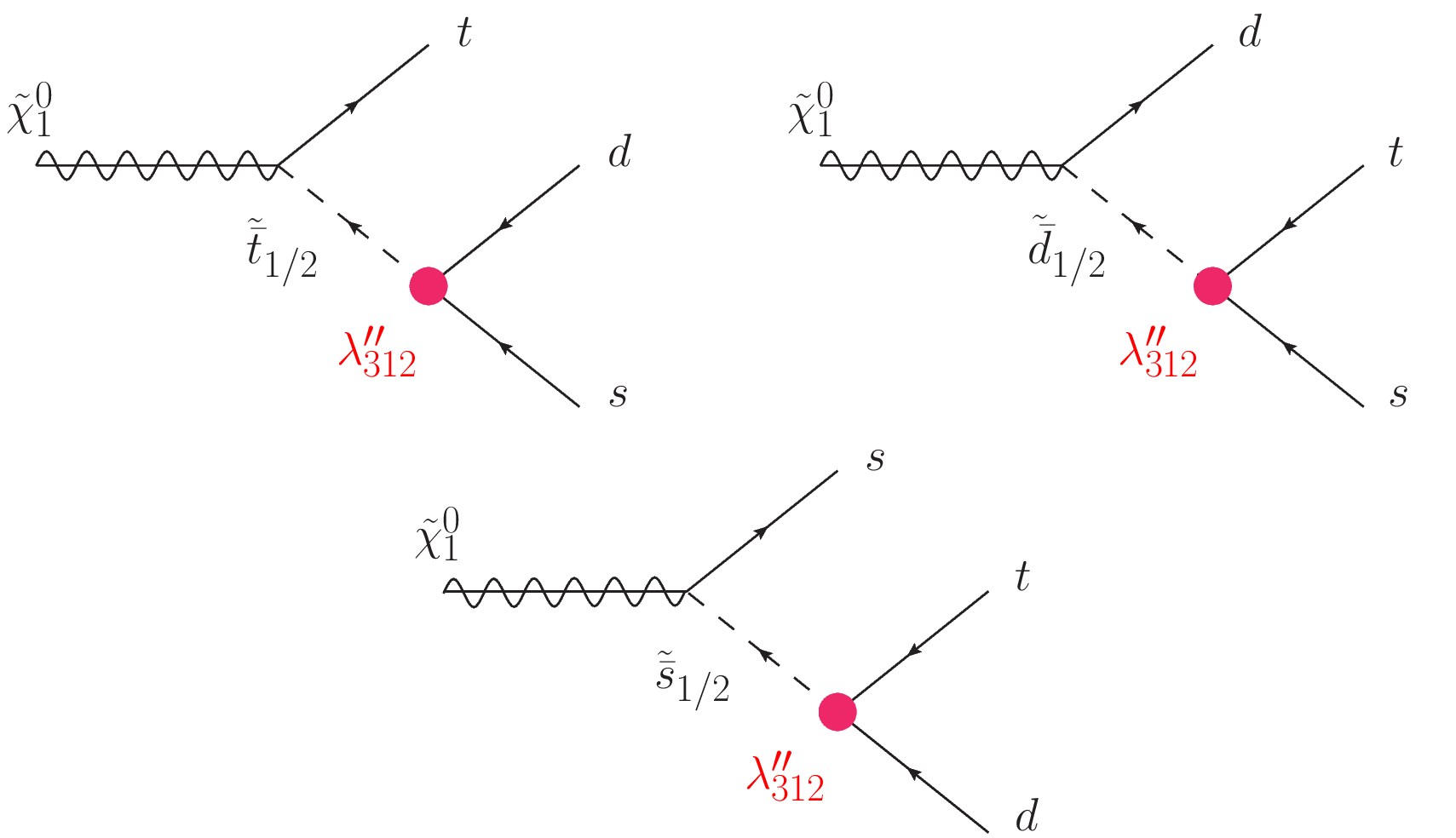}
    \caption{Neutralino decay diagrams involving the studied R-parity violated couplings.}
    \label{fig:neutralino-decay}
\end{figure}

These couplings are assumed to be small as expected from the experimental limits~\cite{RPV_MSSM}. Their contributions to the production of supersymmetric particles at LHC are therefore negligible. However, they allow the LSP (Lightest Supersymmetric Particle) to decay into standard particles and to fly in the detector volume according to the low value of the trilinear coupling. Defining the neutralino as the LSP, only a non-null value for the couplings $\lambda'_{i3k}$ or $\lambda''_{312}=-\lambda''_{321}$ can lead to one top quark in the decay products of the neutralino and only a non-null value for the coupling $\lambda''_{313}=-\lambda''_{331}$ or $\lambda''_{323}=-\lambda''_{332}$ can give two top quarks. Figure~\ref{fig:neutralino-decay} illustrates this decay channel in the case of a non-null coupling $\lambda''_{312}$. Note that since the LSP is not stable, there is no dark matter candidate in this model.\\

The first simplified model is built by taking all R-parity violated coupling as zero except $\lambda''_{312}$ (noted in the following $\lambda''$). We only study extreme cases where the neutralino is a pure state, \textsl{i.e.} it can be a bino $\tilde{B}$, wino $\tilde{W}^3$ or a higgsino ($\tilde{h}_1$ or $\tilde{h}_2$). Mixing matrices for sleptons and squarks are chosen such that there is no mixing between particles of different generations, and the mass states are in an equal proportion of left-handed and right-handed states. Finally, the stop squark mass is fixed to $1.0\ \text{TeV}$ (from typical LHC constraints, see Section \ref{sec:stop} for more informations), while the other squarks are set at higher masses and so neglected.

\subsubsection{Neutralino production and cross section}

To produce the long-lived neutralinos, two hard processes are presupposed at LO QCD: the direct neutralino-pair production or the smuon-pair production decaying into neutralino and muon. The additional supersymmetric Higgs bosons contributions have been ignored, assuming a high mass for such particles.

\begin{itemize}
\item[$\bullet$] Figure~\ref{fig:smuon-prod} displays a smuon production ($pp\rightarrow \tilde{\mu}^+ \tilde{\mu}^-$) where the smuon decays into a muon and a neutralino ($\tilde{\mu}^\pm \rightarrow \tilde{\chi}^0_1 \mu^\pm$). We set a branching fraction of 100\% by assuming that the smuon $\tilde{\mu}$ is the Next-to-Lightest-Supersymmetric-Particle (NLSP). The muons in the final states are very convenient for triggering the events and facilitating the identification of such a signature.
 Concerning the definition of the neutralino, the pure $\tilde{h}_2$ state is forbidden to make the smuon decay possible. Consequently, only a bino-like neutralino $\tilde{B}$ can decay via the RPV-coupling $\lambda''$. The smuon mass is then the only free parameter for the cross section computation. The coupling between the Z boson and the smuons is fixed by the choice of mixing between left and right smuons. Note also that this analysis can be easily extended to the case $pp\rightarrow \tilde{\ell}\bar{\tilde{\ell}}\rightarrow \tilde{\chi}^0_1\tilde{\chi}^0_1 \ell\bar{\ell}$, with the lepton $\ell$ including electrons and taus. 
\begin{figure}[htbp]
    \centering
    \includegraphics[scale=0.15]{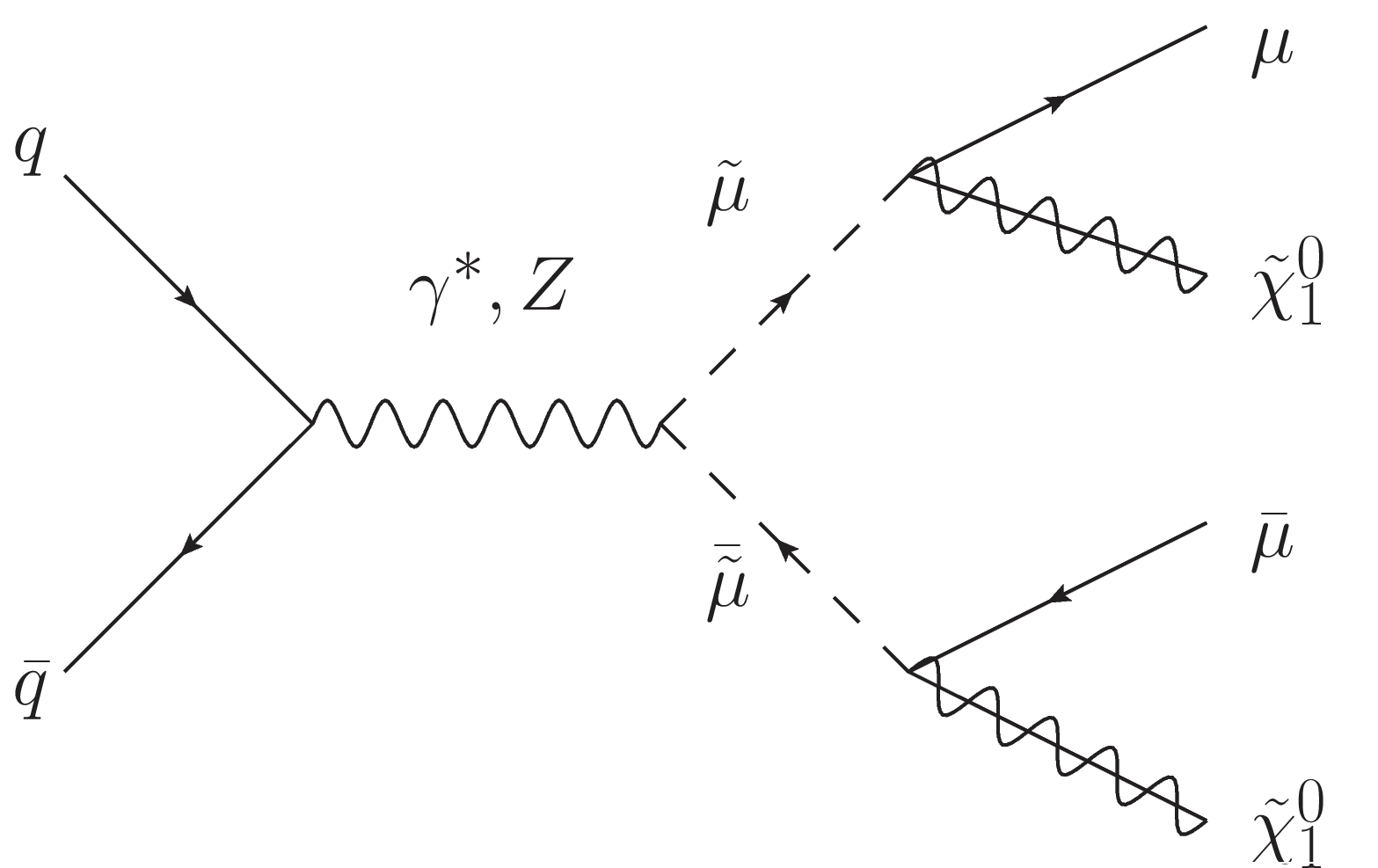}
    \caption{Feynman diagram (at LO QCD) related to the indirect production of neutralinos by smuons.}
    \label{fig:smuon-prod}
\end{figure}
\item[$\bullet$]  The second process is the direct neutralino production ($pp\rightarrow \tilde{\chi}_1^0 \tilde{\chi}_1^0$) in the s-channel and the t-channel (see Fig.~\ref{fig:neutralino-prod}). The squark exchange only contributes to a few {percent} of the production cross section. The only types of neutralino which allow such processes are $\tilde{h}_1$ and $\tilde{h}_2$. Due to the form of the couplings, a pure $\tilde{h}_1$ or $\tilde{h}_2$ state leads to the same cross section. Thus the cross section depends only on the neutralino mass. 
\begin{figure}[htbp]
    \centering
    \includegraphics[scale=0.17]{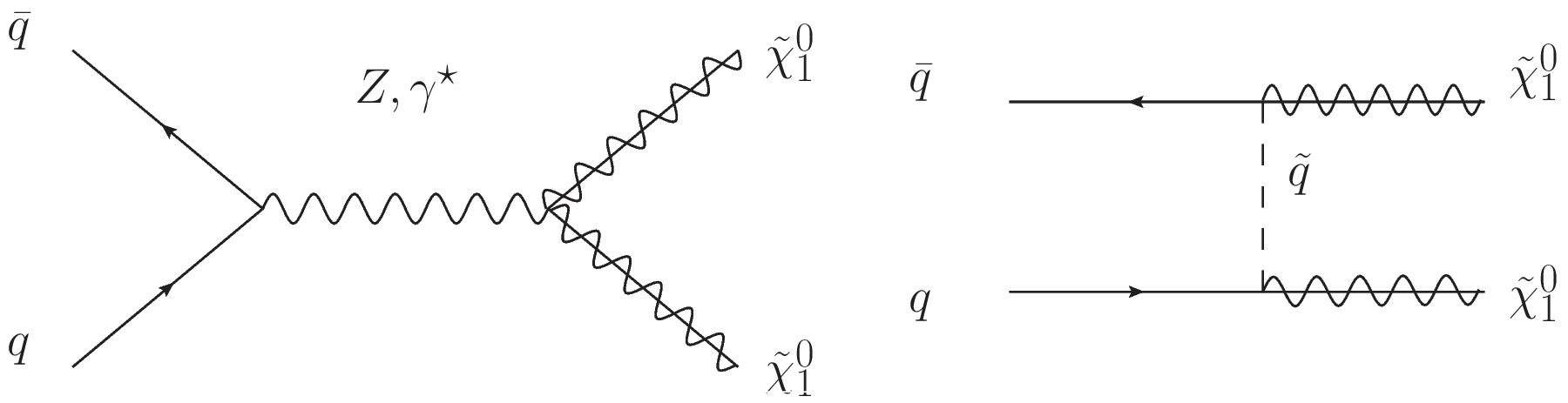}
    \caption{Feynman diagrams (at LO QCD) related to the direct production of neutralinos.}
    \label{fig:neutralino-prod}
\end{figure}
\end{itemize}

\begin{figure}
{\centering
    \includegraphics[scale=0.43]{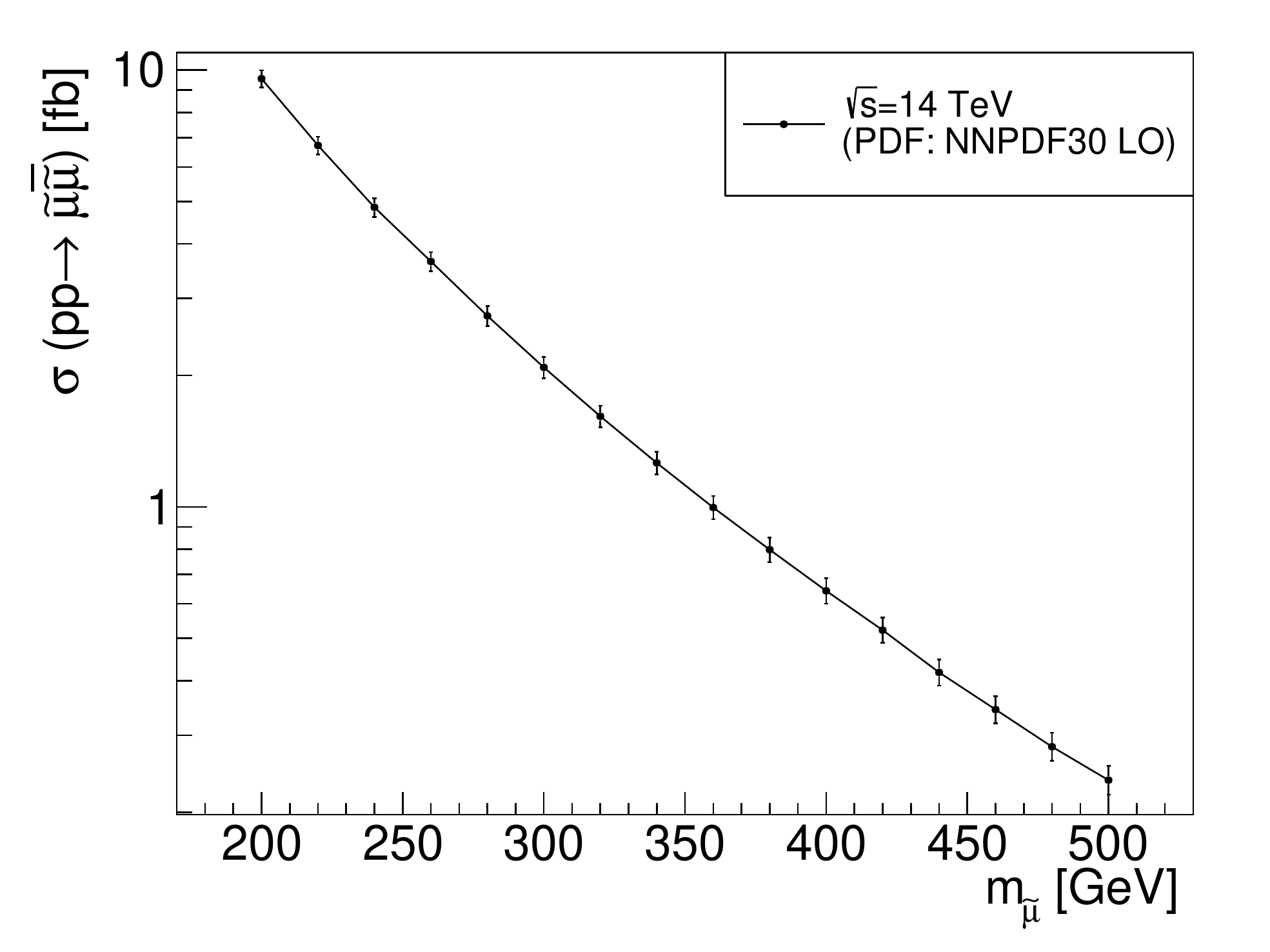}
    \caption{Cross section of the smuon production at leading order as a function of the smuon mass $m_{\tilde{\mu}}$. The uncertainties are related to PDF variations and the factorisation and renormalisation scales.}
    \label{fig:smuon-xsection-new}}
\end{figure}

\begin{figure}
    {\centering
\includegraphics[scale=0.43]{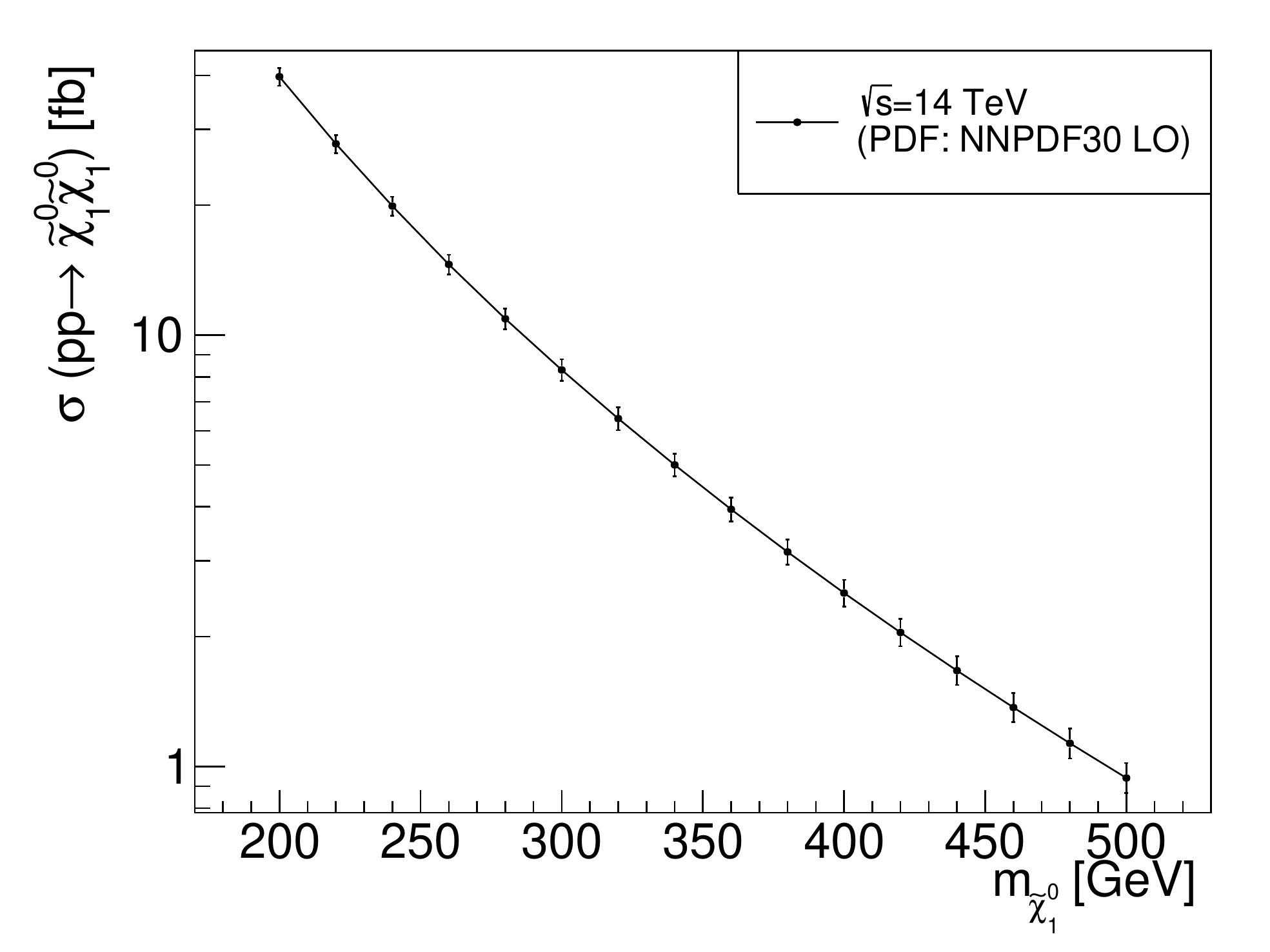}
    \caption{Cross section of the direct neutralino production at leading order as a function of the neutralino mass $m_{\tilde{\chi}_1^0}$. The uncertainties are related to PDF variations and the factorisation and renormalisation scales.}
    \label{fig:neutralino-xsection-new}}
\end{figure}

Another related signal can be investigated, the production of an LSP neutralino and an NLSP chargino leading to a higher production cross section and one lepton in the final state ($\tilde{\chi}^0_1 \tilde{\chi}^\pm_1\rightarrow \tilde{\chi}^0_1\tilde{\chi}^0_1 \ell^\pm$). However, we do not consider such signal contributions since the chargino mass and mixing matrix will provide supplementary free parameters to fix, which makes the benchmarks more complicated.
 Figures~~\ref{fig:smuon-xsection-new} and \ref{fig:neutralino-xsection-new} illustrate the dependence of the neutra\-lino-pair production cross section as a function of the smuon mass (for the bino case) and the neutralino mass (for the higgsino case). These cross sections decrease with the mass of the supersymmetric particles. A more complete analysis of a neutralino pair production at $\sqrt{s}=14\ \text{TeV}$ including QCD-corrections can be found in \cite{NLOdirectneutralino}. Concerning the smuon production, the parameter region is limited to a smuon mass of  $350~\textrm{GeV}$ in order to collect and analyse a sufficient amount of data with a Run~3 integrated luminosity of $\mathcal{L}=300~\textrm{fb}^{-1}$. A more precise theoretical Next-to-Leading Order (NLO) calculation of the cross section can be found in~\cite{Baer_1998}. The K-factor, $K=\sigma_{NLO}/ \sigma_{LO}$, for both processes is around 1.4. For the direct neutralino production, we expect to probe a neutralino mass range up to $500~\textrm{GeV}$. The associated uncertainties reflect variations on the PDF, the factorisation and the renormalisation scales at leading order.
 

\subsubsection{Neutralino decay and flight distance}

\begin{figure}[htbp]
    \centering
    \includegraphics[scale=0.44]{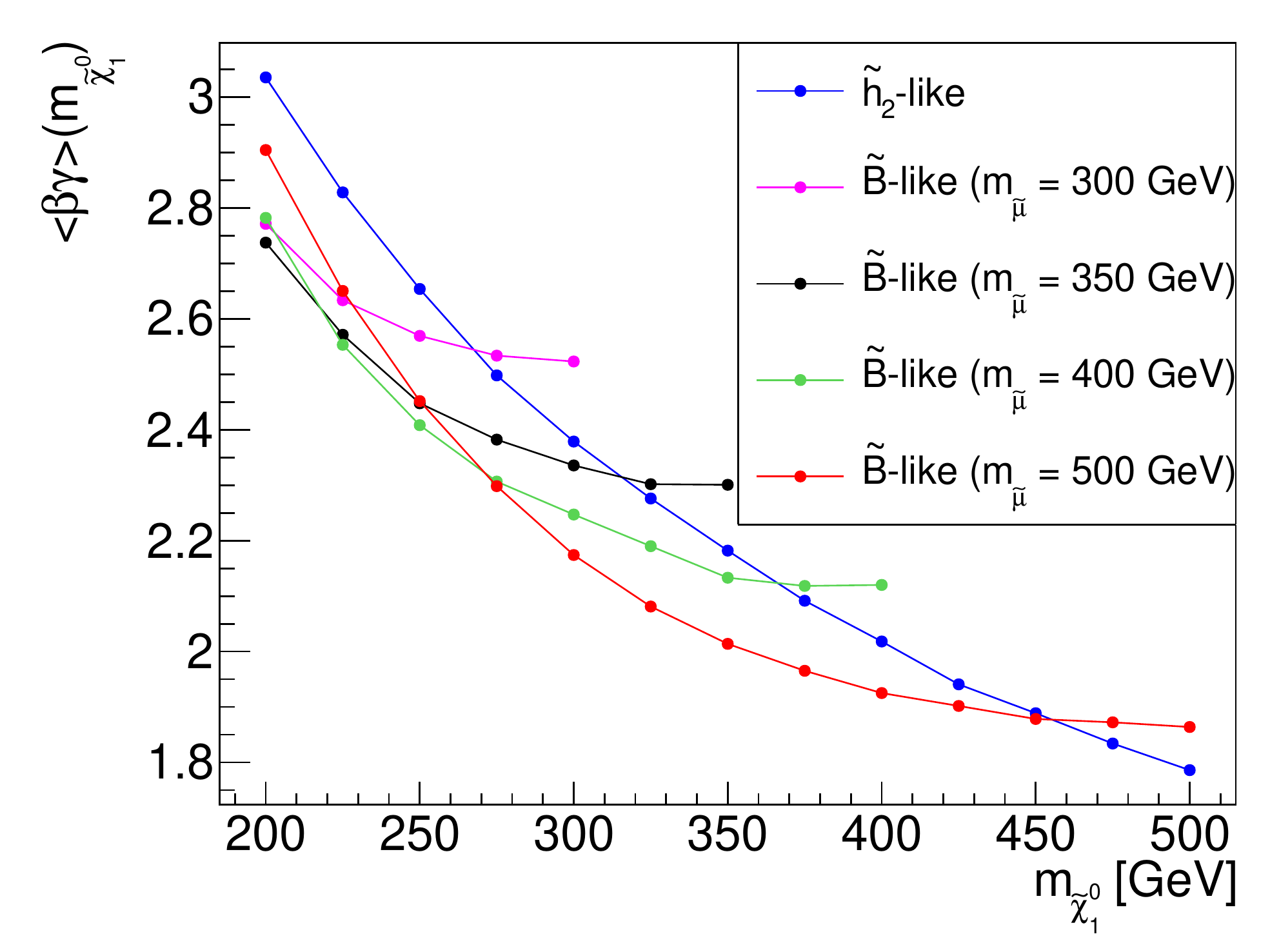}
    \caption{Mean value of the Lorentz factor $\langle \beta\gamma\rangle $ for the flying neutralino in both production cases as a function of the neutralino mass $m_{\tilde{\chi}^0_1}$. }
    \label{fig:slepton-neutralino-betagamma-new}
\end{figure}

 Together with the long-lived particle lifetime, the Lorentz factor $\beta\gamma$ of the neutralino in the two processes has been taken into account as a requirement to estimate the flight distance in the detector frame. Figure~\ref{fig:slepton-neutralino-betagamma-new} illustrates the variation of the average of the mean value of $\beta\gamma$ as a function of the mass of the neutralino. Several values for the smuon mass $m_{\tilde{\mu}}$ have been investigated for the slepton production. For $m_{\tilde{\chi}^0_1}	\gtrsim 250\ \text{GeV}$, the mean value $\langle \beta\gamma\rangle $ in the bino case is higher for a lower smuon mass $m_{\tilde{\mu}}$. It has to be noted that we have the explicit constraint $m_{\tilde{\mu}}>m_{\tilde{\chi}_1^0}$. The mean Lorentz factor $\langle \beta \gamma \rangle_{\tilde{\chi}_1^0}$ varies from 1.8 to 3.0 and so affects the flight distance $\langle d^{\text{flight}}_{\tilde{\chi}^0_1}\rangle$.  \\

\begin{figure}[htbp]
{\centering
    \includegraphics[scale=0.42]{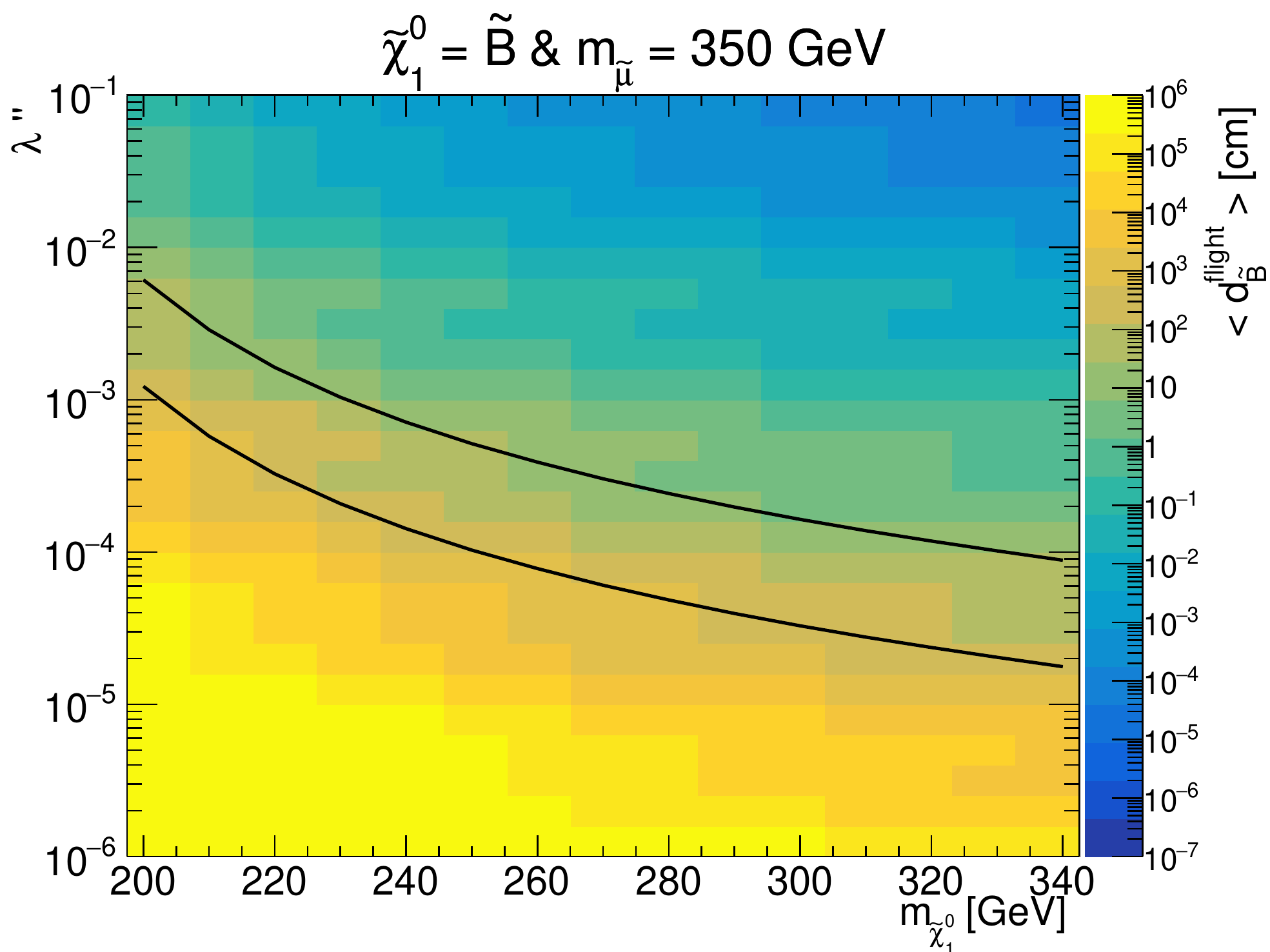}
    \caption{Average flight distance $\langle d^{\text{flight}}_{\tilde{B}}\rangle $ of the neutralino $\tilde{B}$-like in the laboratory frame as a function of $\lambda''$ and the neutralino mass $m_{\tilde{\chi}^0_1}$. We assume a smuon mass of $m_{\tilde{\mu}}=350\ \text{GeV}$. The black lines represent the geometrical limits of the tracker volume.}
    \label{fig:b-bgctau}}
\end{figure}

\begin{figure}[htbp]
    {\centering
    \includegraphics[scale=0.42]{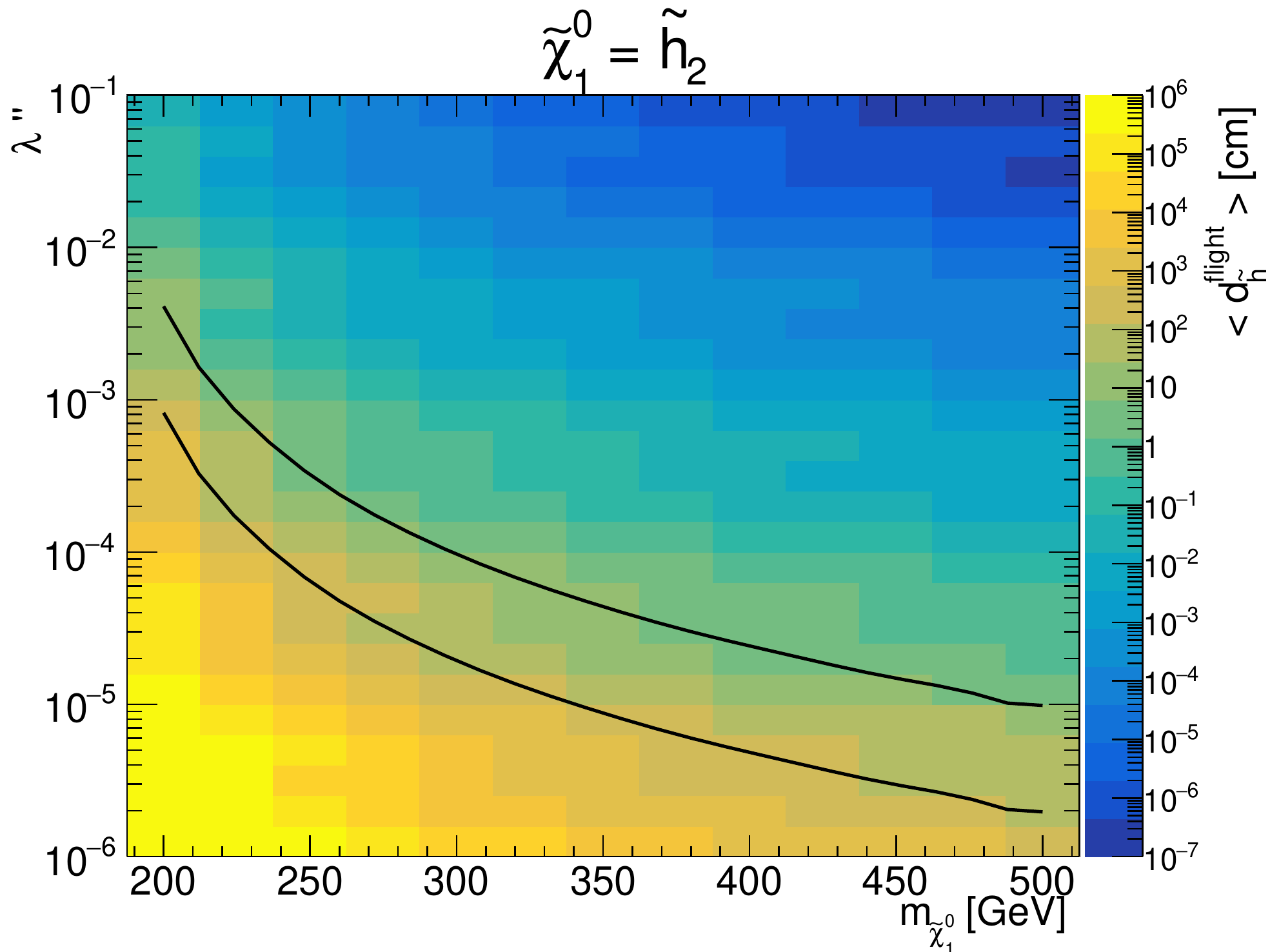}
    \caption{Average flight distance $\langle d^{\text{flight}}_{\tilde{h}}\rangle $ of the neutralino $\tilde{h}_2$-like in the laboratory frame as a function of $\lambda''$ and the neutralino mass $m_{\tilde{\chi}^0_1}$. The black lines represent the geometrical limits of the tracker volume.}
    \label{fig:h-bgctau}}
\end{figure}

The mean flight distance of the neutralino $\langle d^{\textrm{flight}}_{\tilde{\chi}_1^0} \rangle$ in the laboratory frame can be then computed for different values of the R-parity violated coupling $\lambda''$ and neutralino masses, as shown in Figs.~\ref{fig:b-bgctau} and \ref{fig:h-bgctau}, where we fix $m_{\tilde{\mu}}=350\ \text{GeV}$ in the case of the smuon pair production. The dependency in $m_{\tilde{\mu}}$ comes only from the kinetic factor $\langle \beta\gamma \rangle$. Thus, a lower mass $m_{\tilde{\mu}}$ does not highly modify the flight distance, see Fig.~\ref{fig:slepton-neutralino-betagamma-new}. As expected, the flight distance is inversely proportional to $\lambda''^2$ and, for $\lambda''$ fixed, it {decreases} with the mass of the neutralino. We also notice that an RPV coupling higher than $10^{-2}$ is not allowed to avoid a long-lived particle from decaying outside the tracker volume. Nonetheless, all neutralino masses in the range $[200,475]\ \text{GeV}$ can be scrutinised with ATLAS or CMS detectors at LHC Run~3.


\subsubsection{Benchmarks definition}\label{rpv_benchmark}

\begin{table*}[!htbp]
\begin{tabular*}{\textwidth}{@{\extracolsep{\fill}}lcccccccc@{}}
\hline
Name & $\tilde{B}_{0.2}^{10}$ & $\tilde{B}_{0.2}^{30}$ & $\tilde{B}_{0.2}^{50}$  & $\tilde{B}_{0.2}^{70}$ & $\tilde{B}_{0.3}^{10}$ & $\tilde{B}_{0.3}^{30}$ & $\tilde{B}_{0.3}^{50}$  & $\tilde{B}_{0.3}^{70}$ \\
\hline
$\langle d^\textrm{flight}_{\tilde{B}}\rangle~[\textrm{cm}]$ & 10 & 30 & 50 & 70 & 10 & 30 & 50 & 70 \\ 
$m_{\tilde{\chi}_1^0}~[\textrm{TeV}]$ & 0.20 & 0.20 & 0.20 & 0.20 & 0.30 & 0.30 & 0.30 & 0.30 \\
\hline
$m_{\tilde{t}}~[\textrm{TeV}]$ & 1.00 & 1.00 & 1.00 & 1.00 & 1.00 & 1.00 & 1.00 & 1.00 \\
$m_{\tilde{\mu}}~[\textrm{TeV}]$ & 0.35 & 0.35 & 0.35 & 0.35 & 0.35 & 0.35 & 0.35 & 0.35 \\
$\lambda''~[10^{-6}]$ & $6407$ & $3699$ & $2865$ & $2422$ & $158$ & $91$ & $71$ & $60$  \\
\hline
\end{tabular*}
\caption{Definition of the eight benchmarks for the smuon production case $pp\rightarrow \tilde{\mu}\bar{\tilde{\mu}}$.}
\label{tab:bino_bcmk}
\end{table*}
\begin{table*}[!htbp]
\begin{tabular*}{\textwidth}{@{\extracolsep{\fill}}lcccccccc@{}}
\hline
Name & $\tilde{h}_{0.2}^{10}$ & $\tilde{h}_{0.2}^{30}$ & $\tilde{h}_{0.2}^{50}$  & $\tilde{h}_{0.2}^{70}$ & $\tilde{h}_{0.4}^{10}$ & $\tilde{h}_{0.4}^{30}$ & $\tilde{h}_{0.4}^{50}$  & $\tilde{h}_{0.4}^{70}$ \\
\hline
$\langle d^\textrm{flight}_{\tilde{h}}\rangle~[\textrm{cm}]$ & 10 & 30 & 50 & 70 & 10 & 30 & 50 & 70 \\ 
$m_{\tilde{\chi}^0_1}~[\textrm{TeV}]$ & 0.20 & 0.20 & 0.20 & 0.20 & 0.40 & 0.40 & 0.40 & 0.40 \\
\hline
$m_{\tilde{t}}~[\textrm{TeV}]$ & 1.00 & 1.00 & 1.00 & 1.00 & 1.00 & 1.00 & 1.00 & 1.00 \\
$\lambda''~[10^{-6}]$ & $2599$ & $1501$ & $1162$ & $983$ & $15$ & $8.8$ & $6.8$ & $5.8$ \\
\hline
\end{tabular*}
\caption{Definition of the eight benchmarks for the direct neutralino production $pp\rightarrow \tilde{\chi}^0_1\tilde{\chi}^0_1$.}
\label{tab:higgsino_bcmk}
\end{table*}

From the previous results, we define benchmarks by making a compromise between a large cross section and a relevant flight distance for the neutralino. The cross section in Fig.~\ref{fig:smuon-xsection-new} limits the choice of the smuon mass. In order to get a sufficient amount of data, we fix the mass of the smuon to $350\ \text{GeV}$, leading to approximately 300 events for $\mathcal{L}_{int.}=300\ \text{fb}^{-1}$. A choice of $\Delta m = m_{\tilde{\mu}} - m_{\tilde{\chi}^0_1}$ will mainly impact the particle's transverse momentum in the final state, whereas $m_{\tilde{\mu}}$ influences the global observables such as THT. Note that the cross section of the direct neutralino production is sufficiently large to not constrain the choice of $m_{\tilde{\chi}^0_1}$ in the range $[200,500]\ \text{GeV}$. Two values of neutralino mass are chosen for both processes, and four mean flight distances (10, 30, 50 and $70\ \text{cm}$) are considered to cover all the tracker volume geometry. A neutralino mass of $200$ and $400\ \text{GeV}$ has been taken in the direct neutralino production. For the slepton channel, we assume $m_{\tilde{\chi}^0_1}=200$ and $300\ \text{GeV}$. The RPV coupling $\lambda''$ is fixed through the choice of the two mentioned parameters and varies between $10^{-3}$ and $10^{-6}$. We summarise the definition of these benchmarks in Table~\ref{tab:bino_bcmk} and Table~\ref{tab:higgsino_bcmk}.\\


No exclusive analysis has yet been designed for the signal studied here. Indeed related signals have already been investigated by the CMS and ATLAS collaborations: some processes including R-Parity conserved (RPC) slepton and direct neutralino production (see \cite{RPCslepton1,RPCslepton2,RPCslepton3,RPCslepton4,RPCslepton5,RPCslepton6,RPCslepton7} with exclusion from $m_{\tilde{\mu}}>200\ \text{GeV}$ to $m_{\tilde{\mu}}>700\ \text{GeV}$ with massless neutralino by usually considering pure left or right smuon states), and other RPV decays (see \cite{RPVprompt1,RPVprompt2,RPVprompt3,RPVprompt4,RPVprompt5,RPVprompt6} for prompt and \cite{PhysRevD.104.052011,RPVllp1,RPVllp2, RPVllp3,RPVllp4} for non-prompt decays) with constraints like $m_{\tilde{\mu}}>1.2\ \text{TeV}$, $m_{\tilde{\chi}^0_1}$ excluded between 200 and 300 GeV or other limits on RPV couplings, gluino and stop masses. It can be noticed that another analysis~\cite{PhysRevD.102.032006} achieved by the ATLAS collaboration has similarities with the sleptons case. Indeed, the analysis deals the production of non-prompt top quark by a different RPV couplings and takes advantage of the presence of two non-prompt muons in the final state.

\subsubsection{{Event} kinematics}\label{rpv_kinematics}

\begin{figure*}[!htbp]%
\begin{minipage}[t]{0.45\linewidth}
    \centering
    \includegraphics[scale=0.4]{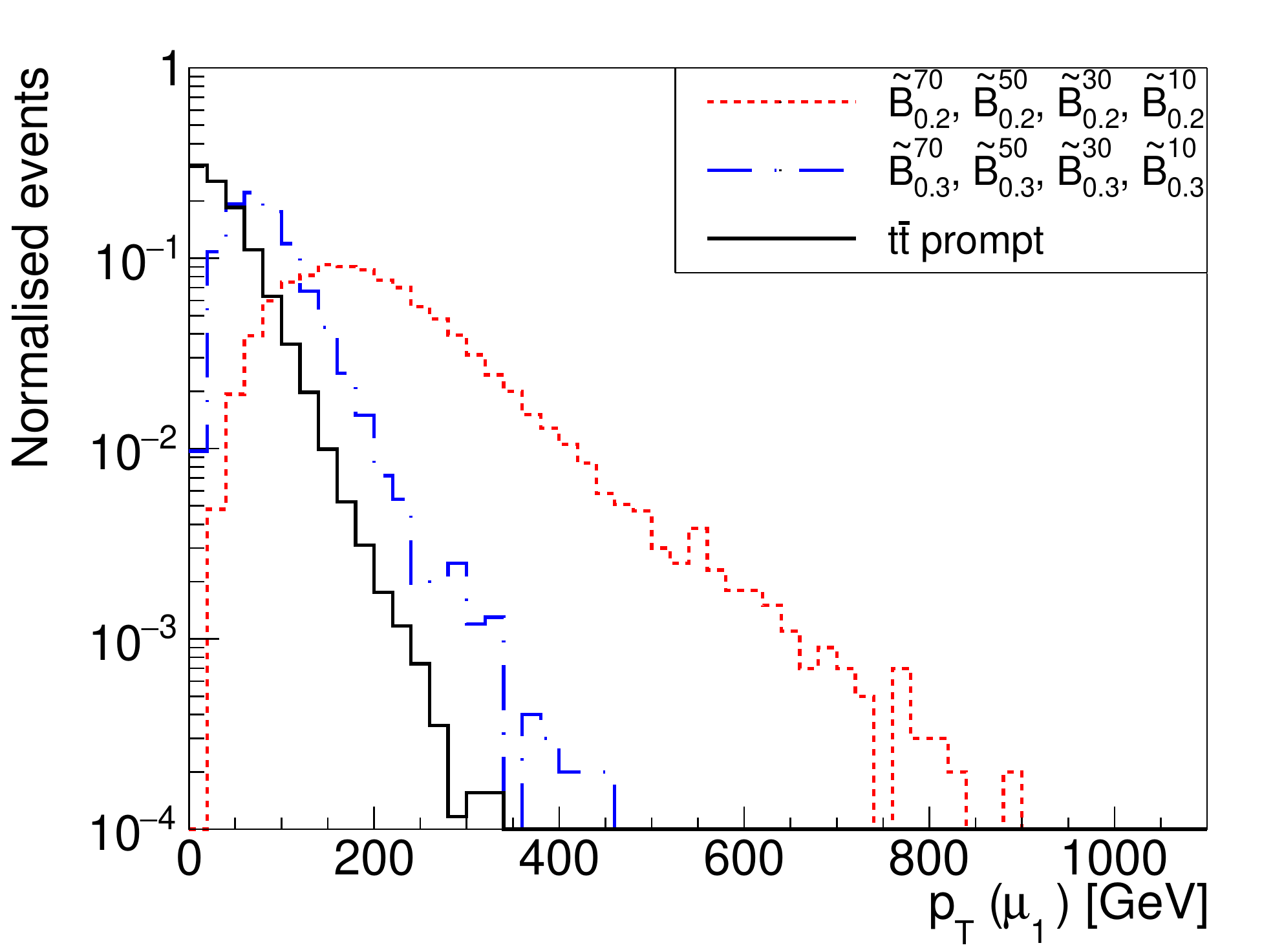}
    \caption{Transverse momentum $p_T$ of the {leading} muon of the event in the $\tilde{B}$-like case.}
    \label{fig:bino_ptmu}
\end{minipage}
\hspace*{1cm}
\begin{minipage}[t]{0.45\linewidth}
    \centering
    \includegraphics[scale=0.4]{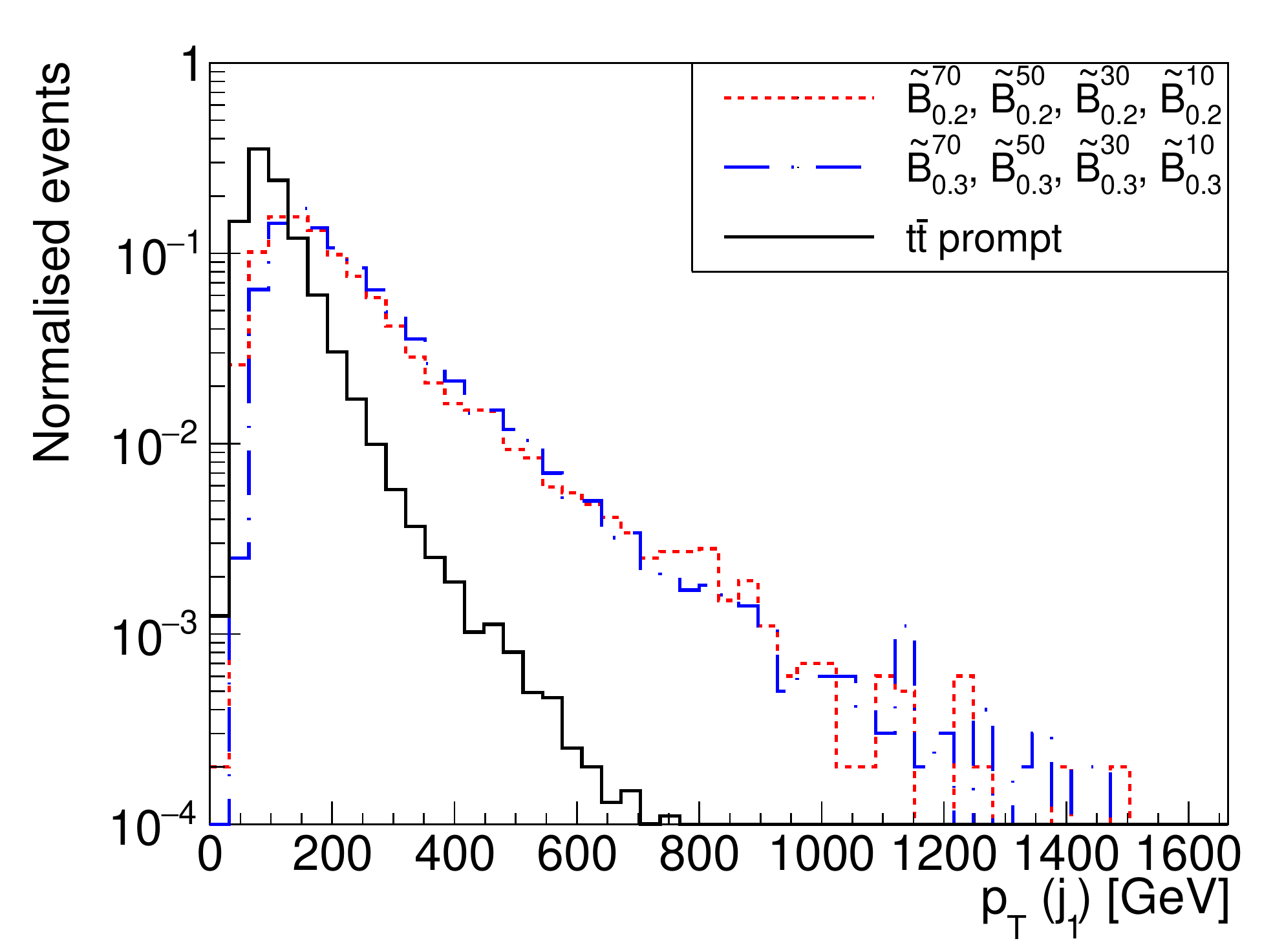}
    \caption{Transverse momentum $p_T$ of the {leading} jet of the event in the $\tilde{B}$-like case.}
    \label{fig:bino_ptj}
\end{minipage}
\begin{minipage}[t]{0.45\linewidth}
    \centering
    \includegraphics[scale=0.4]{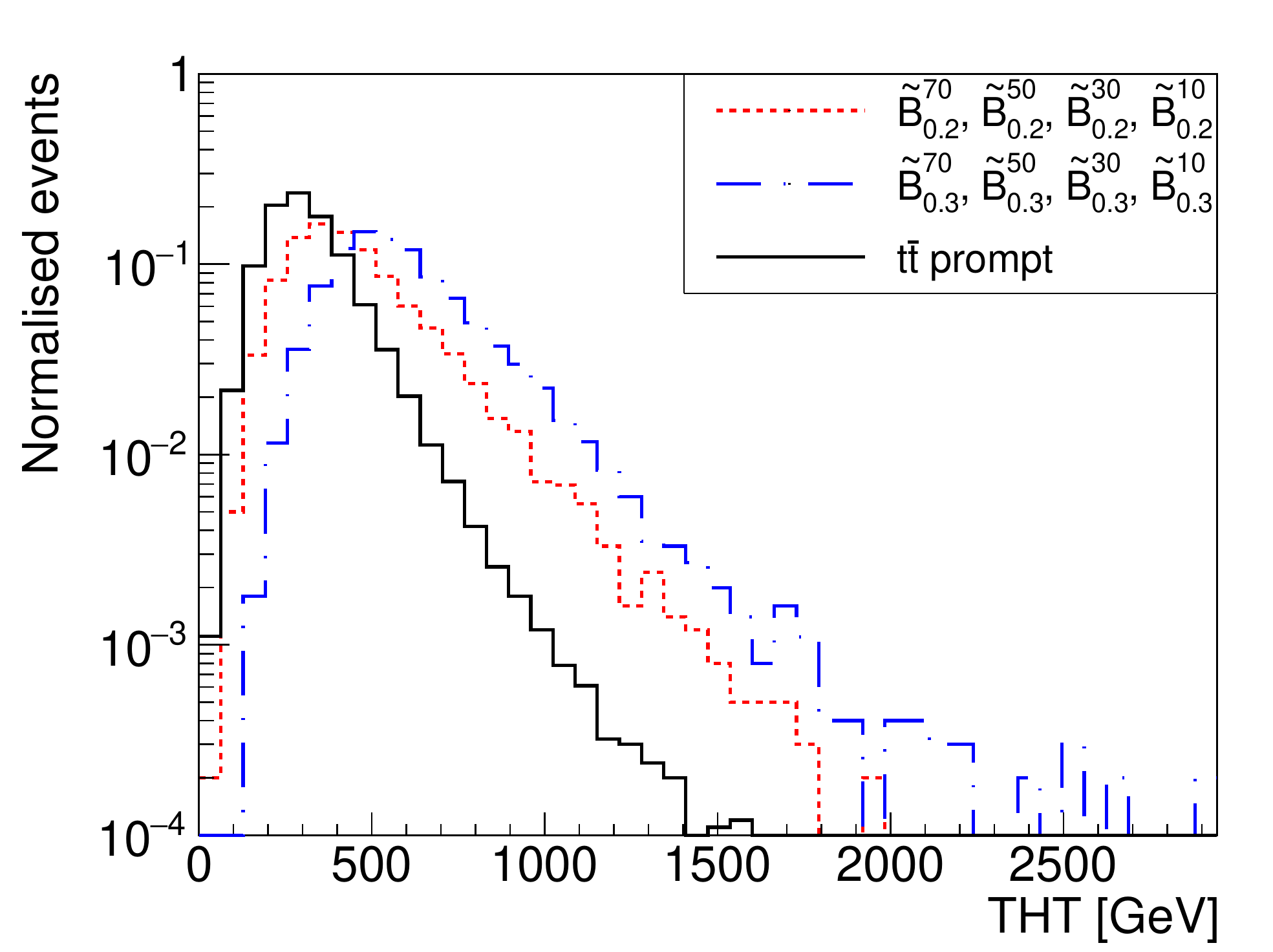}
    \caption{Total transverse hadronic energy THT in the $\tilde{B}$-like case.}
    \label{fig:bino_THT}
\end{minipage}
\hspace*{1cm}
\begin{minipage}[t]{0.45\linewidth}
    \centering
    \includegraphics[scale=0.4]{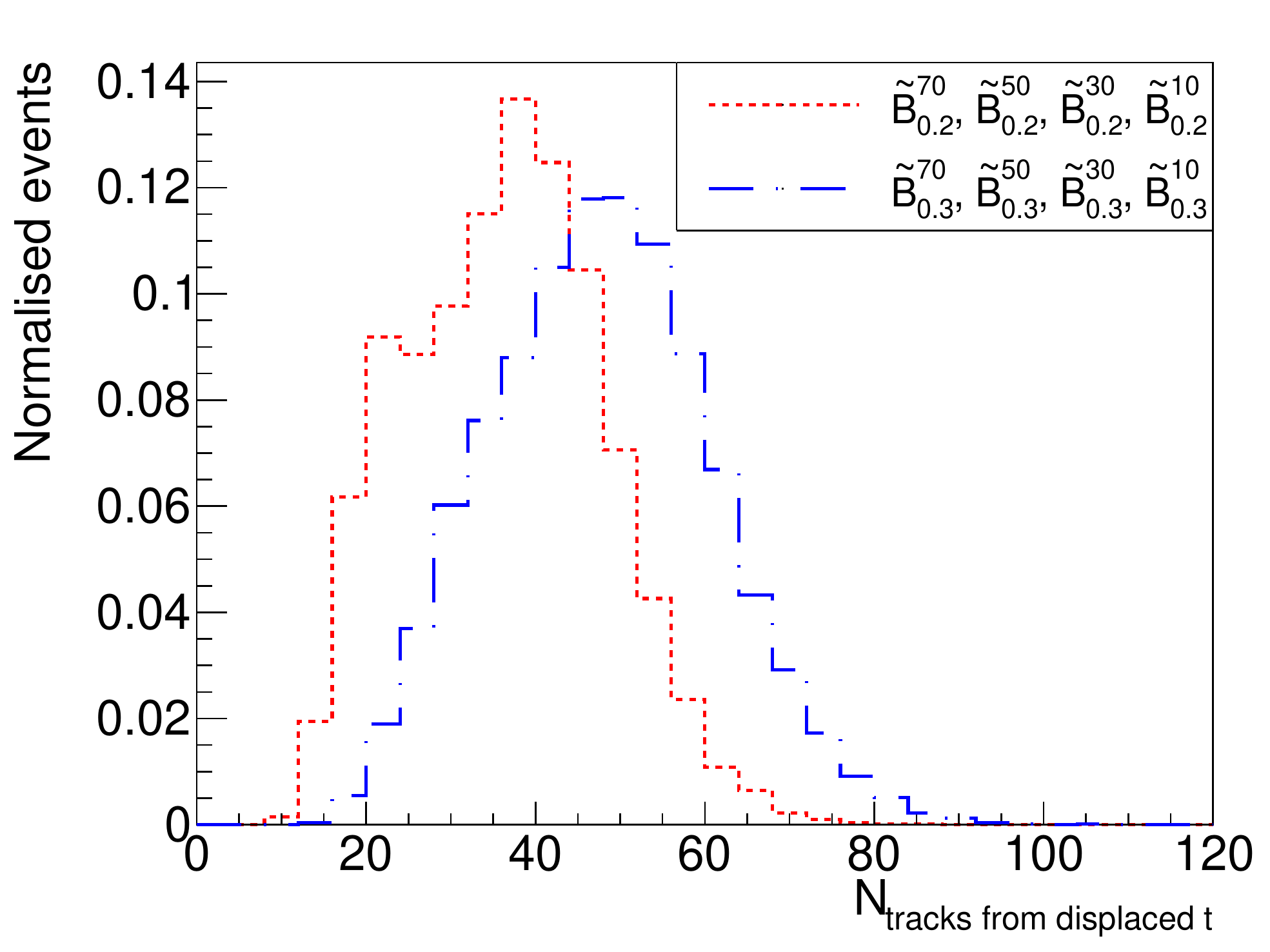}
    \caption{Track multiplicity associated with displaced vertices in the $\tilde{B}$-like case.}
    \label{fig:bino_ntrack}
\end{minipage}
\begin{minipage}[t]{0.45\linewidth}
    \centering
\includegraphics[scale=0.4]{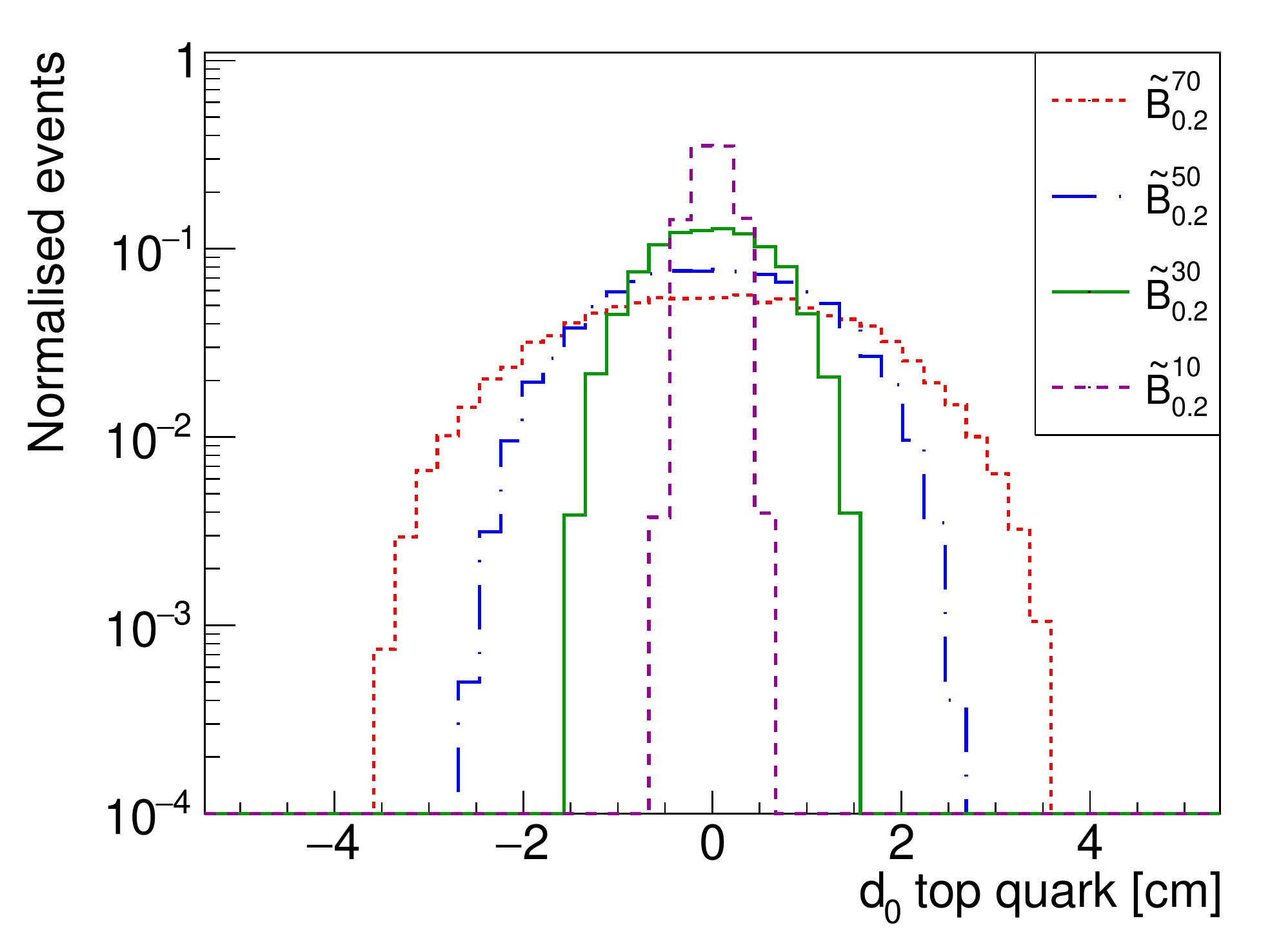}
    \caption{Transverse impact parameter $d_0$ of the top quarks in the $\tilde{B}$-like case with $m_{\tilde{\chi}^0_1}=200\ \text{GeV}$.}
    \label{fig:bino_d0_200}
\end{minipage}
\hspace*{1.4cm}
\begin{minipage}[t]{0.45\linewidth}
    \centering
    \includegraphics[scale=0.4]{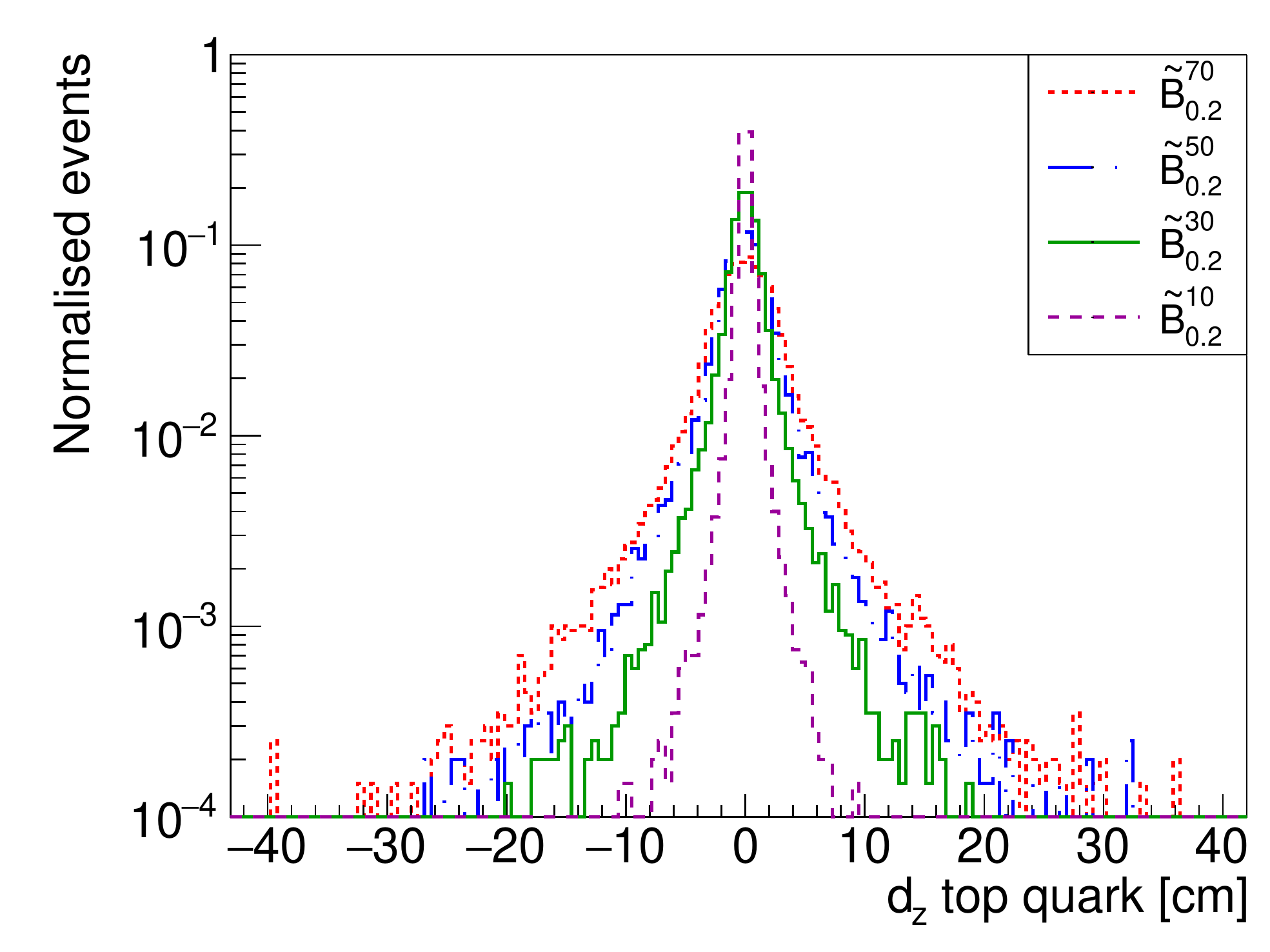}
    \caption{Longitudinal impact parameter $d_z$ of the top quarks in the $\tilde{B}$-like case with $m_{\tilde{\chi}^0_1}=200\ \text{GeV}$.}
    \label{fig:bino_dz_200}
\end{minipage}
\end{figure*}

\begin{figure*}[htbp]
\begin{minipage}[t]{0.45\linewidth}
    \centering
\includegraphics[scale=0.4]{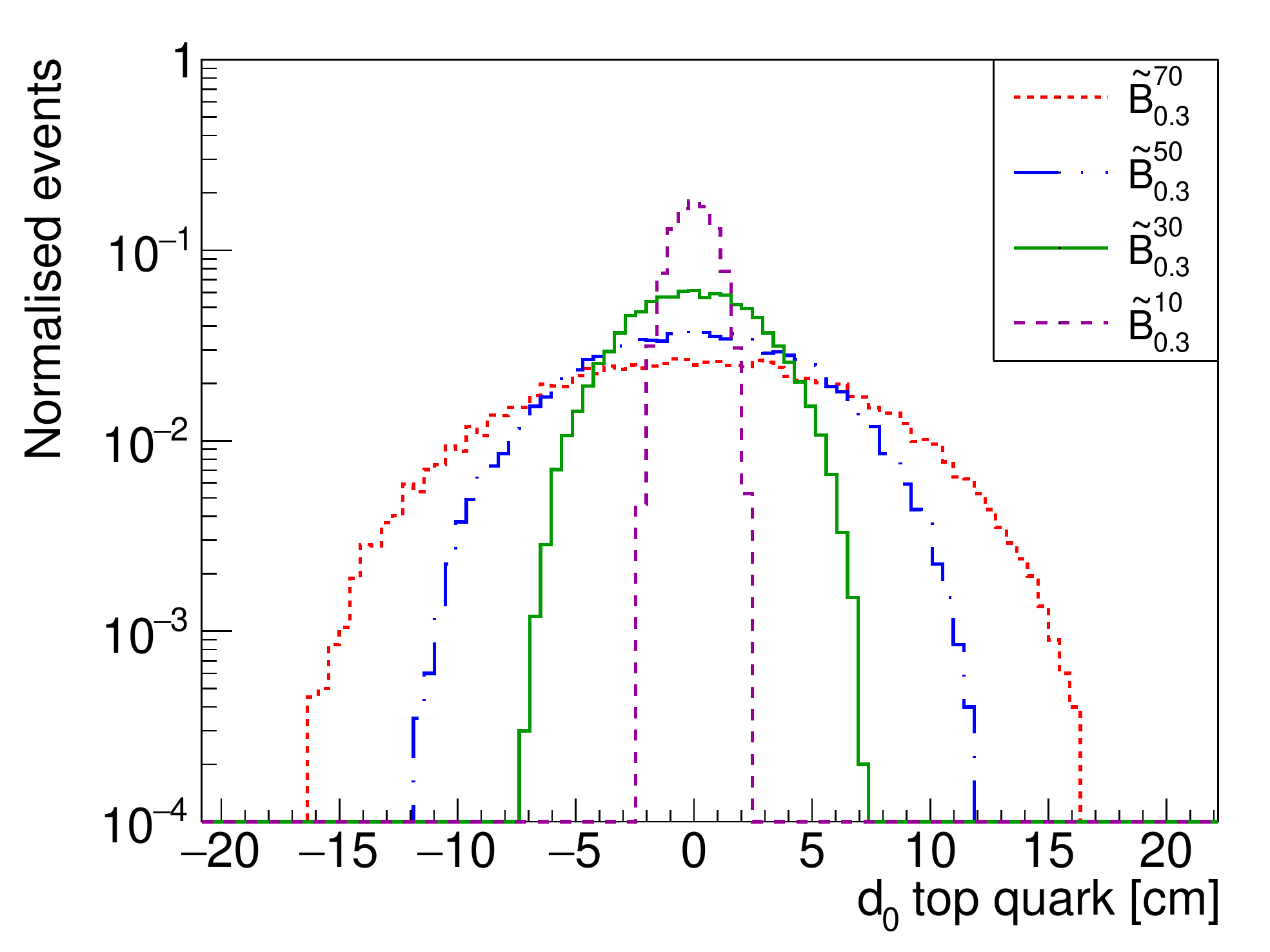}
    \caption{Transverse impact parameter $d_0$ of the top quarks in the $\tilde{B}$-like case with $m_{\tilde{\chi}^0_1}=300\ \text{GeV}$.}
    \label{fig:bino_d0_300}
\end{minipage}
\hspace*{1.4cm}
\begin{minipage}[t]{0.45\linewidth}
    \centering
    \includegraphics[scale=0.4]{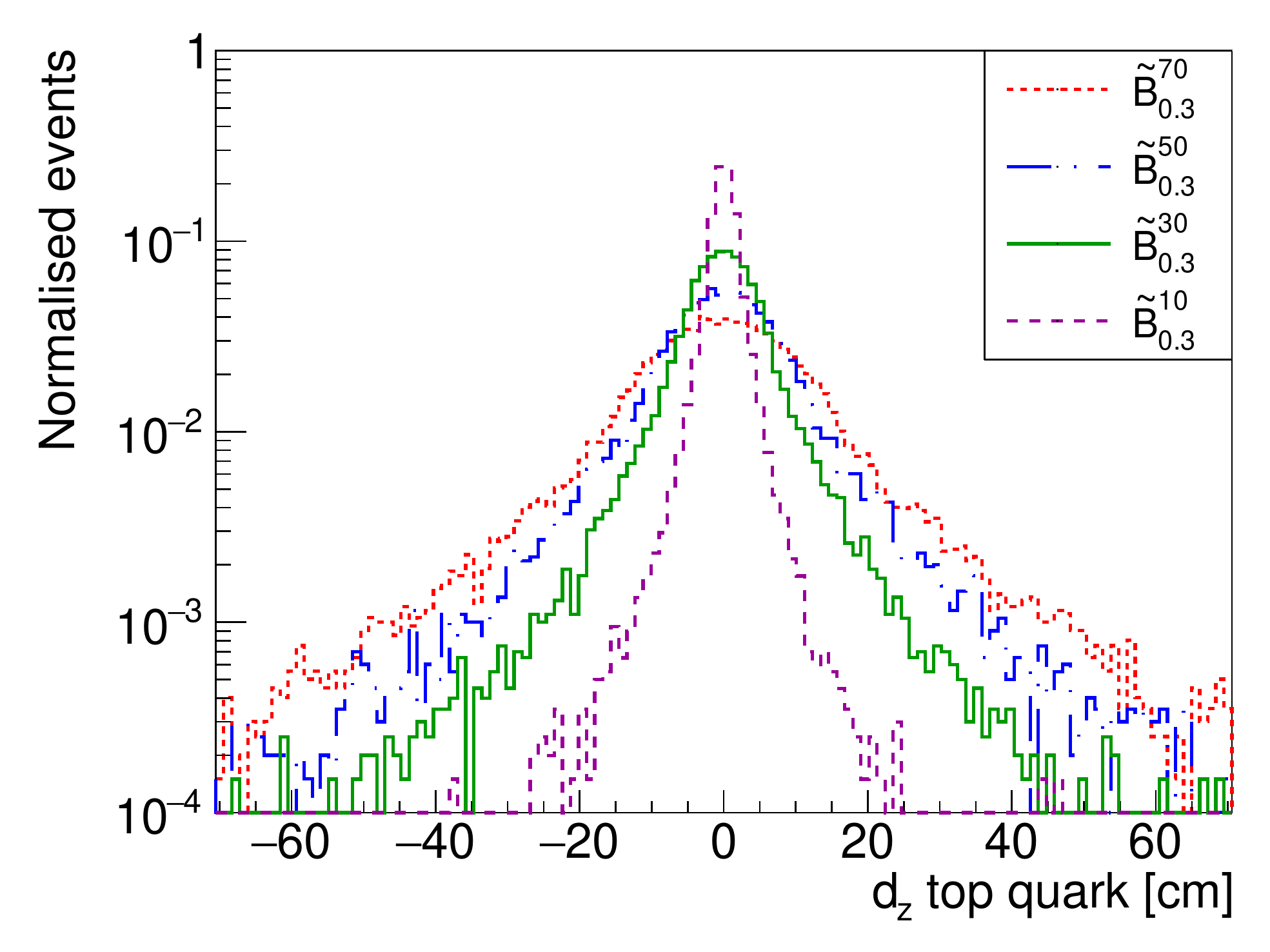}
    \caption{Longitudinal impact parameter $d_z$ of the top quarks in the $\tilde{B}$-like case with $m_{\tilde{\chi}^0_1}=300\ \text{GeV}$.}
    \label{fig:bino_dz_300}
\end{minipage}
\begin{minipage}[t]{0.45\linewidth}
    \centering
\includegraphics[scale=0.4]{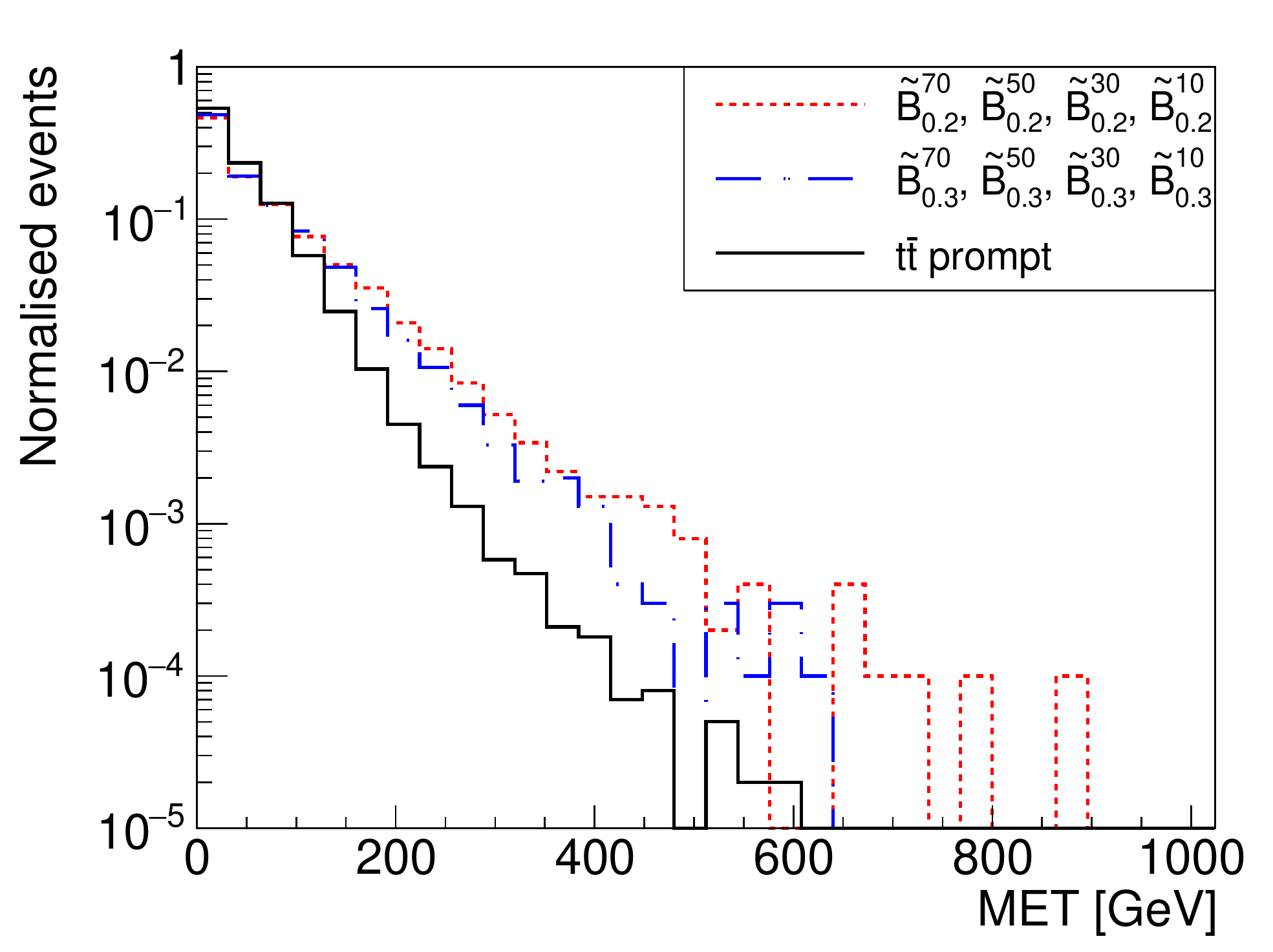}
    \caption{Missing transverse energy MET in the $\tilde{B}$-like case.}
    \label{fig:bino_met}
\end{minipage}
\hspace*{1.4cm}
\begin{minipage}[t]{0.45\linewidth}
    \centering
    \includegraphics[scale=0.4]{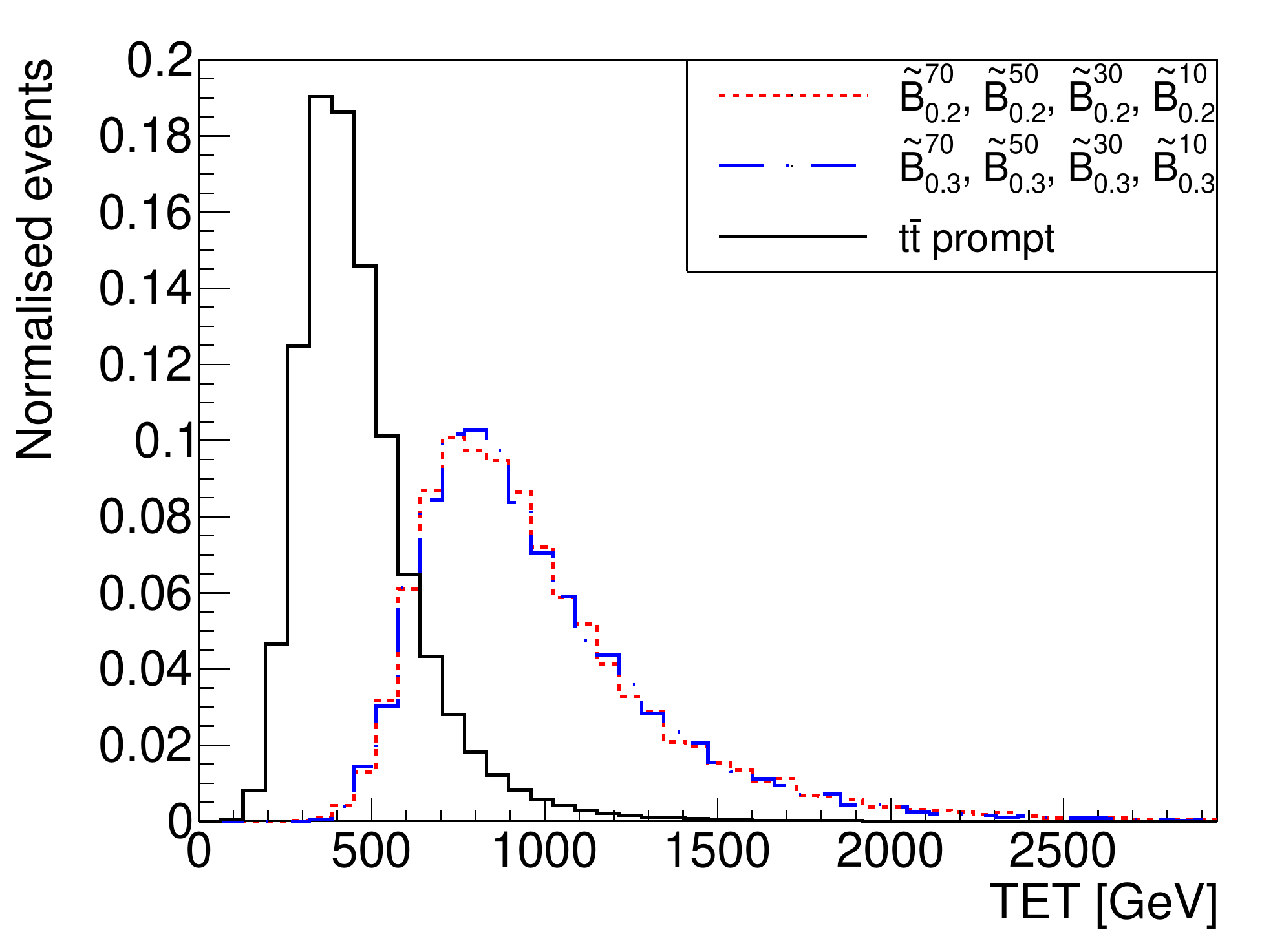}
    \caption{Total transverse energy TET in the $\tilde{B}$-like case.}
    \label{fig:bino_tet}
\end{minipage}
\begin{minipage}[t]{0.45\linewidth}
    \centering
\includegraphics[scale=0.4]{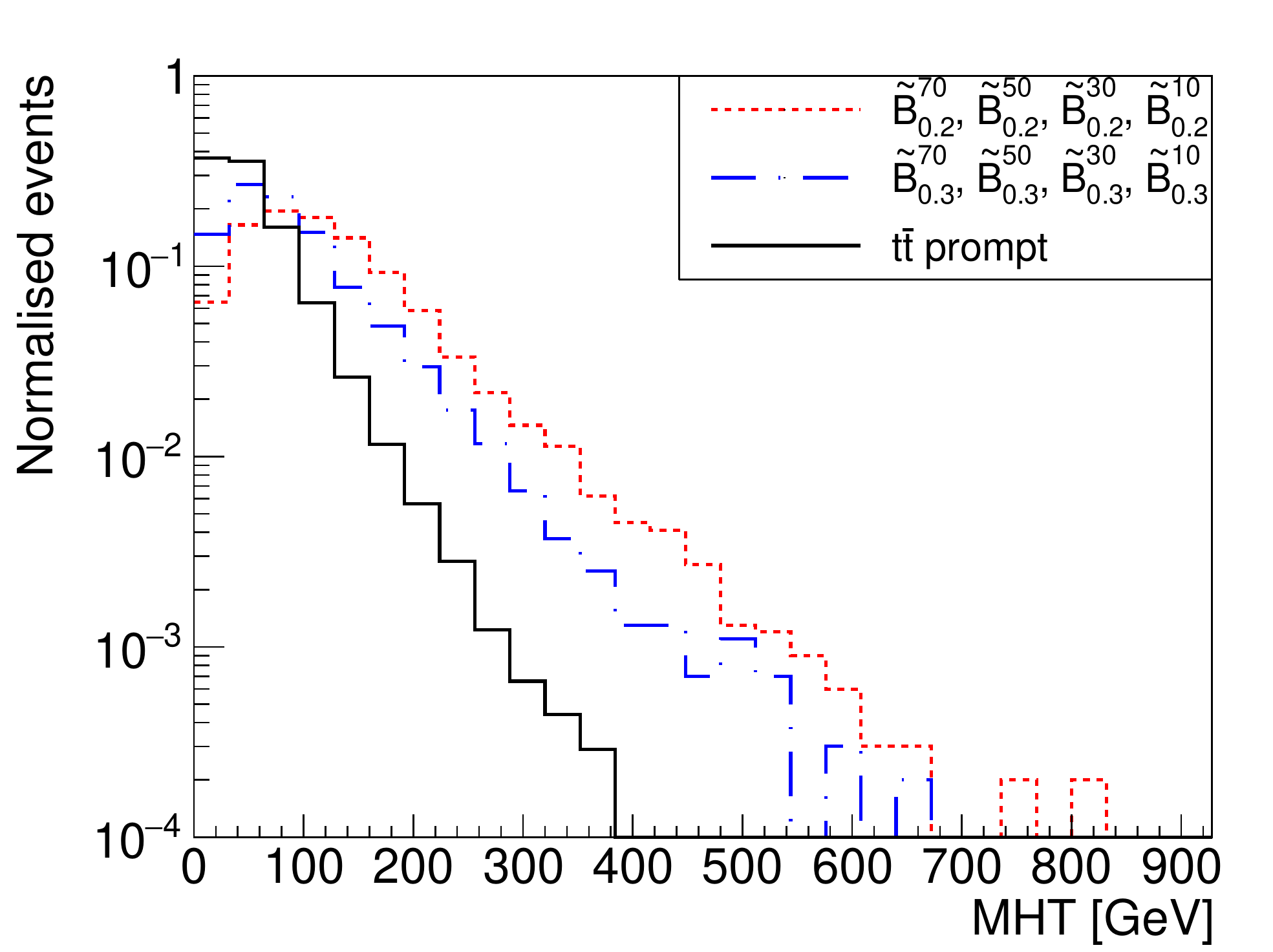}
    \caption{Missing transverse hadronic energy MHT in the $\tilde{B}$-like case.}
    \label{fig:bino_mht}
\end{minipage}
\hspace*{1.4cm}
\begin{minipage}[t]{0.45\linewidth}
    \centering
    \includegraphics[scale=0.4]{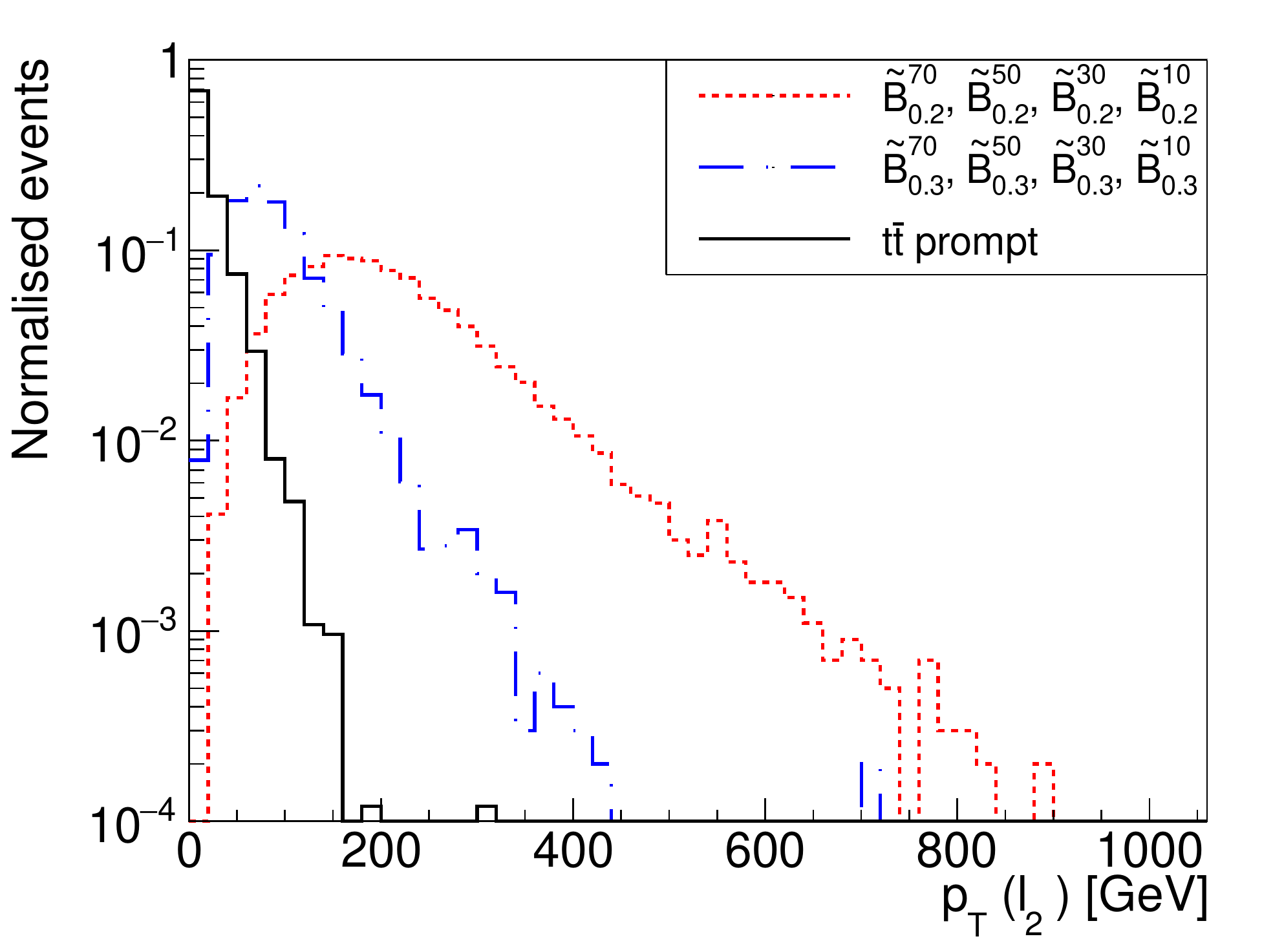}
    \caption{Transverse momentum $p_{T}$ of the second leading jet in the $\tilde{B}$-like case.}
    \label{fig:bino_ptj2}
\end{minipage}
\end{figure*}

\begin{figure*}[htbp]
\begin{minipage}[t]{0.45\linewidth}
    \centering
\includegraphics[scale=0.4]{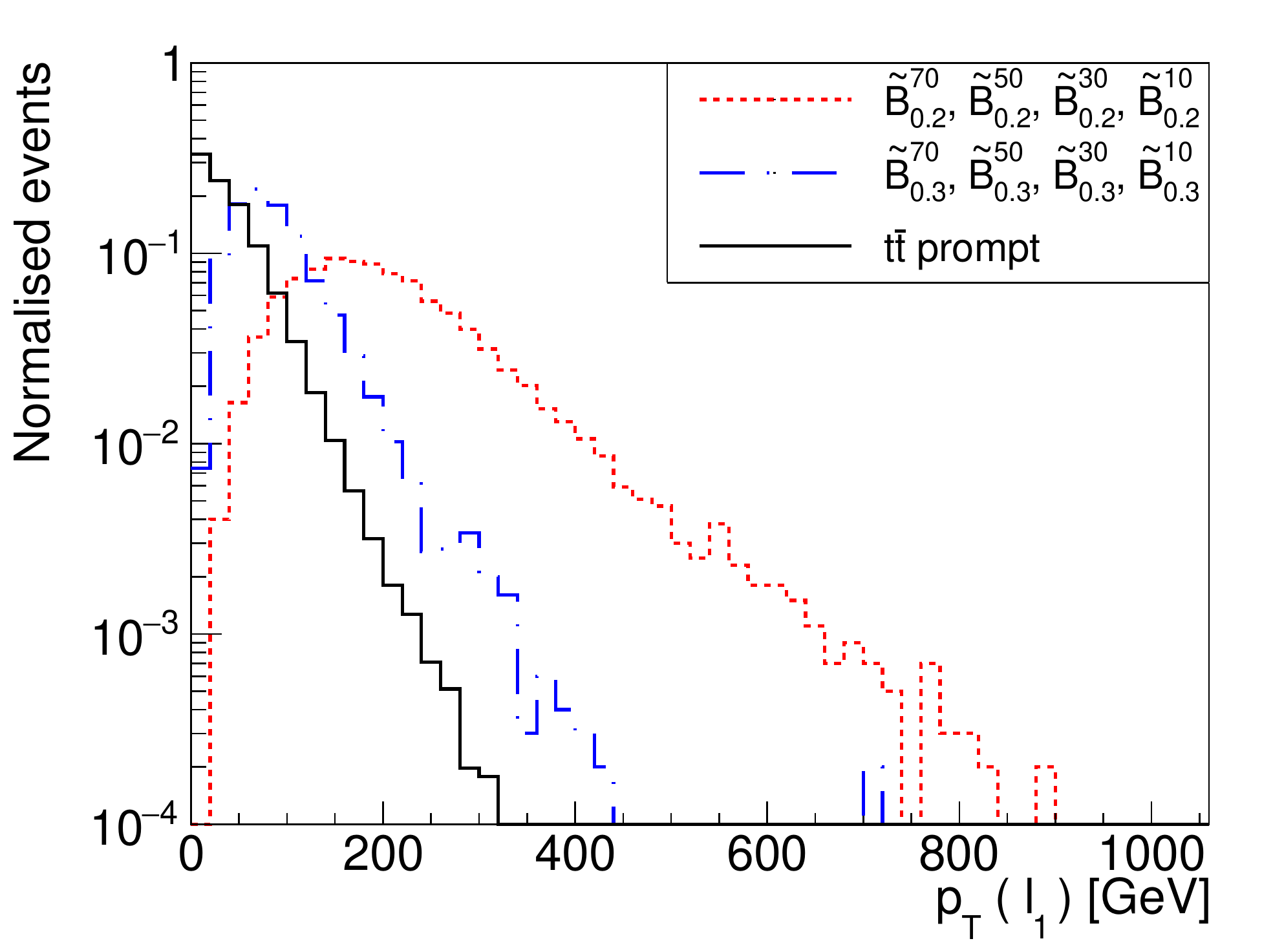}
    \caption{Transverse momentum of the leading lepton in the $\tilde{B}$-like case.}
    \label{fig:bino_ptl1}
\end{minipage}
\hspace*{1.4cm}
\begin{minipage}[t]{0.45\linewidth}
    \centering
    \includegraphics[scale=0.4]{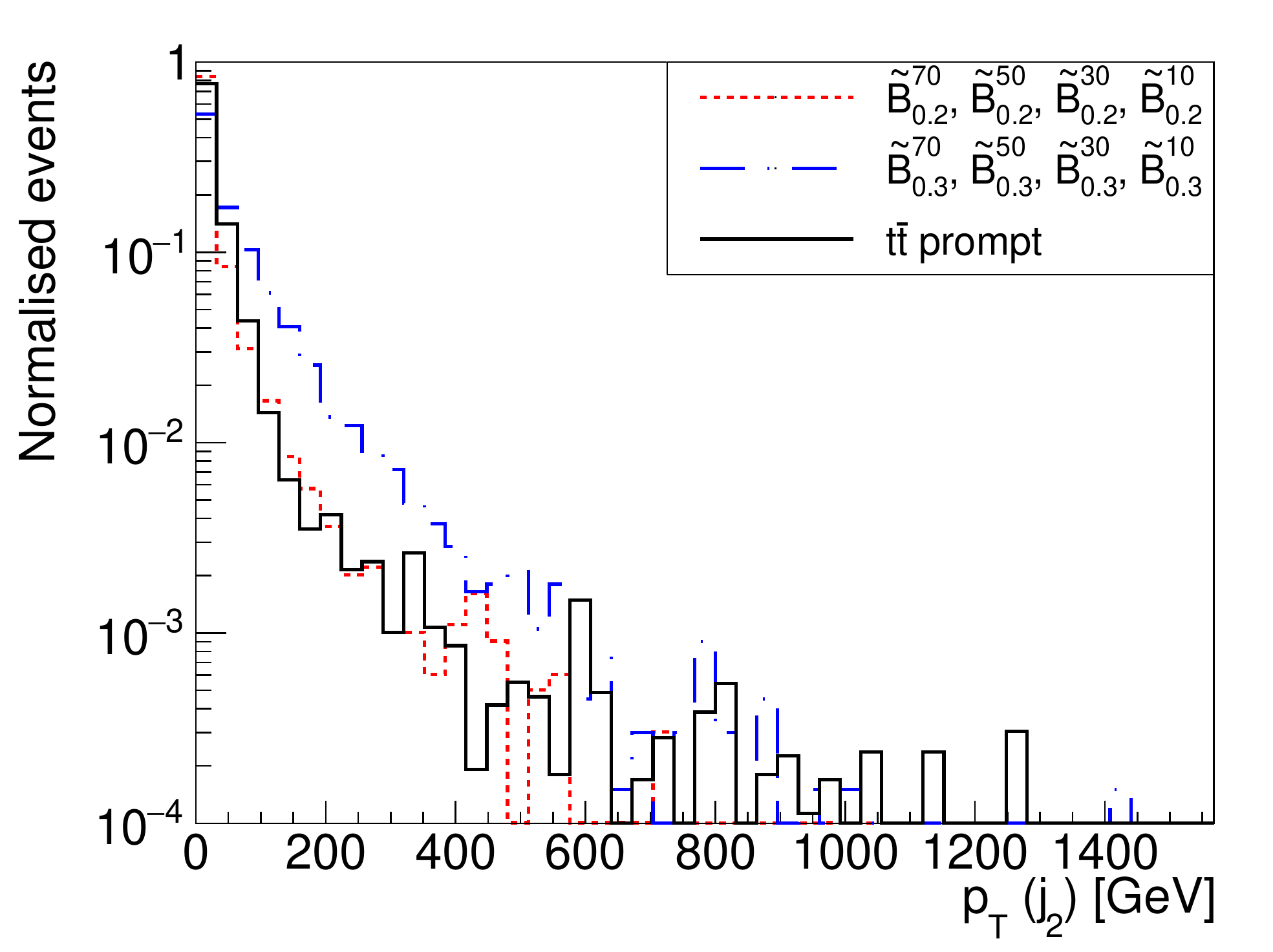}
    \caption{Transverse momentum of the second leading lepton in the $\tilde{B}$-like case.}
    \label{fig:bino_ptl2}
\end{minipage}
\begin{minipage}[t]{0.45\linewidth}
    \centering
\includegraphics[scale=0.4]{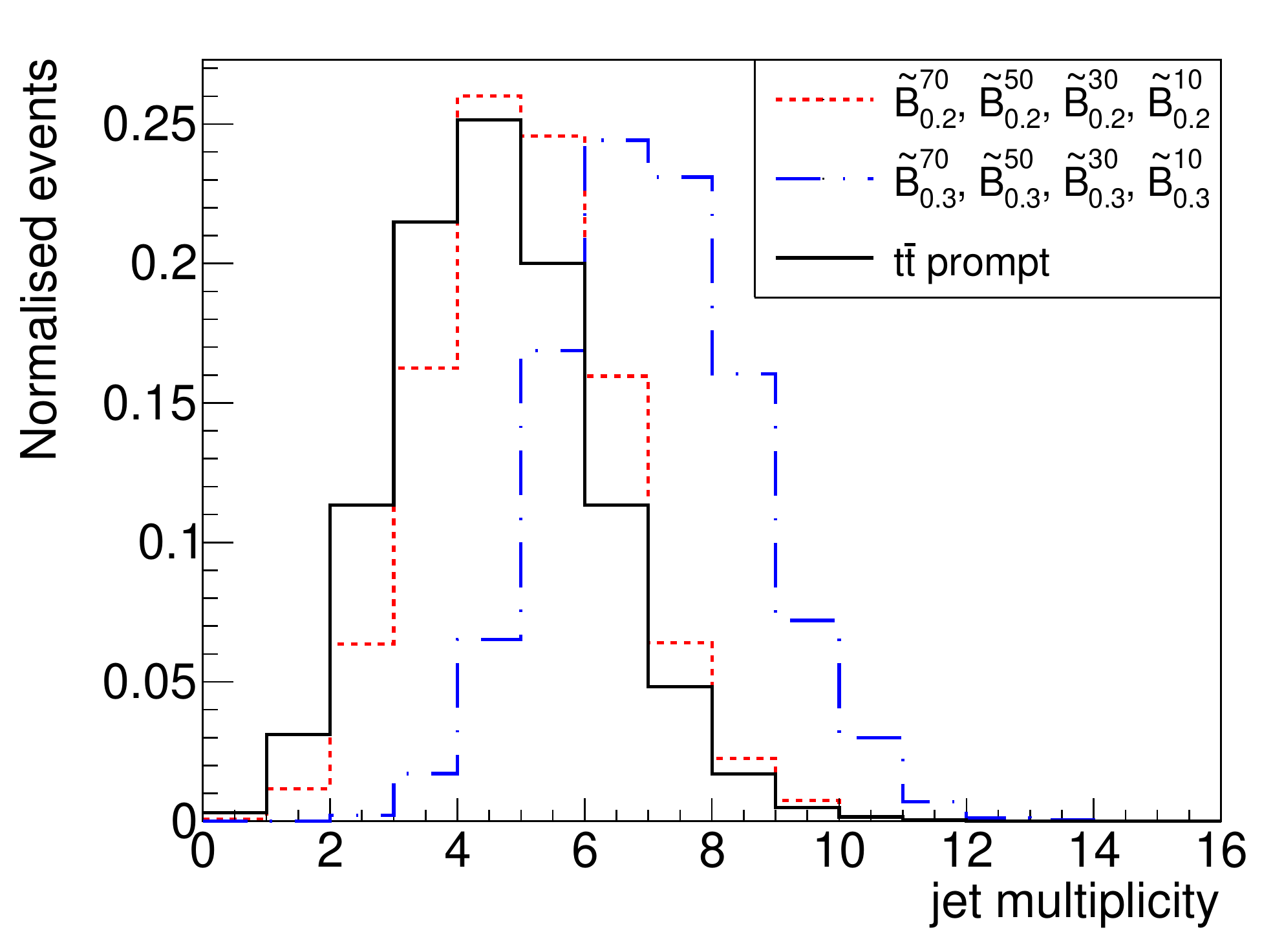}
    \caption{Jet multiplicity $N_J$ in the $\tilde{B}$-like case.}
    \label{fig:bino_nj}
\end{minipage}
\hspace*{1.4cm}
\begin{minipage}[t]{0.45\linewidth}
    \centering
    \includegraphics[scale=0.4]{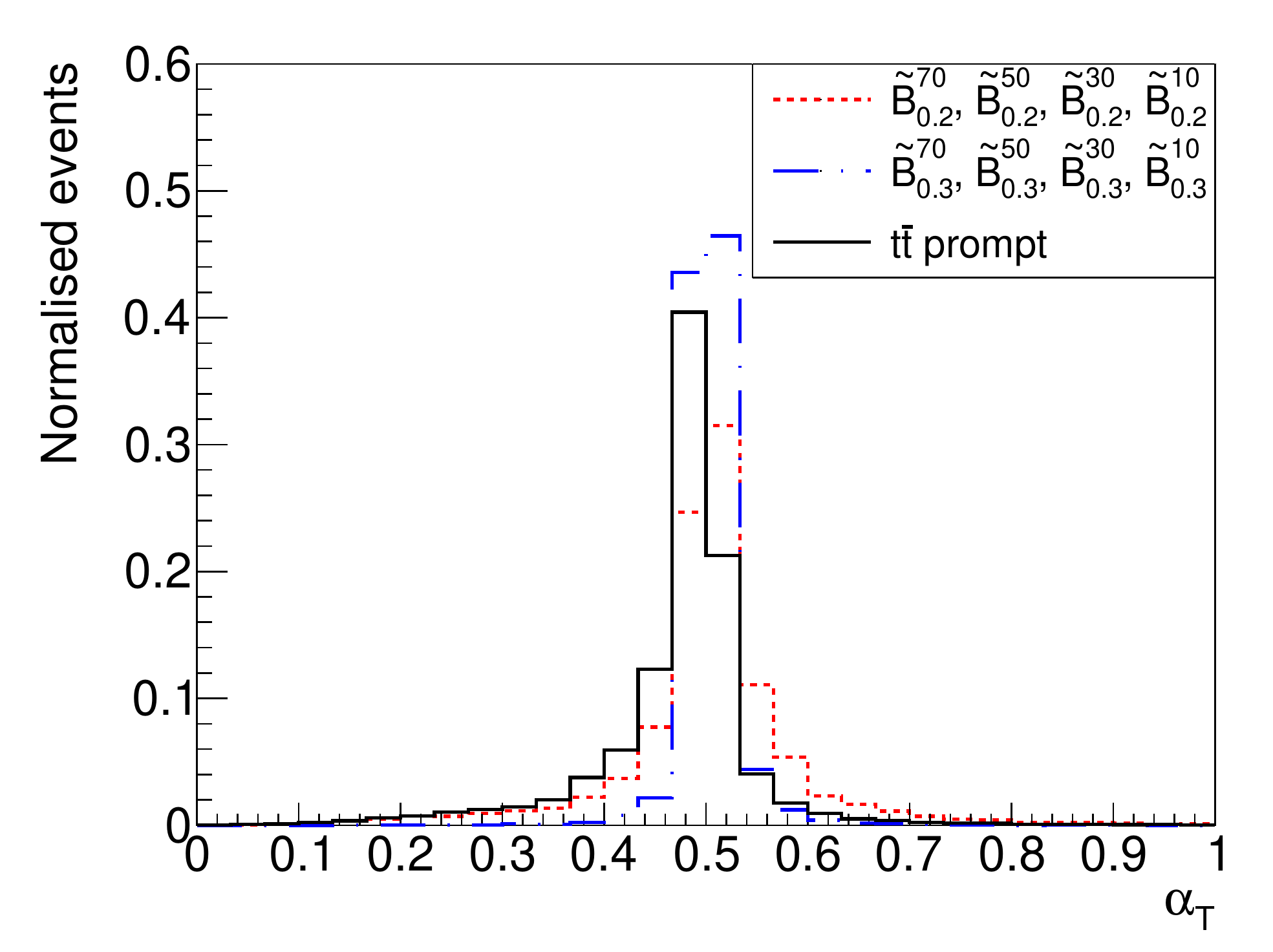}
    \caption{$\alpha_T$ distributions in the $\tilde{B}$-like case.}
    \label{fig:bino_at}
\end{minipage}
\end{figure*}

\begin{figure*}[htbp]
\begin{minipage}[t]{0.45\linewidth}
    \centering
    \includegraphics[scale=0.4]{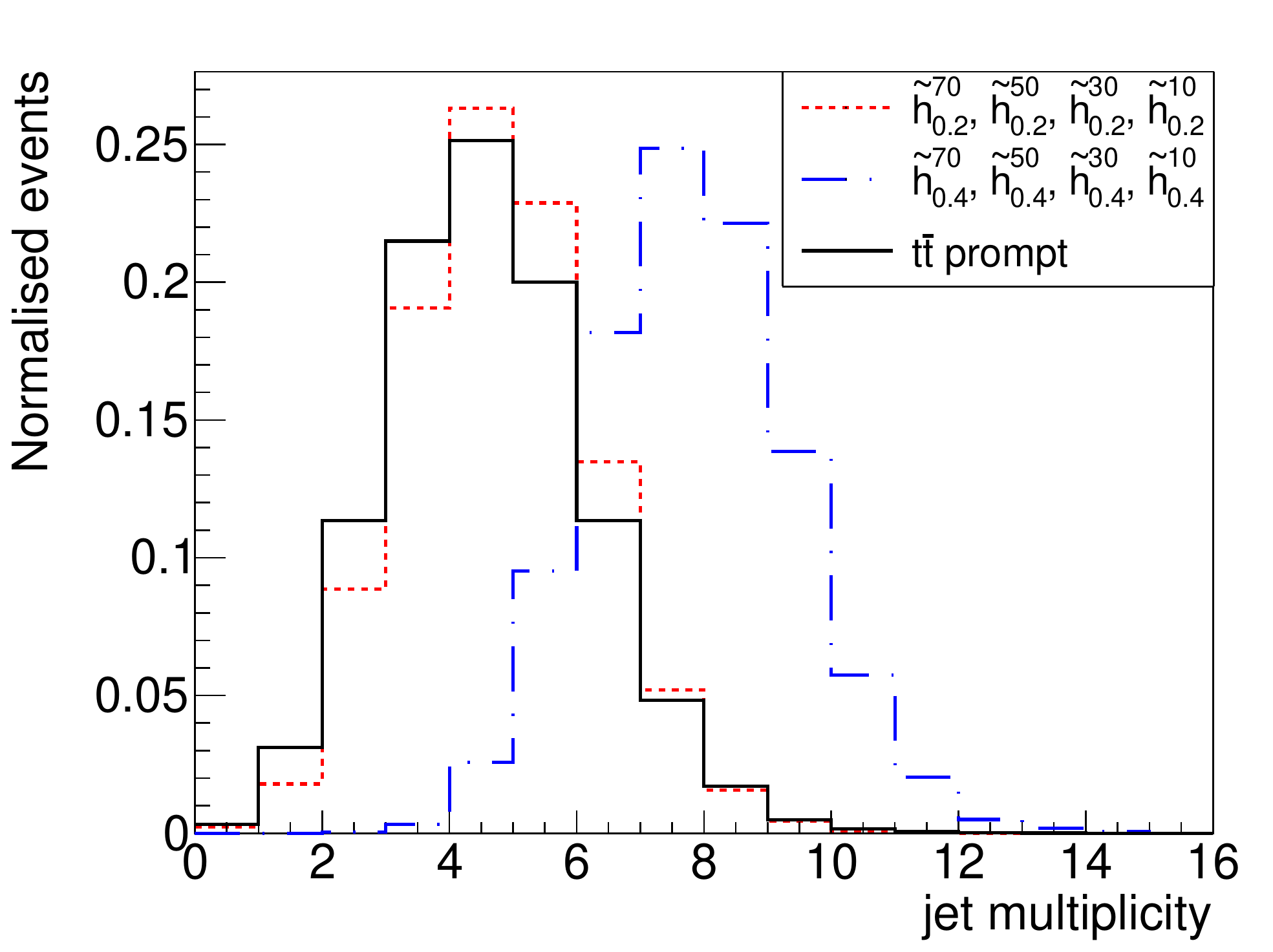}
    \caption{Jet multiplicity $N_j$ in the $\tilde{h}_2$-like case.}
    \label{fig:h_nj}
\end{minipage}
\hspace*{1cm}
\begin{minipage}[t]{0.45\linewidth}
    \centering
    \includegraphics[scale=0.4]{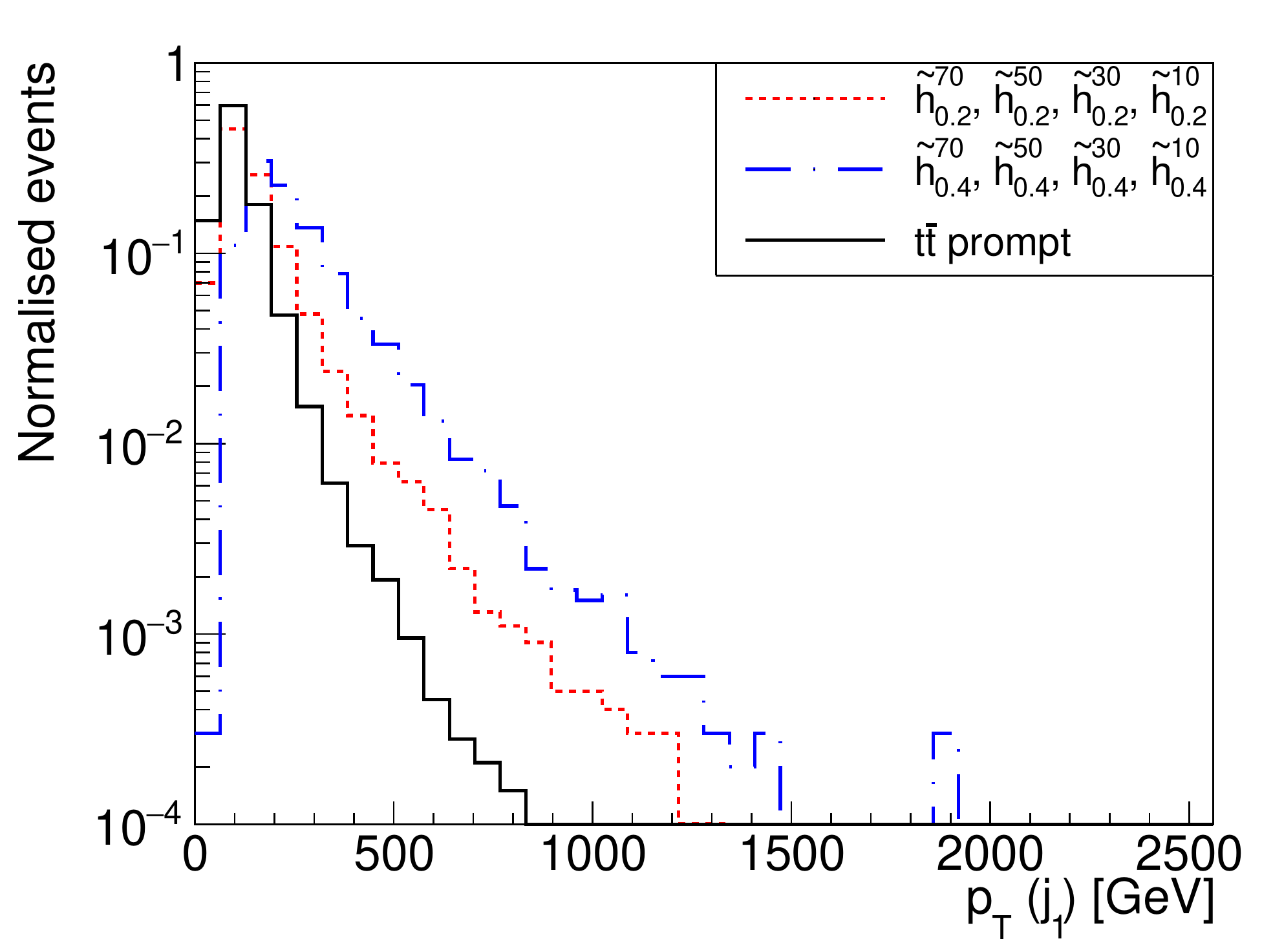}
    \caption{Transverse momentum of the {leading jet of the event} in the $\tilde{h}_2$-like case.}
    \label{fig:h_ptj}
\end{minipage}
\begin{minipage}[t]{0.45\linewidth}
    \centering
    \includegraphics[scale=0.4]{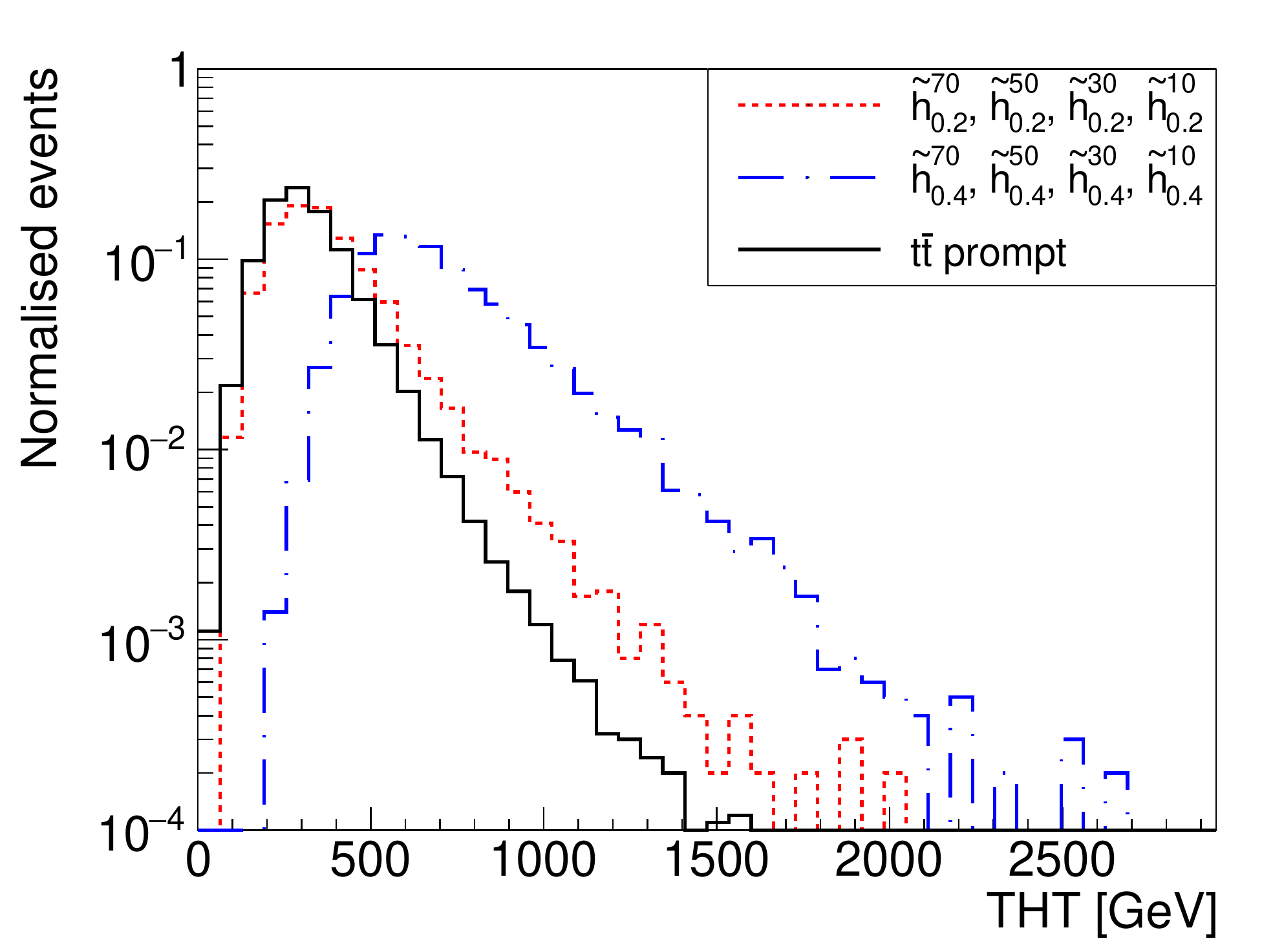}
    \caption{Total transverse hadronic energy $THT$ in the $\tilde{h}_2$-like case.}
    \label{fig:h_THT}
\end{minipage}
\hspace*{1cm}
\begin{minipage}[t]{0.45\linewidth}
    \centering
    \includegraphics[scale=0.4]{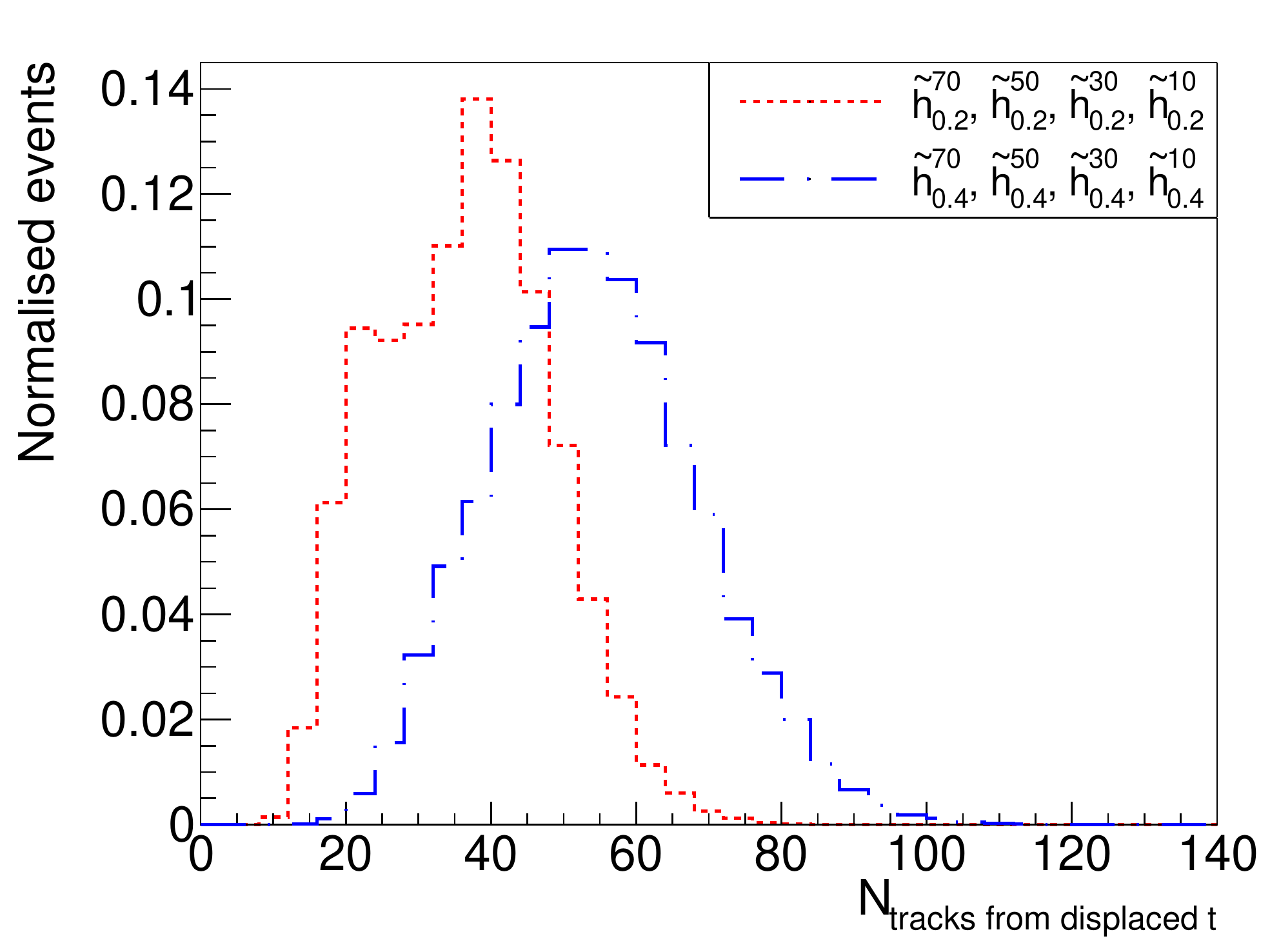}
    \caption{Track multiplicity associated with displaced vertices in the $\tilde{h}_2$-like case.}
    \label{fig:h_ntrack}
\end{minipage}
\begin{minipage}[t]{0.45\linewidth}
    \centering
    \includegraphics[scale=0.4]{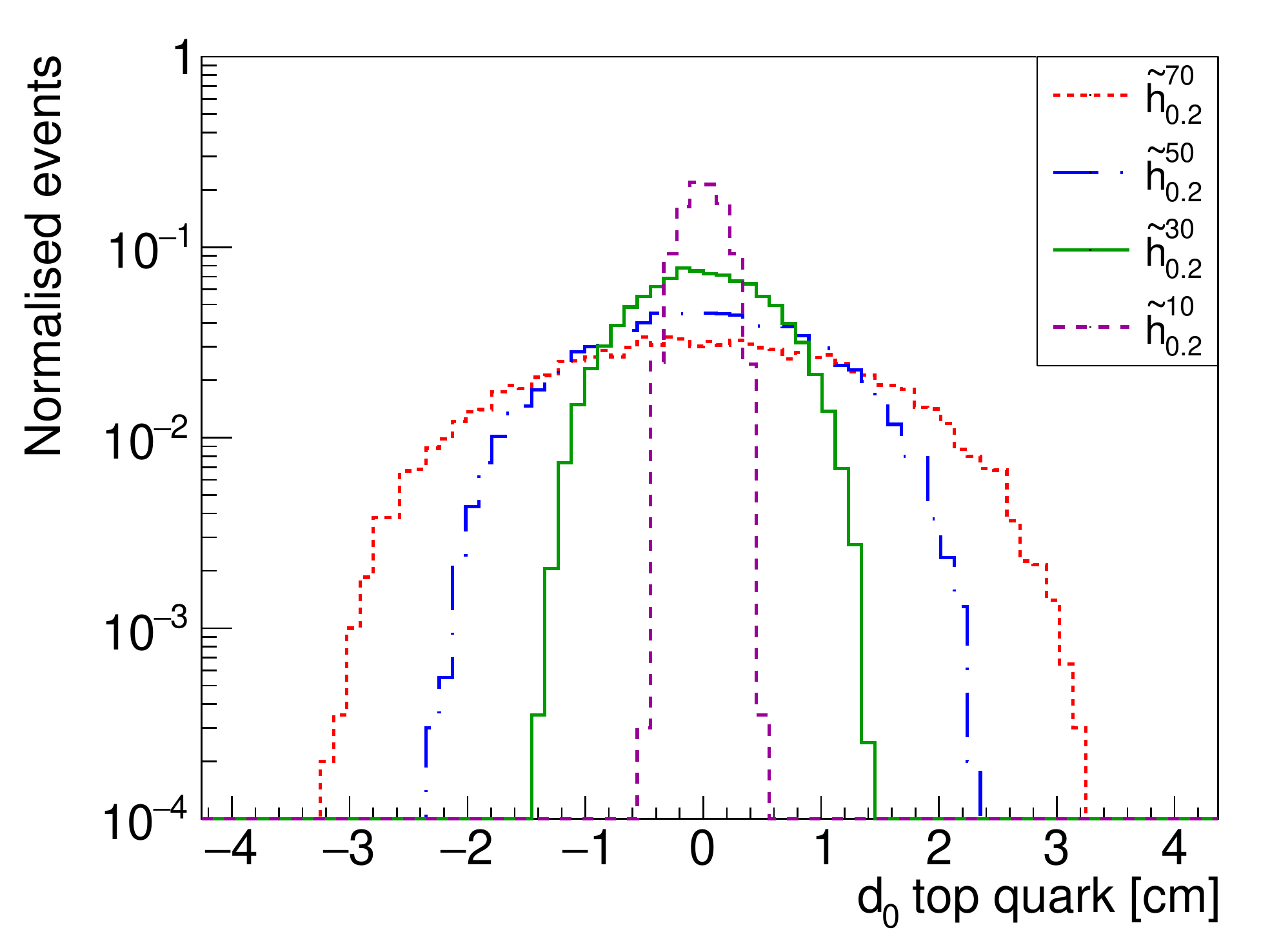}
    \caption{Transverse impact parameter $d_0$ of the top quarks in the $\tilde{h}_2$-like case with $m_{\tilde{\chi}^0_1}=200\ \text{GeV}$.}
    \label{fig:h_d0_200}
\end{minipage}
\hspace*{1.4cm}
\begin{minipage}[t]{0.45\linewidth}
    \centering
    \includegraphics[scale=0.4]{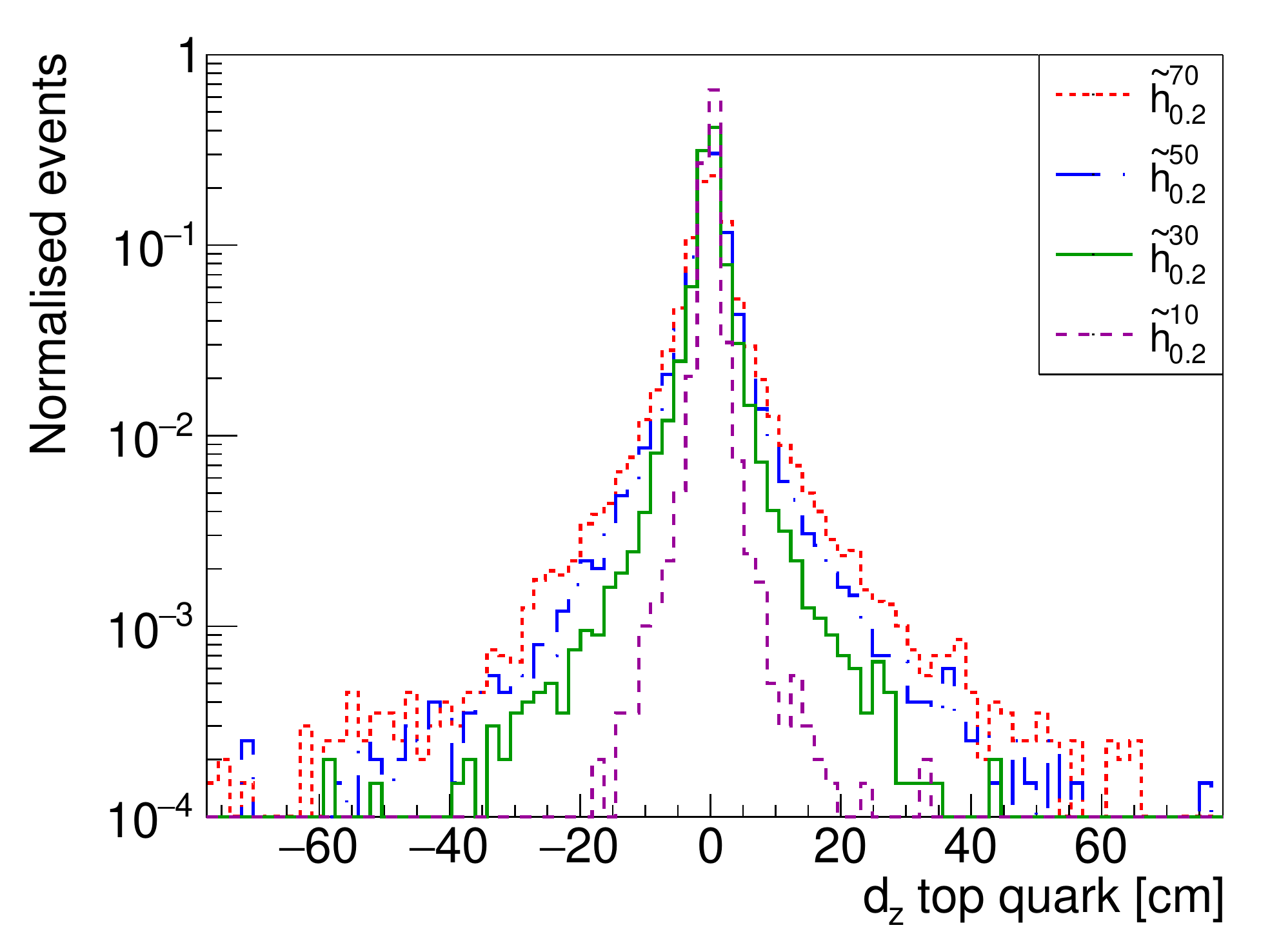}
    \caption{Longitudinal impact parameter $d_z$ of the top quarks in the $\tilde{h}_2$-like case with $m_{\tilde{\chi}^0_1}=200\ \text{GeV}$.}
    \label{fig:h_dz_200}
\end{minipage}
\end{figure*}

\begin{figure*}[htbp]
\begin{minipage}[t]{0.45\linewidth}
    \centering
    \includegraphics[scale=0.4]{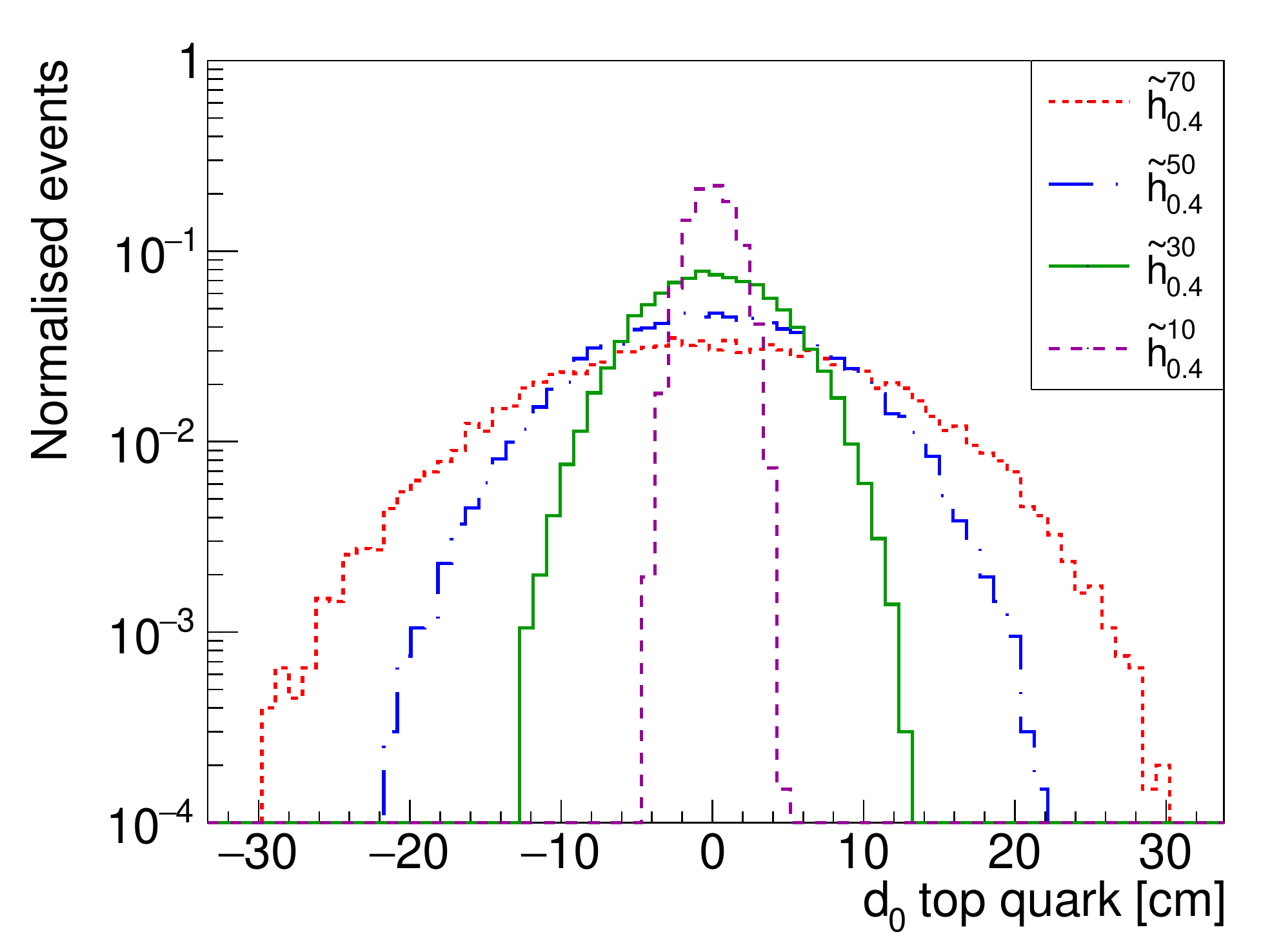}
    \caption{Transverse impact parameter $d_0$ of the top quarks in the $\tilde{h}_2$-like case with $m_{\tilde{\chi}^0_1}=400\ \text{GeV}$.}
    \label{fig:h_d0_400}
\end{minipage}
\hspace*{1.4cm}
\begin{minipage}[t]{0.45\linewidth}
    \centering
    \includegraphics[scale=0.4]{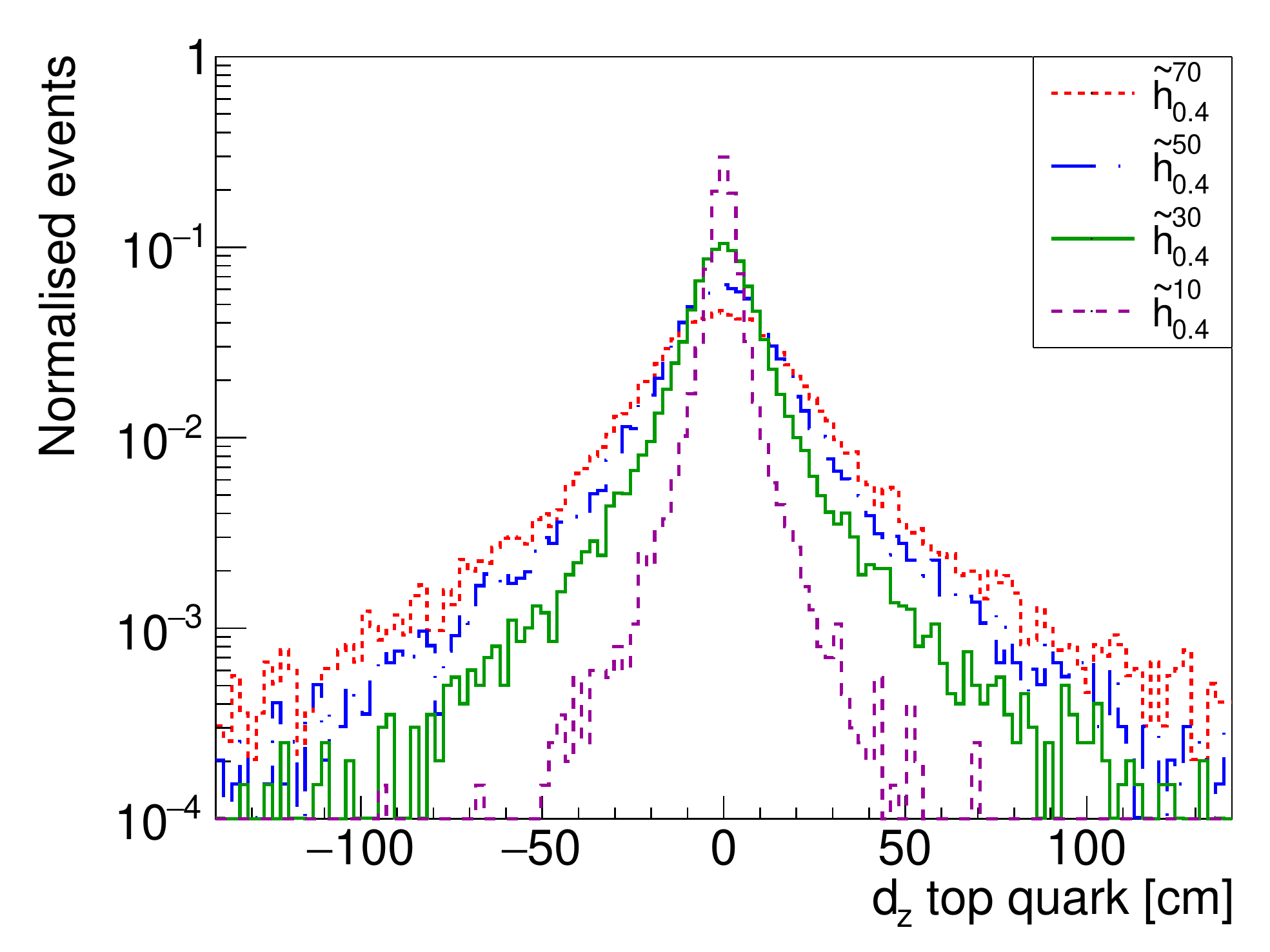}
    \caption{Longitudinal impact parameter $d_z$ of the top quarks in the $\tilde{h}_2$-like case with $m_{\tilde{\chi}^0_1}=400\ \text{GeV}$.}
    \label{fig:h_dz_400}
\end{minipage}
\begin{minipage}[t]{0.45\linewidth}
    \centering
    \includegraphics[scale=0.4]{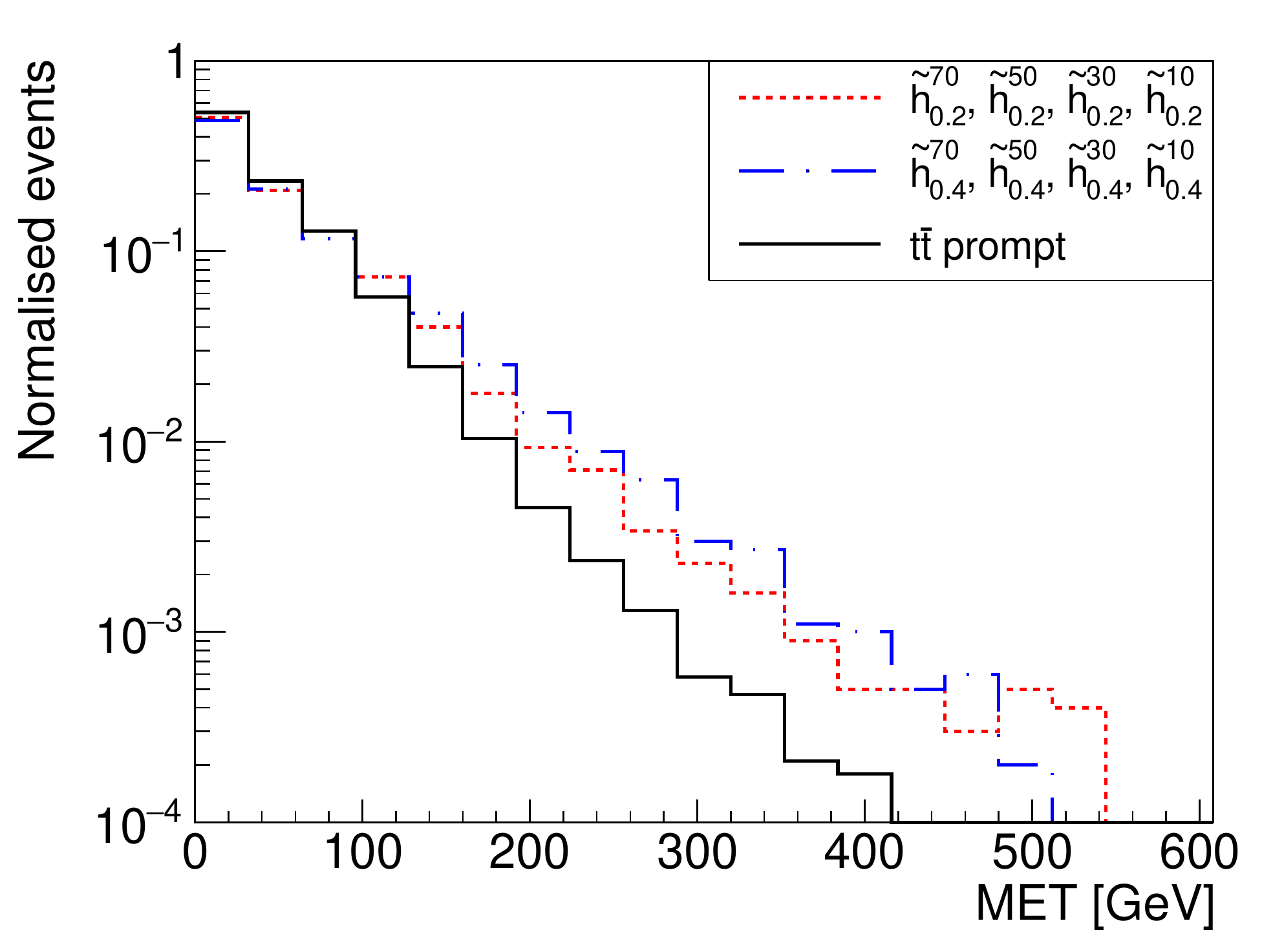}
    \caption{Missing transverse energy MET in the $\tilde{h}_2$-like case.}
    \label{fig:h_met}
\end{minipage}
\hspace*{1.4cm}
\begin{minipage}[t]{0.45\linewidth}
    \centering
    \includegraphics[scale=0.4]{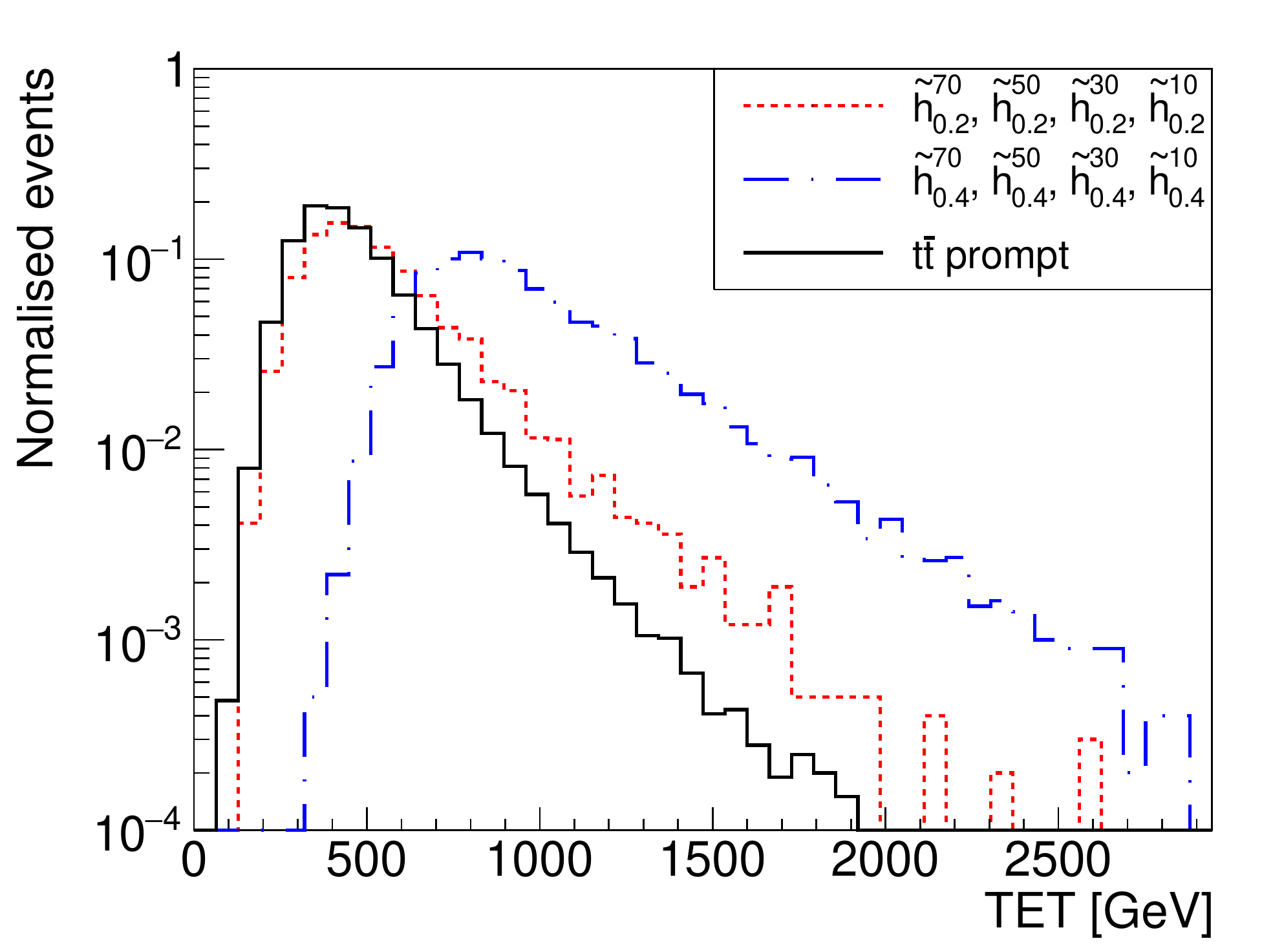}
    \caption{Total transverse energy TET in the $\tilde{h}_2$-like case.}
    \label{fig:h_tet}
\end{minipage}
\begin{minipage}[t]{0.45\linewidth}
    \centering
    \includegraphics[scale=0.4]{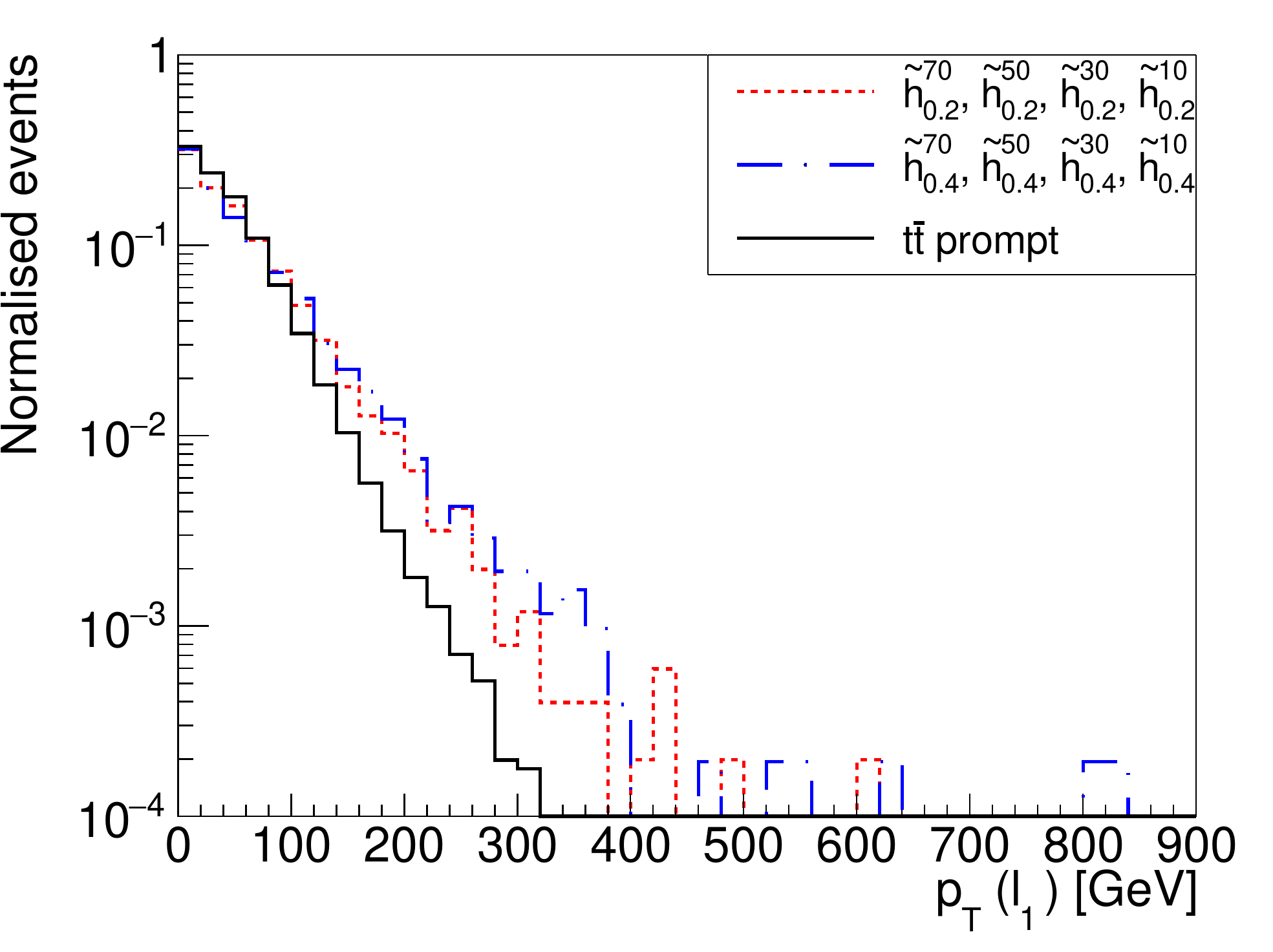}
    \caption{Transverse momentum $p_T$ of the leading lepton in the $\tilde{h}_2$-like case.}
    \label{fig:h_ptl1}
\end{minipage}
\hspace*{1.4cm}
\begin{minipage}[t]{0.45\linewidth}
    \centering
    \includegraphics[scale=0.4]{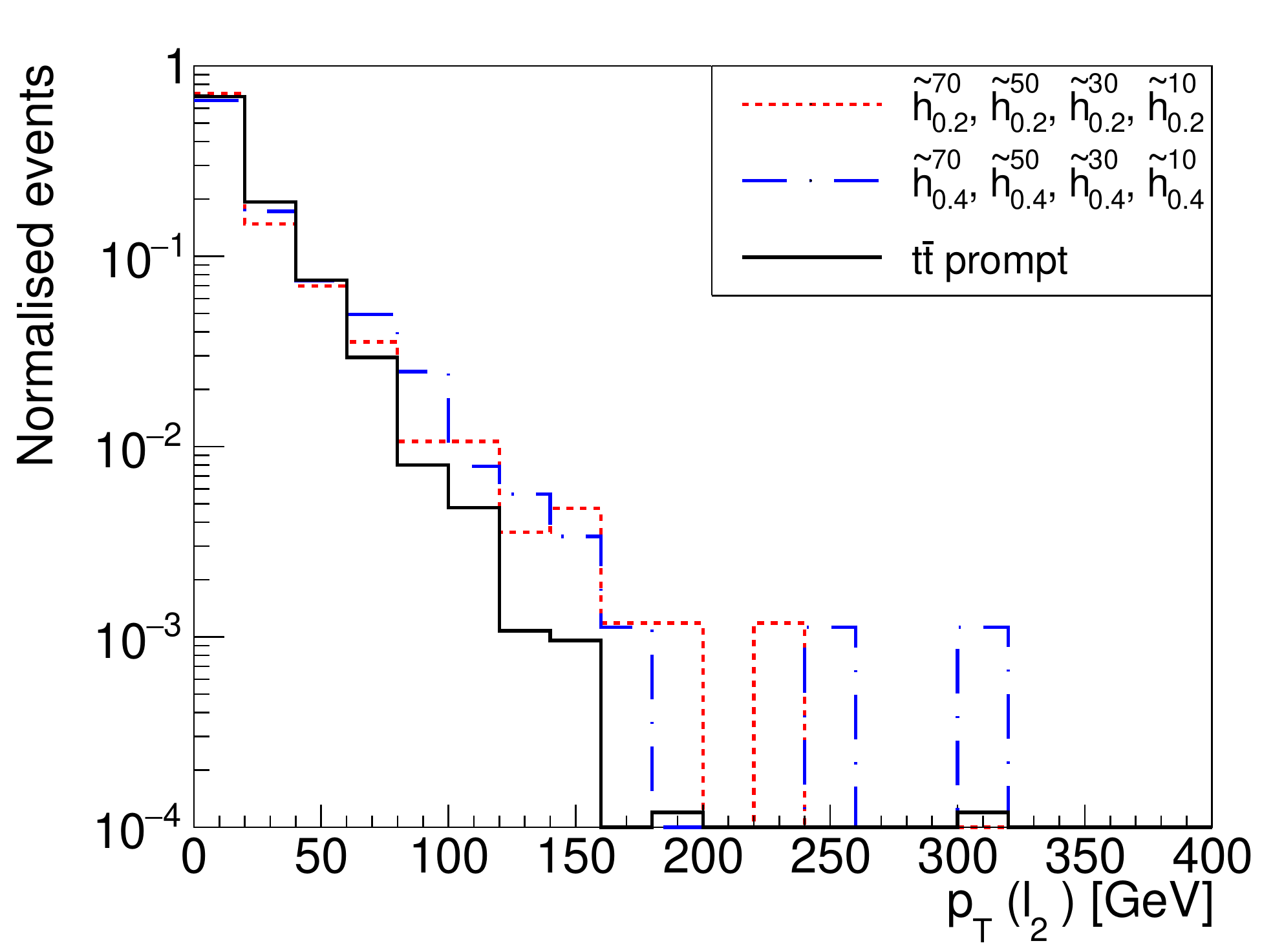}
    \caption{Transverse momentum $p_T$ of the second leading lepton in the $\tilde{h}_2$-like case.}
    \label{fig:h_ptl2}
\end{minipage}
\end{figure*}

\begin{figure*}[htbp]
\begin{minipage}[t]{0.45\linewidth}
    \centering
    \includegraphics[scale=0.4]{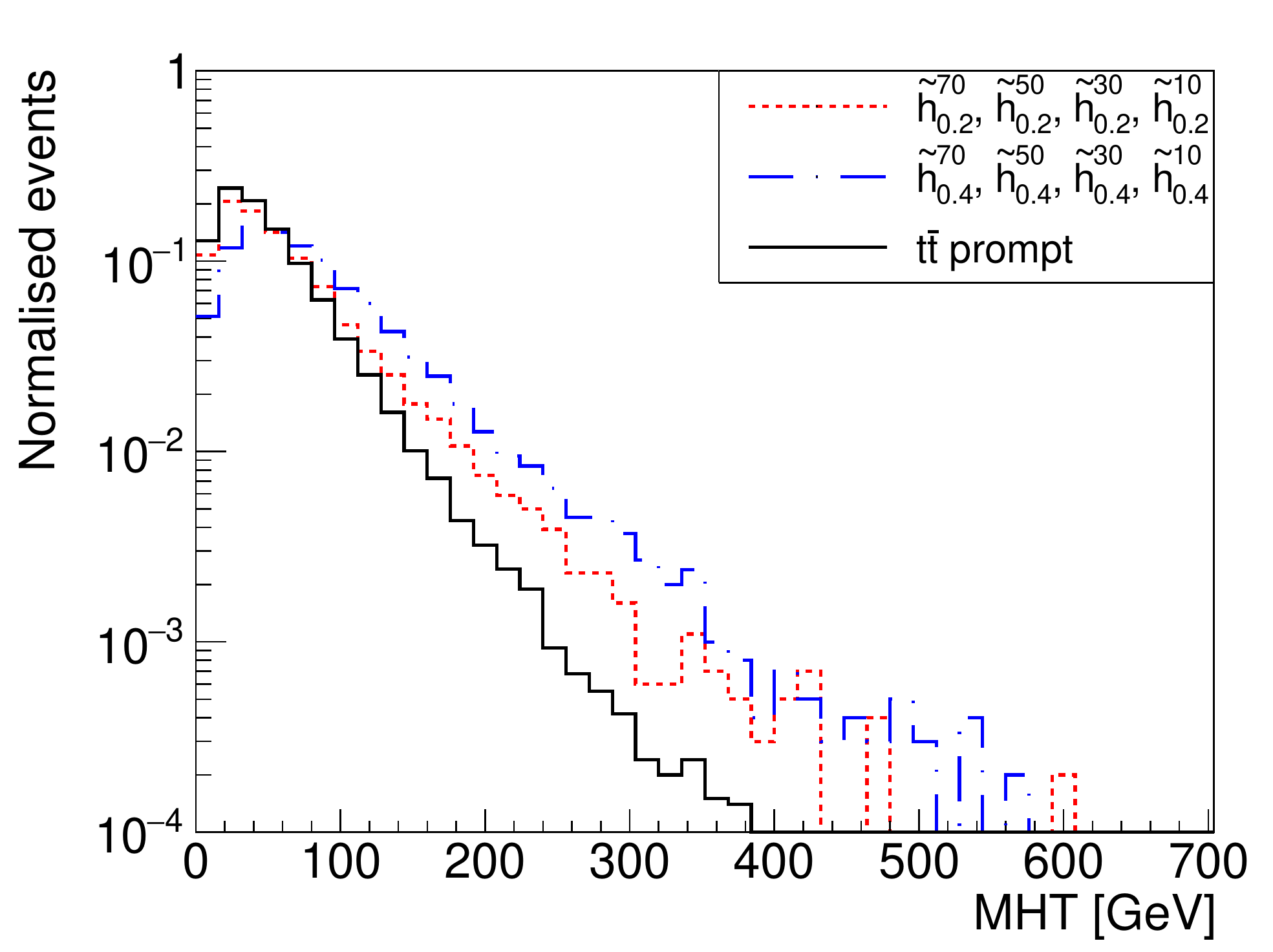}
    \caption{Missing transverse hadronic energy MHT in the $\tilde{h}_2$-like case.}
    \label{fig:h_mht}
\end{minipage}
\hspace*{1.4cm}
\begin{minipage}[t]{0.45\linewidth}
    \centering
    \includegraphics[scale=0.4]{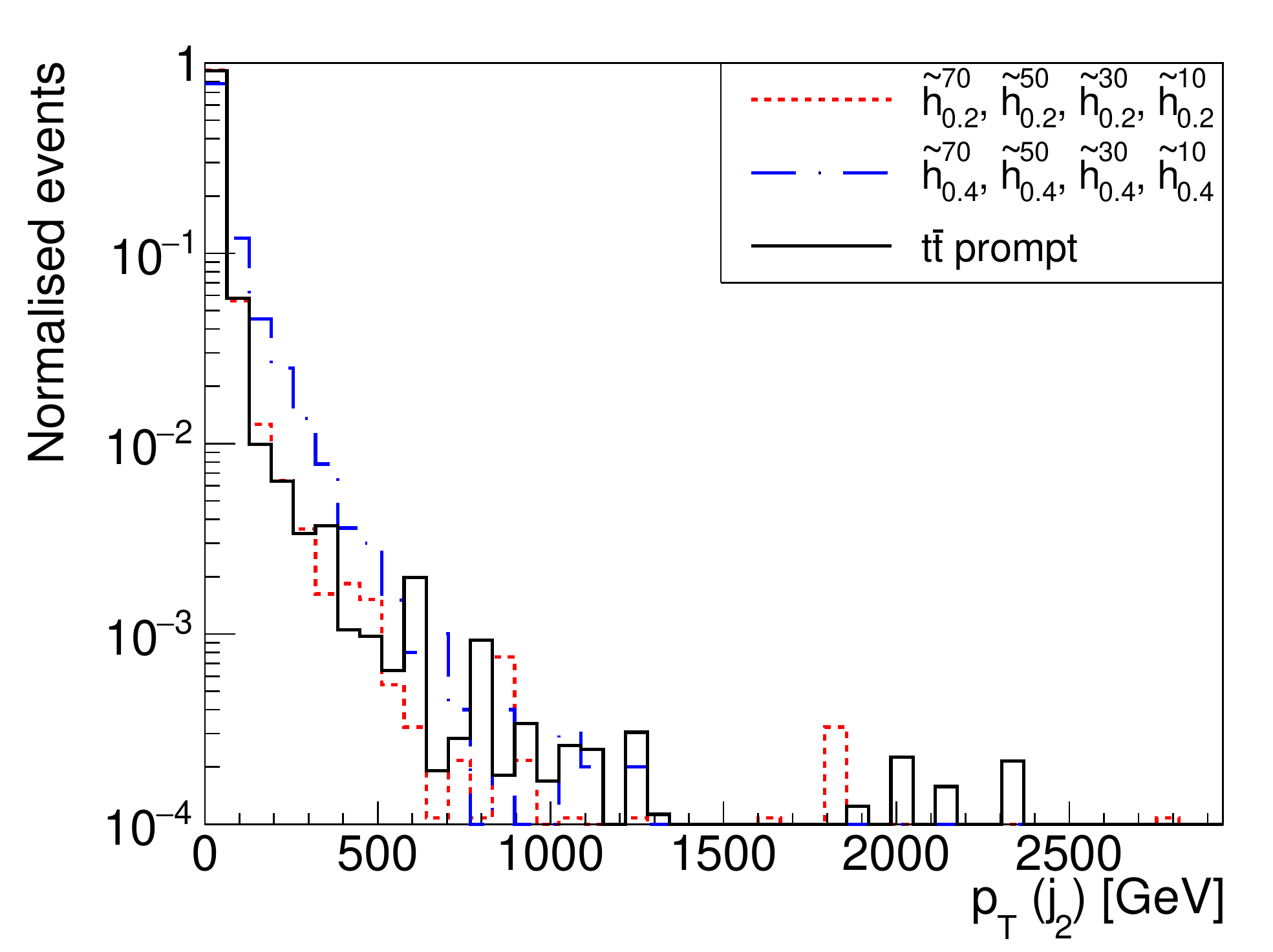}
    \caption{Transverse momentum $p_T$ of the second leading jet in the $\tilde{h}_2$-like case.}
    \label{fig:h_ptj2}
\end{minipage}
\begin{minipage}[t]{0.45\linewidth}
    \centering
    \includegraphics[scale=0.4]{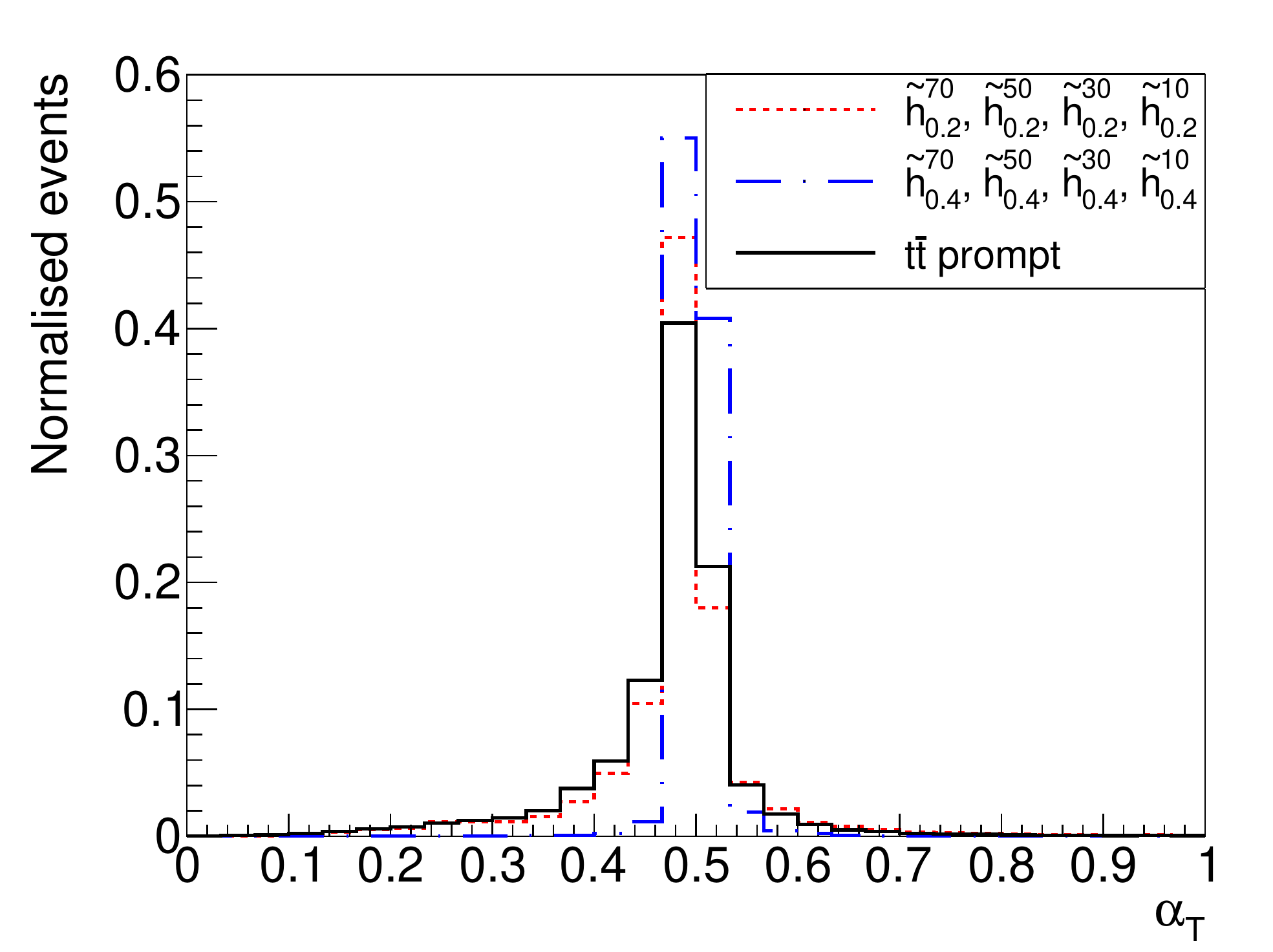}
    \caption{$\alpha_T$ distributions in the $\tilde{h}_2$-like case.}
    \label{fig:h_at}
\end{minipage}
\hspace*{1.4cm}
\begin{minipage}[t]{0.45\linewidth}
    \centering
    \includegraphics[scale=0.4]{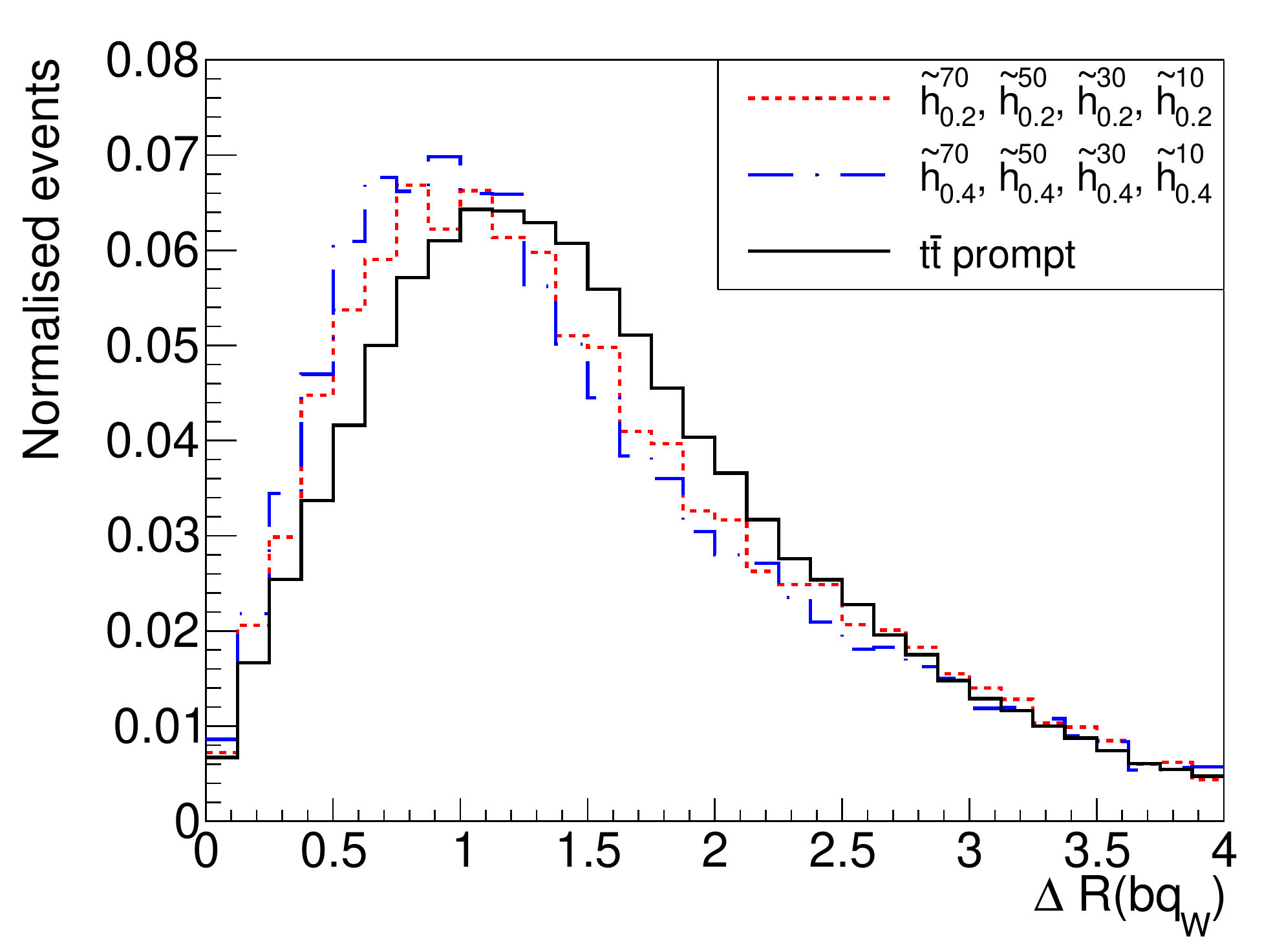}
    \caption{$\Delta R$ distributions in the $\tilde{h}_2$-like case.}
    \label{fig:h_DR}
\end{minipage}
\end{figure*}

\begin{table*}[htbp]
\begin{tabular*}{\textwidth}{@{\extracolsep{\fill}}lcccccccc|c}
\hline
 & $\tilde{B}_{0.2}^{10}$ & 
$\tilde{B}_{0.2}^{30}$ & 
$\tilde{B}_{0.2}^{50}$ & 
$\tilde{B}_{0.2}^{70}$ &
$\tilde{B}_{0.3}^{10}$ & 
$\tilde{B}_{0.3}^{30}$ & 
$\tilde{B}_{0.3}^{50}$ &
$\tilde{B}_{0.3}^{70}$ & $t\bar{t}$\\
\hline
$\sigma_{LO}(pp\rightarrow \tilde{\mu}\bar{\tilde{\mu}})\ [\text{fb}]$ & \multicolumn{8}{c|}{ 1.12} & - \\
\hline
$R_{\ =\ 2\ \textrm{displaced t}}$~[\%] & 38.1 & 87.7 & 94.6 & 94.4 & 47.5 & 95.2 & 98.6 & 98.0 & -  \\ 
$R_{\ \geq\ 1\ \textrm{displaced t}}$~[\%] &  72.7 & 99.0 & 99.8 & 99.3 & 63.6 & 99.0 & 99.8  & 99.4 & -  \\ 
\hline
$\langle \text{MET} \rangle$ [GeV]  & 63 & 65 & 64 & 64 & 58 & 59 & 58 & 58  & 43 \\ 
$\langle \text{TET} \rangle$ [GeV] & 989 & 986 & 990 & 988 & 980 & 982 & 983 & 982 & 465 \\ 
$\langle \text{THT} \rangle$ [GeV]  & 464 & 461 & 464 & 463 & 611 & 612 & 611 & 613 & 327 \\ 
$\langle \text{MHT}\rangle $ [GeV] & 126 & 126 & 127 & 126 & 90 & 92 & 90 &  91 & 52  \\
\hline
$\langle \alpha_T \rangle$& 0.50 & 0.50 & 0.50 & 0.50 & 0.50 & 0.50 & 0.50 &  0.50 & 0.48 \\  
\hline
$\langle p_T (j_1)\rangle $ [GeV] & 219 & 218 & 218 & 218 & 230 & 233 & 231 & 231 & 118 \\ 
$\langle p_T (j_2)\rangle $ [GeV] & 19 & 15 & 15 & 16 & 41 & 36 & 42 & 56 & 23 \\ 
$\langle p_T (\ell_1)\rangle $ [GeV] & 210 & 211 & 211 & 210 & 89 & 90 & 90 &  90  & 32 \\
$\langle p_T (\ell_2)\rangle $ [GeV] & 210 & 211 & 211 & 210 & 88 & 88 & 89 &  90  & 31 \\
\hline
$\langle |d_0(t)|\rangle $ [cm] & 0.17 & 0.50 & 0.83 & 1.16 & 0.73 & 2.19 & 3.65 & 5.10 &  - \\
$\langle |d_z(t)|\rangle $ [cm] & 0.47 & 1.37 & 2.26 & 3.11 & 1.91 & 5.64 & 9.30 &  12.9 & - \\
\hline
$N_{\text{jets}}$ & 4.6 & 4.6 & 4.6 & 4.6 & 6.6 & 6.6 & 6.6 & 6.6 & 4.3\\
$N_{\text{tracks from displaced t}}$ & 35.6 & 35.4 & 35.4 & 35.4 & 47.0 & 46.9 & 46.8 & 46.8 & - \\
\hline
\end{tabular*}
\caption{Mean values of some observables and number of events with top quarks decaying in the tracker volume for the eight benchmarks in the bino $\tilde{B}$ case. We remind that $R_{\geq \textrm{1 displaced t}}$ corresponds to the number of events with at least one top quark decaying in the tracker volume.}
\label{table:eventninvol1_B}
\end{table*}

\begin{table*}[htbp]
\begin{tabular*}{\textwidth}{@{\extracolsep{\fill}}lcccccccc|c}
\hline
 & $\tilde{h}_{0.2}^{10}$ & 
$\tilde{h}_{0.2}^{30}$ & 
$\tilde{h}_{0.2}^{50}$ & 
$\tilde{h}_{0.2}^{70}$ &
$\tilde{h}_{0.4}^{10}$ & 
$\tilde{h}_{0.4}^{30}$ & 
$\tilde{h}_{0.4}^{50}$ & 
$\tilde{h}_{0.4}^{70}$ & $t\bar{t}$ \\
\hline
$\sigma_{LO}(pp\rightarrow \tilde{\chi}^0_1\tilde{\chi}^0_1)\ [\text{fb}]$ & \multicolumn{4}{c|}{39.7} & \multicolumn{4}{c|}{2.52} &  - \\
\hline 
$R_{\ =\ 2\ \textrm{displaced t}}$~[\%] & 25.1 & 81.8 & 92.3 & 93.5 & 44.5 & 90.4 & 96.3 & 96.8 &  - \\ 
$R_{\ \geq\ 1\ \textrm{displaced t}}$~[\%] & 33.1 & 89.7 & 96.7 & 98.3 & 53.1 & 94.7 & 98.5 & 98.7& -  \\ 
\hline
$\langle \text{MET} \rangle$ [GeV] & 51 & 51 & 51  & 51 & 56 & 56 & 56 &  56 & 43 \\    
$\langle \text{TET} \rangle$ [GeV] & 549 & 549 & 551 & 548 & 995 & 992 & 996 & 995 & 465 \\ 
$\langle \text{THT} \rangle$ [GeV] & 384 & 384 & 385 & 384 & 718 & 713 & 718 & 716 & 327 \\ 
$\langle \text{MHT}\rangle $ [GeV] & 63 & 64 & 64 & 64 & 84 & 85 & 85 & 85 & 52 \\
\hline
$\langle \alpha_T \rangle$ & 0.48 & 0.48 & 0.48 & 0.48 & 0.50 & 0.50 & 0.50 & 0.50 & 0.48  \\  
\hline
$\langle p_T (j_1)\rangle $ [GeV] & 156 & 155 & 156 & 155 & 257 & 256 & 258 & 258 & 118\\ 
$\langle p_T (j_2)\rangle $ [GeV] & 32 & 25 & 14 & 20 & 46 & 37 & 39 & 44 & 23 \\ 
$\langle p_T (\ell_1)\rangle $ [GeV] & 50 & 51 & 52 & 52 & 53 & 56 & 54 & 55 & 32 \\
$\langle p_T (\ell_2)\rangle $ [GeV] & 19 & 19 & 19 & 19 & 21 & 21 & 19 & 21 & 31 \\
\hline
$\langle |d_0(t)|\rangle $ [cm] & 0.15 & 0.44 & 0.74 & 1.03 & 1.19 & 3.57&5.94 & 8.34 & - \\
$\langle |d_z(t)|\rangle $ [cm] & 0.69 & 2.04 & 3.40 & 4.62 & 3.85 & 11.12 & 17.88 & 24.04 & - \\
\hline
$N_{\text{jets}}$ & 4.3 & 4.3 & 4.3 & 4.3 & 7.3 & 7.3 & 7.3 & 7.3 & 4.3 \\
$N_{\text{tracks from displaced t}}$ & 35.6 & 35.5 & 35.5 & 35.4 & 53.5 & 53.5 & 53.4 & 53.4 & -\\
\hline
\end{tabular*}
\caption{Mean values of some observables and number of events with top quarks decaying in the tracker volume for the eight benchmarks in the higgsino $\tilde{h}_2$ case. We remind that $R_{\geq \textrm{1 displaced t}}$ corresponds to the number of events with at least one top quark decaying in the tracker volume. }
\label{table:eventninvol1_h}
\end{table*}

Distributions of some relevant observables have been computed for each benchmark. The results of the analysis are summarised in Table~\ref{table:eventninvol1_B} and Table~\ref{table:eventninvol1_h}\footnote{Despite the fact that all the distributions are not Gaussian, we still present the mean values to summarise the information of all observables. This remark applies to each analysis.}. Some distributions are presented in Figs.~\ref{fig:bino_ptmu} to 
\ref{fig:bino_at} for the $\tilde{B}$-case and in Figs.~\ref{fig:h_nj} to 
\ref{fig:h_DR}, for the $\tilde{h}_2$-case. Standard top quark pair (prompt) production has also been added to the distributions to compare our signals with a typical source of background. The distributions are all normalised to unity.\\ 

First, considering the bino-like configuration, we expect at least two muons in the final state with jets coming from the top quarks. 
The distributions of transverse momentum of muons are shown in Fig.~\ref{fig:bino_ptmu}. On average, the muons coming from the smuons are more energetic in the signal than in the $t\bar{t}$ background events (the muons coming essentially from the top quark leptonic decay). 
The same comment remains valid for the transverse momentum of the leading jets $p_T(j_1)$ according to Figure~\ref{fig:bino_ptj} as the top quarks in the signal are harder.\smallskip

One of the main differences between the benchmarks and the $t\bar{t}$ prompt signal is the presence of down and strange quarks coming from the neutralino decay which increases the jet multiplicity in the events. 
An analysis at the partonic level showed that the d and s quarks coming from the neutralino decay do not have a large $p_T$, thus not passing the requirement of the anti-$k_T$ algorithm: $p_T > 30\ \text{GeV}$.\smallskip

 Unlike the RPC studies mentioned in Section \ref{rpv_benchmark}, only a moderate amount of MET is expected from the leptonic W boson decays. However, with a mean value of $65$ GeV, this observable may still be of interest to
reduce the QCD background. Conversely, the mean values of THT (see Fig.~\ref{fig:bino_THT}) are approximately $460$ and $610\ \text{GeV}$ for $m_{\tilde{\chi}^0_1}=200$ and $300\ \text{GeV}$, respectively. \smallskip

Finally, more than 90\% of the events include at least one top quark decaying inside the simplified-tracker acceptance for $\langle d^{\textrm{flight}}_{\tilde{\chi}^0_1}\rangle  \geq 30\ \text{cm}$ (see Table \ref{table:eventninvol1_B}) producing on average between 35 and 47 displaced potential tracks (Fig.~\ref{fig:bino_ntrack}). As {shown} Figures~\ref{fig:bino_d0_200} to 
\ref{fig:bino_dz_300}, the transverse and longitudinal impact parameters $d_0$ and $d_z$ of the top quarks depends strongly on the neutralino mass. Their mean values do not exceed 13~cm in absolute although the $d_z$ parameter can reach values between -40 cm and 40 cm.

Concerning the higgsino-like process, the experimental signature is similar to the bino case {other than the absence of a muon} with high $p_T$ coming from the smuon decay in the final state. 
It can be noted in Figs.~\ref{fig:h_nj} and \ref{fig:h_THT} (jet multiplicity and THT) that distributions of the prompt $t\bar{t}$ and the higgsino direct production with $m_{\tilde{\chi}^0_1}=200\ \text{GeV}$ are quite similar and so could make difficult the identification of the signal with such observables.

The transverse momentum of jets is presented in Fig.~\ref{fig:h_ptj} with a mean value of $160$ and $260\ \text{GeV}$ for $m_{\tilde{\chi}^0_1}=200$ and $400\ \text{GeV}$, respectively. Higher neutralino masses naturally lead to a higher $p_{T}$ of the jets, and it affects the jet multiplicity (see Fig.~\ref{fig:h_nj}). 
The multiplicity for $m_{\tilde{\chi}^0_1}=200\ \text{GeV}$ is equivalent to the prompt $t\bar{t}$ signal with a mean value of $4.3$, whereas for $m_{\tilde{\chi}^0_1}=400\ \text{GeV}$ the number of jets is higher with a mean value of $7.3$. The first value remains low compared to the expected one (between 6 and 10 quarks in the final state) but is explained by the selection cut of 30~GeV on the jet transverse momentum.\smallskip

Considering the global variables, the MET in the event (with a mean value of $50$ and $55\ \text{GeV}$ for $m_{\tilde{\chi}^0_1}= 200$ and $400\ \text{GeV}$, respectively, see Table \ref{table:eventninvol1_h}) originates from the neutrino in the W decay. The total hadronic energy THT is presented in Fig.~\ref{fig:h_THT}. The $m_{\tilde{\chi}^0_1}=400\ \text{GeV}$ case stands out with a mean value of $710\ \text{GeV}$ whereas at $m_{\tilde{\chi}^0_1}=200\ \text{GeV}$, the THT distribution is close to the prompt $t\bar{t}$ one.\smallskip

One main difference with the bino-like process is the number of events with at least one top quark decaying inside the ``tracker volume'' {(see $R_{\ \geq\ 1\ \textrm{displaced t}}$ observable in Table \ref{table:eventninvol1_h})}. For $\langle d^\textrm{flight}_{\tilde{h}}\rangle = 10\ \text{cm}$, only $33\%$ and $53\%$ (for $m_{\tilde{\chi}^0_1}=200$ and $400\ \text{GeV}$, respectively) of the events match such a configuration. These values correspond to $73\%$ and $64\%$ for the bino case. This difference can be explained by the fact that the top quarks are produced closer to the beam in the bino case, leading to fewer top quarks in the tracker volume for small flight distances. 


\subsection{Mediation by a coloured long-lived particle (R-hadron based on a squark)} \label{sec:stop}


A particular simplified model where the long-lived particle is a R-hadron containing a squark is designed, inspired by the R-parity conserved MSSM where supersymmetry is {broken} by the GMSB mechanism (for Gauge Mediated Supersymmetry Breaking \cite{GaugeMediated1,GaugeMediated2}). This model introduces a top squark, and a gravitino as an electrically neutral LSP.

\subsubsection{Model description}

The model is based on the Lagrangian density of the MSSM to which is added the kinetic part of the gravitino as well as its interactions with the fermions and bosons in the MSSM. These interactions can be expressed as:

\begin{equation}
\begin{aligned}
\mathcal{L}_{3/2}^{int} = -\frac{\sqrt{4\pi}}{m_p} \Big( \mathcal{D}_\nu \phi^{i}&\bar{\chi}_{L i}\gamma^{\mu}\gamma^{\nu}\Psi_{\mu}^{(M)} \\
& + \mathcal{D}_{\nu}\phi^{\dagger}_i\bar{\Psi}_{\nu}^{(M)}\gamma^{\nu}\gamma^{\mu}\chi_R^i\Big)\label{eq:gmu_int}
\end{aligned}
\end{equation}

\noindent where $\mathcal{D}_\nu \phi^{i}$ is the covariant derivative of the scalar field $\phi$, $\Psi_\mu^{(M)}$ the gravitino field, $\bar{\chi}_{L}$ and $\chi_R$ the left and right-handed fermions corresponding to the superpartner of $\phi$, and $m_p$ the Planck mass. In this context, the gravitino mass is naturally suppressed by the Planck scale, as shown in the equation below, and leads to a natural LSP gravitino:
\begin{gather}
m_{3/2} \propto \frac{F_0}{m_p}  \nonumber\end{gather}
with $\sqrt{F_0}$ the fundamental scale of supersymmetry breaking.\\

\noindent In order to build a simplified model, only the superpartners of the third generation of quarks are involved. Like in the previous section, we define the stop squark as an equal mixing of the left-handed and right-handed stop states. This stop squark is taken as the NLSP and therefore the decay mode $\tilde{t}\rightarrow t \psi_{\mu}$ is the only possible. Removing this assumption would merely consist in specifying a branching fraction to the latter decay mode. 

\subsubsection{Cross sections and decays}

\begin{figure}[htbp]
    \centering
   \includegraphics[scale=0.5]{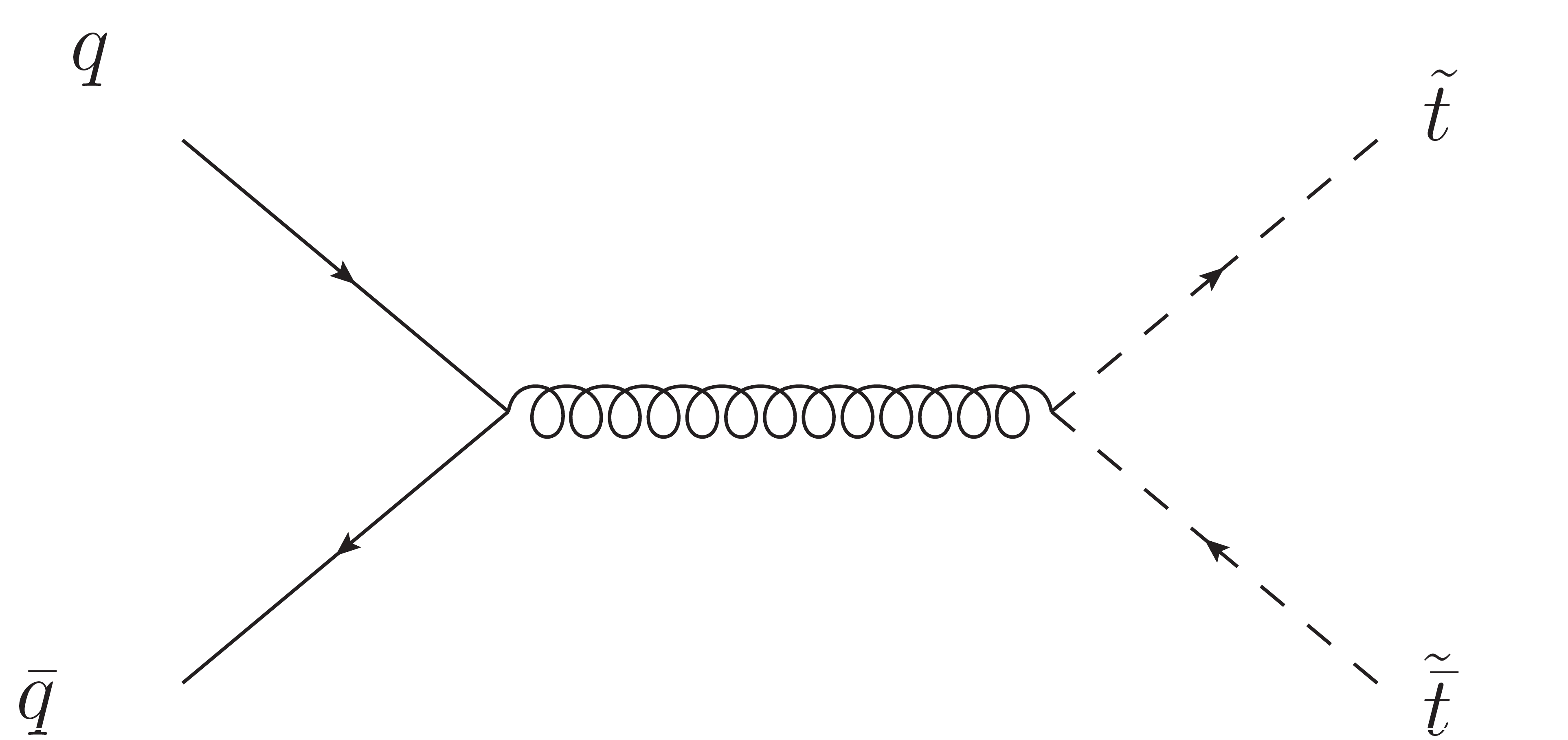}
    \caption{Feynman diagram of a stop squark pair production by quark annihilation.}
    \label{fig:stop-prod1}
\end{figure}

\begin{figure}[htbp]
    \vspace{-0.5cm}
    \centering
    \includegraphics[scale=0.22]{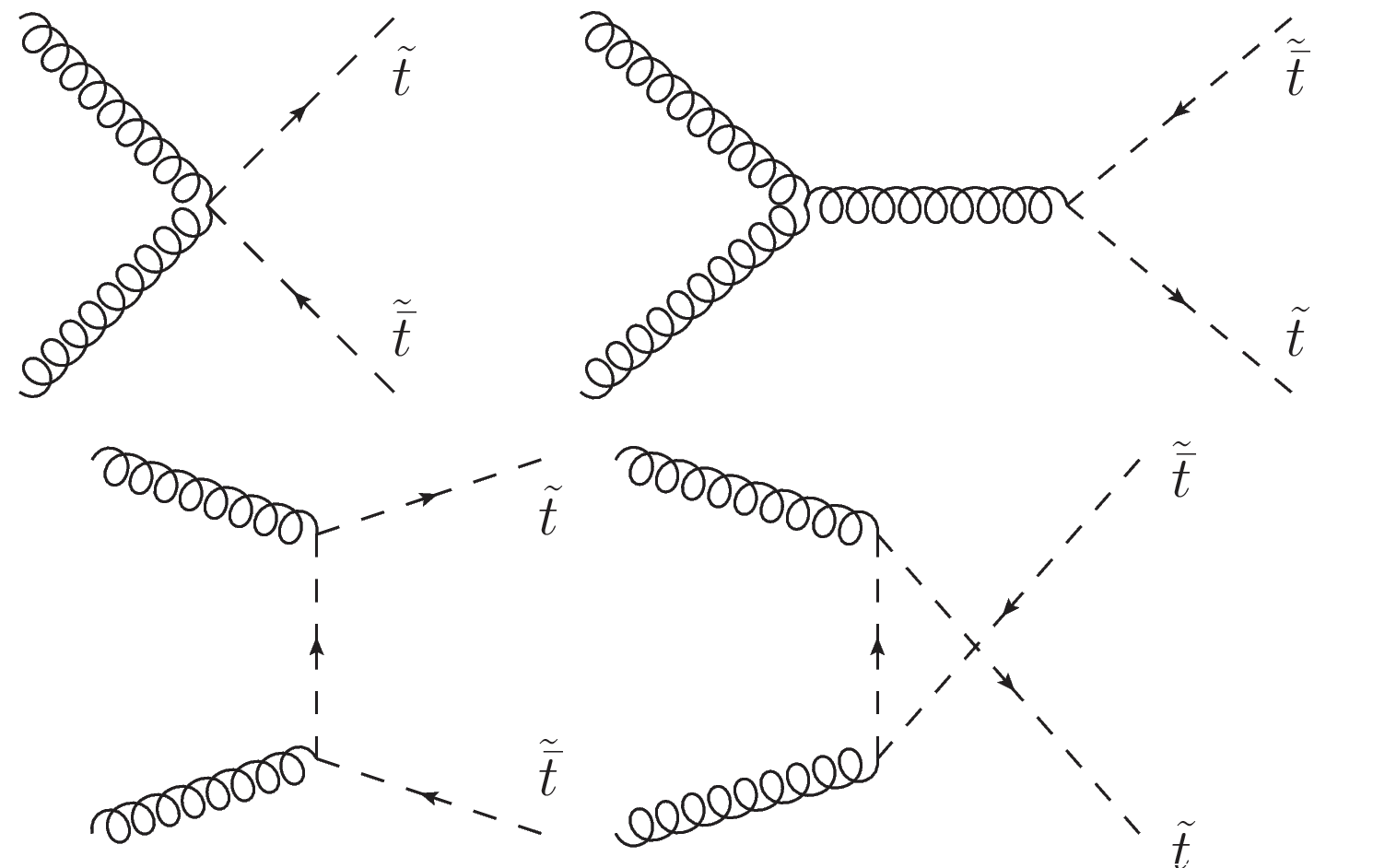}
    \caption{Feynman diagrams of a stop squark pair production by gluon fusion.}
    \label{fig:stop-prod2}
\end{figure}

\begin{figure*}[htbp]
\begin{minipage}[t]{0.45\linewidth}
    {\centering
    \includegraphics[scale=0.43]{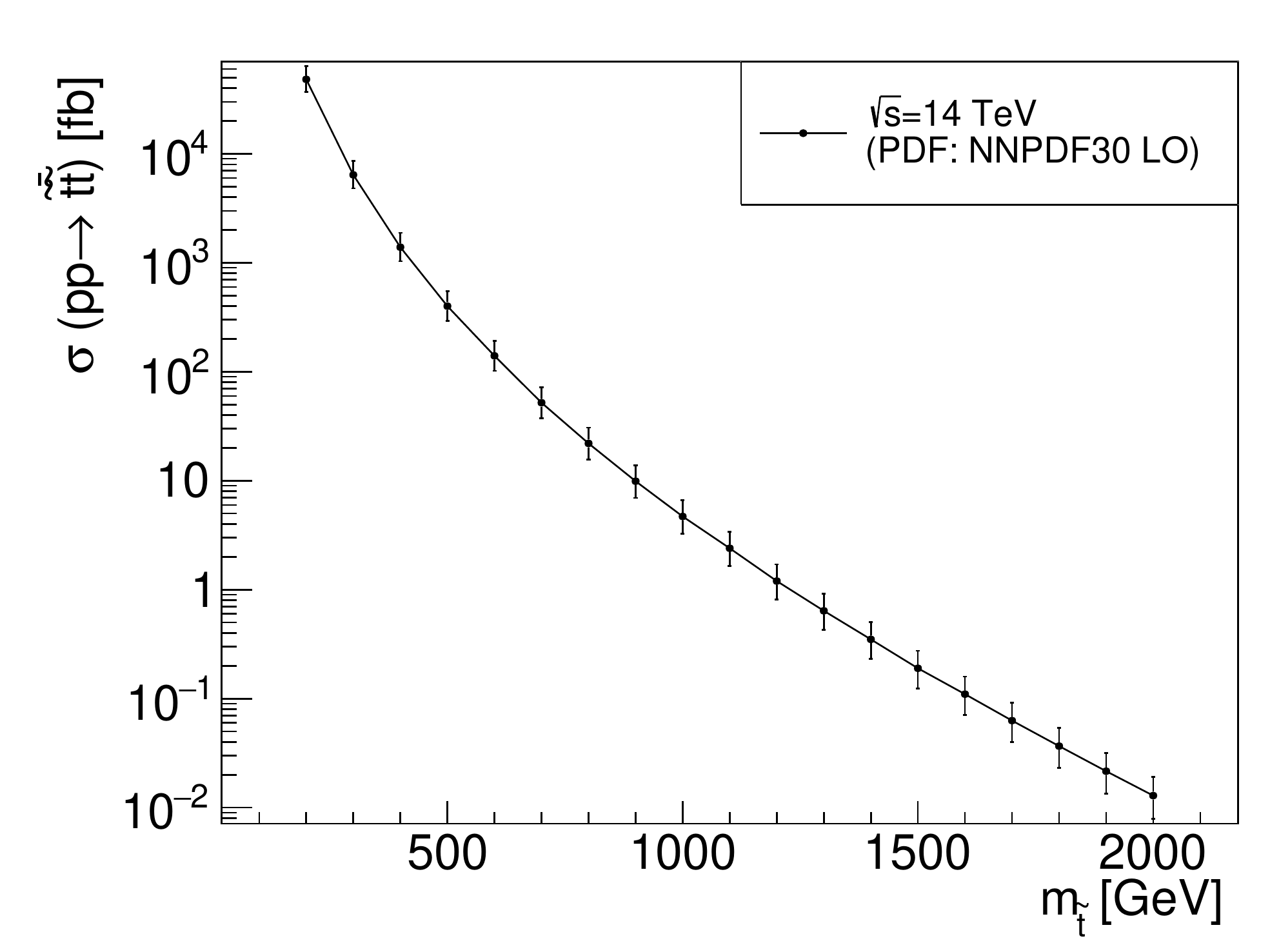}
    \caption{Leading order cross section $\sigma (pp\rightarrow \tilde{t}\tilde{\bar{t}})$ at $\sqrt{s}=14\ \text{TeV}$ as a function of the stop squark mass $m_{\tilde{t}}$. The uncertainties are related to PDF variations and the factorisation and renormalisation scales.}
    \label{fig:stop-xsection}}
\end{minipage}
\hspace{1 cm}
\begin{minipage}[t]{0.45\linewidth}
{\centering
     \includegraphics[scale=0.43]{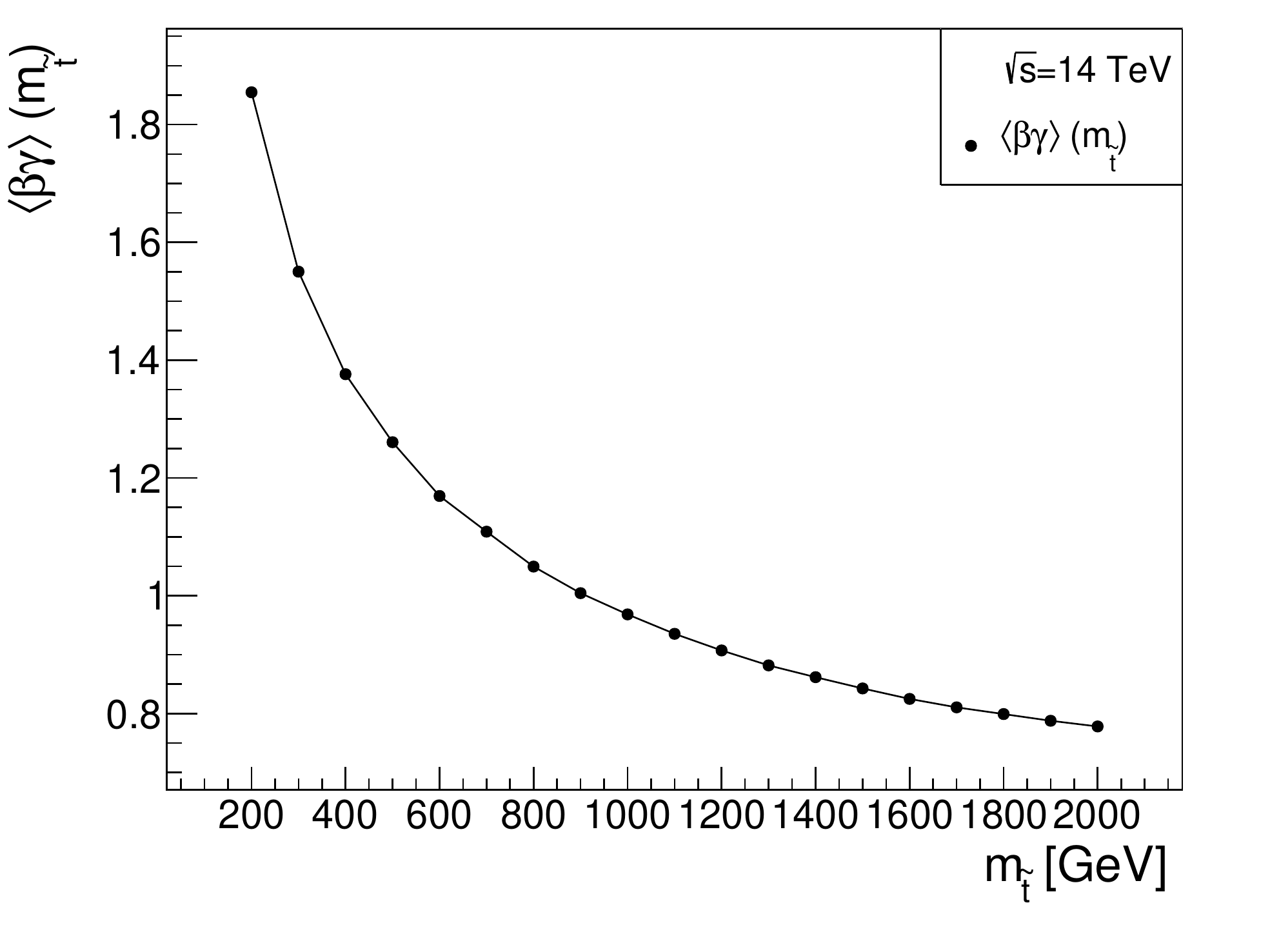}
    \caption{Average value of the Lorentz factor $\beta\gamma$ of the stop squark.}
    \label{fig:stop-betagamma}}
\end{minipage}
\end{figure*}

At the LO QCD, the stop squarks can be produced at the LHC by quark annihilation (Fig.~\ref{fig:stop-prod1}) or by gluon fusion (Fig.~\ref{fig:stop-prod2}). The cross section depends only on the stop mass $m_{\tilde{t}}$ and has already been calculated in more extensive studies (up to NLO calculation, with $K=\sigma_{NLO}/\sigma_{LO}\approx 1.3$, see~\cite{Beenakker:1996ch,Beenakker:1997ut}). Figures~\ref{fig:stop-xsection} and \ref{fig:stop-betagamma} display the variation of the production cross section and the mean Lorentz factor $\langle \beta\gamma \rangle$ of the stop squark according to its mass.  The uncertainties in Fig.~\ref{fig:stop-xsection} are related to PDF  and factorisation and renormalisation scale variations. The cross section decreases with the stop mass and, assuming a Run~3 integrated luminosity of $\mathcal{L}= 300~\textrm{fb}^{-1}$, gives an upper limit on the stop mass that we can reach with a reasonable number of events. Concerning the Lorentz factor $\langle \beta\gamma \rangle$ (see Fig.~\ref{fig:stop-betagamma}), the mean value is between 0.8 and 1.9 for the investigated masses, so the flight distance of the stop is not highly modified for a stop mass higher than approximately $600\ \text{GeV}$.\medskip

\noindent The total decay width of the stop can be calculated analytically from the {interaction} terms present in the Lagrangian density (see Eq.~\ref{eq:gmu_int}) and is expressed as:
\begin{eqnarray}
\Gamma(\tilde{t}\rightarrow \psi_{\mu}t)=&&\frac{1}{6m_p^2 m_{\tilde{t}}^3 m_{3/2}^2} \lambda^{3/2}\left(m^2_{\tilde{t}},m^2_t,m^2_{3/2}\right)\nonumber\\
&&\times\left(m_{\tilde{t}}^2-(m_t-m_{3/2})^2\right)\nonumber
\end{eqnarray}
with the K\"{a}llén function $\lambda(x,y,z)=x^2+y^2+z^2-2xy-2xz-2yz$.\\

\begin{figure}[htbp]
    \centering
    \includegraphics[scale=0.19]{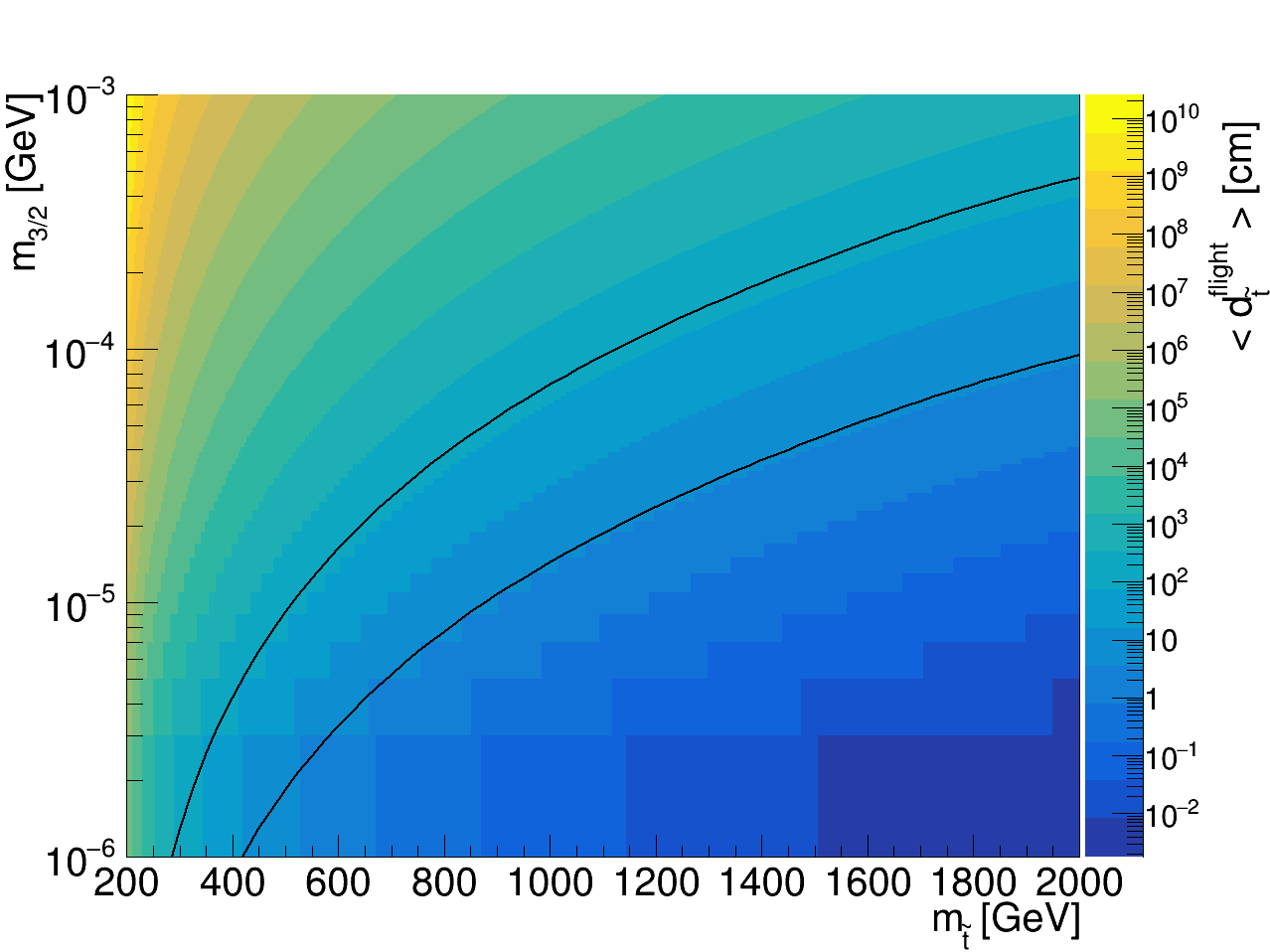}
    \caption{Mean flight distance of the stop in the laboratory frame as a function of the stop and the gravitino masses. The black lines represent the geometrical limits of the tracker volume, \textit{i.e.}, $4\ \text{cm}$ and $100\ \textrm{cm}$.}
    \label{fig:stop_flightdistance}
\end{figure}

\noindent Taking into account the mean Lorentz factor $\langle \beta\gamma \rangle$, the mean flight distance of the stop in the laboratory frame $\langle d^{\text{flight}}_{\tilde{t}} \rangle$ is shown in Fig.~\ref{fig:stop_flightdistance}. This observable is totally driven by the stop and the gravitino masses. If we focus on the geometrical boundaries of $4\ \textrm{cm}$ and  $100\ \textrm{cm}$, a long-lived stop can be observed for a wide range of values of stop mass.


\subsubsection{Benchmarks definition} 
\noindent With the help of Figs.~\ref{fig:stop-xsection}-\ref{fig:stop_flightdistance}, we have identified eight different benchmarks given in Table~\ref{tab:stop_bcmk}. Limits on stop squark decaying to a top quark and a gravitino have not been achieved yet at the LHC. Nevertheless, the equivalent process $\tilde{t}\rightarrow t \tilde{\chi}^0_1$ is widely investigated and can be compared with our signal in the limit $m_{\tilde{\chi}^0_1}\rightarrow 0$ (in the RPC case). The constraints coming from ATLAS and CMS collaborations with prompt stop squarks (see \cite{ATLAS:2020syg,CMS:2021beq,Aaboud_2017,aad:hal-03143576,Aad:2746797,sirunyan:hal-02934073}) give lower bounds between $m_{\tilde{t}}> 0.9$ to $1.9\ \text{TeV}$ in the most restrictive case. Thus two masses are considered in the following benchmarks: $m_{\tilde{t}}=1.0$ and $1.4\ \textrm{TeV}$ with a production cross section of $4.7$ and $0.4\ \text{fb}$, respectively. We also assume a flight distance of $\langle d^{\text{flight}}_{\tilde{t}}\rangle \in [10,30,50,70]\ \text{cm}$. For each benchmark, the gravitino mass $m_{3/2}$ is totally constrained and its value is in a range between $20$ and $140~\textrm{keV}$. Note that these values are consistent with astrophysics and cosmology in some scenarios, see \cite{cosmo1,cosmo2}.\\

It is important to highlight that a long-lived stop, as a coloured particle, must hadronise before decaying. More precisely, two kinds of R-hadrons can be produced: the R-meson $R(\tilde{t}\bar{q})$ and the R-baryon $R(\tilde{t}qq)$ which can be electrically neutral or charged according to the flavour of the constituent quarks. The case of a charged R-hadron offers the possibility to reconstruct its hits in the tracker sensors and thus its track. In the following, we do not take into account this potential signature and, by misuse of language, we refer the R-hadron as the stop squark.

\begin{table*}[htbp]
\begin{tabular*}{\textwidth}{@{\extracolsep{\fill}}lcccccccc@{}}
\hline
Name & $\tilde{t}_{1.0}^{10}$ & $\tilde{t}_{1.0}^{30}$ & $\tilde{t}_{1.0}^{50}$  & $\tilde{t}_{1.0}^{70}$ & $\tilde{t}_{1.4}^{10}$ & $\tilde{t}_{1.4}^{30}$ & $\tilde{t}_{1.4}^{50}$  & $\tilde{t}_{1.4}^{70}$ \\
\hline
$\langle d^\textrm{flight}_{\tilde{t}}\rangle~[\textrm{cm}]$ & 10 & 30 & 50 & 70 & 10 & 30 & 50 & 70 \\
$m_{\tilde{t}}~[\textrm{TeV}]$ & 1.0 & 1.0 & 1.0 & 1.0 & 1.4 & 1.4 & 1.4 & 1.4 \\
\hline
$m_{3/2}~[\textrm{keV}]$ & 22 & 39 & 50 & 59 & 53 & 93 & 120 & 142 \\
\hline
\end{tabular*}
\caption{Definition of the eight benchmarks for the stop squark pair production $pp\rightarrow \tilde{t}\bar{\tilde{t}}$ analysis.}
\label{tab:stop_bcmk}
\end{table*}


\subsubsection{{Event} kinematics}

\begin{figure*}[htbp]
\begin{minipage}[t]{0.45\linewidth}
    \centering
    \includegraphics[scale=0.4]{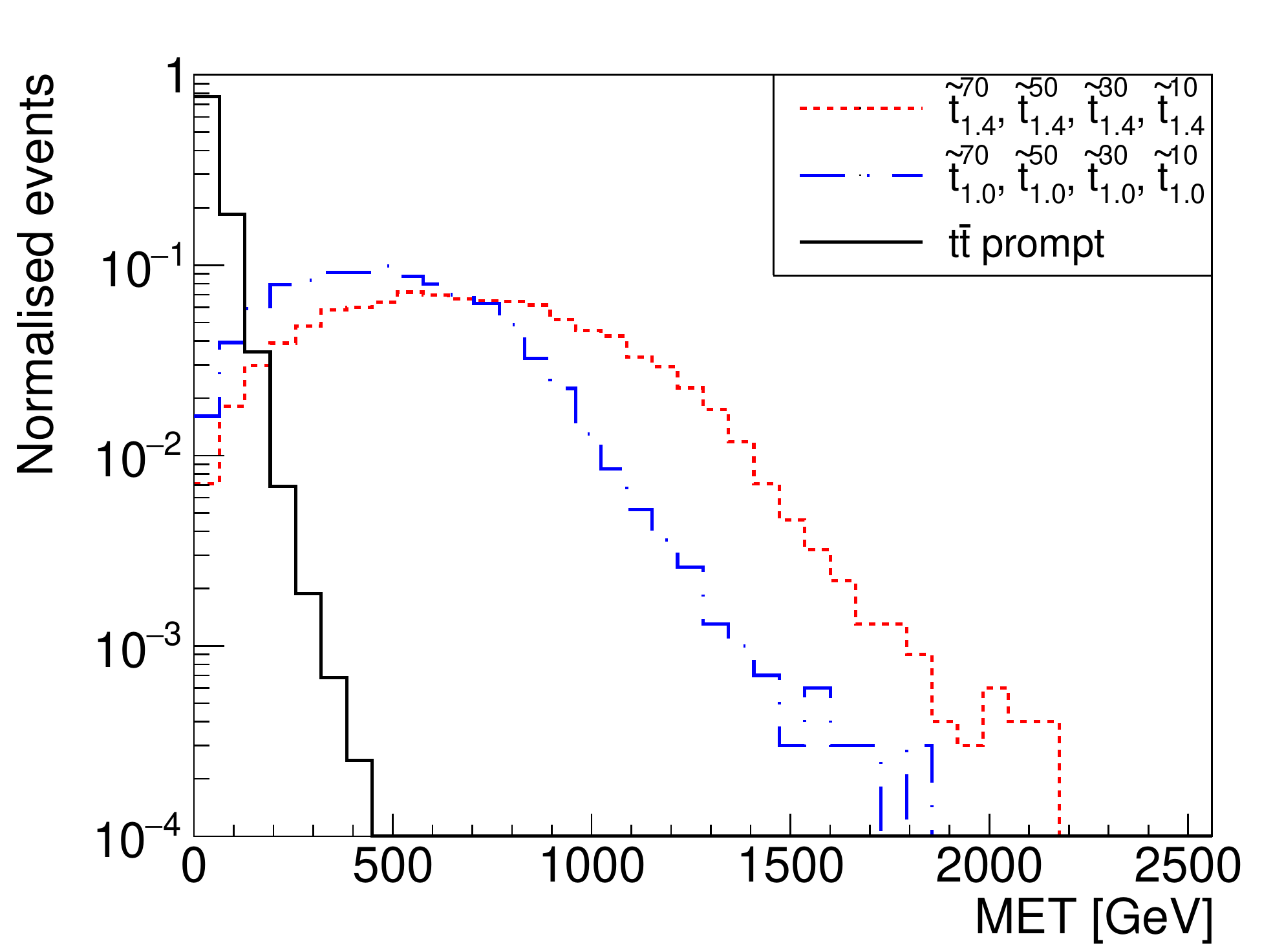}
    \caption{Missing transverse energy MET in the stop squark $\tilde{t}$ case.}
    \label{fig:stop_MET}
\end{minipage}
\hspace*{1cm}
\begin{minipage}[t]{0.45\linewidth}
    \centering
    \includegraphics[scale=0.4]{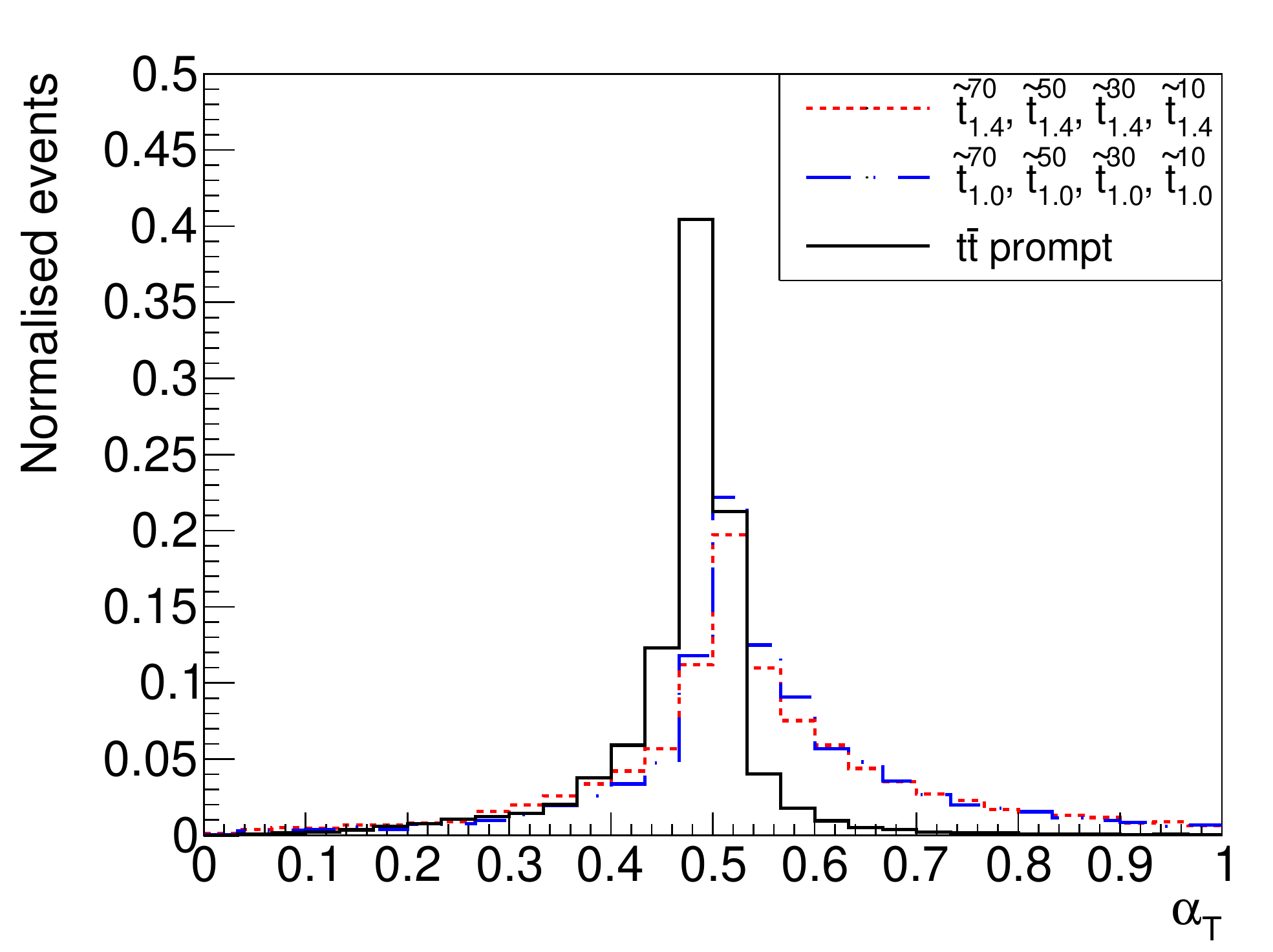}
    \caption{$\alpha_T$ distributions in the stop squark $\tilde{t}$ case.}
    \label{fig:stop_alphaT}
\end{minipage}
\begin{minipage}[t]{0.45\linewidth}
    \centering
    \includegraphics[scale=0.4]{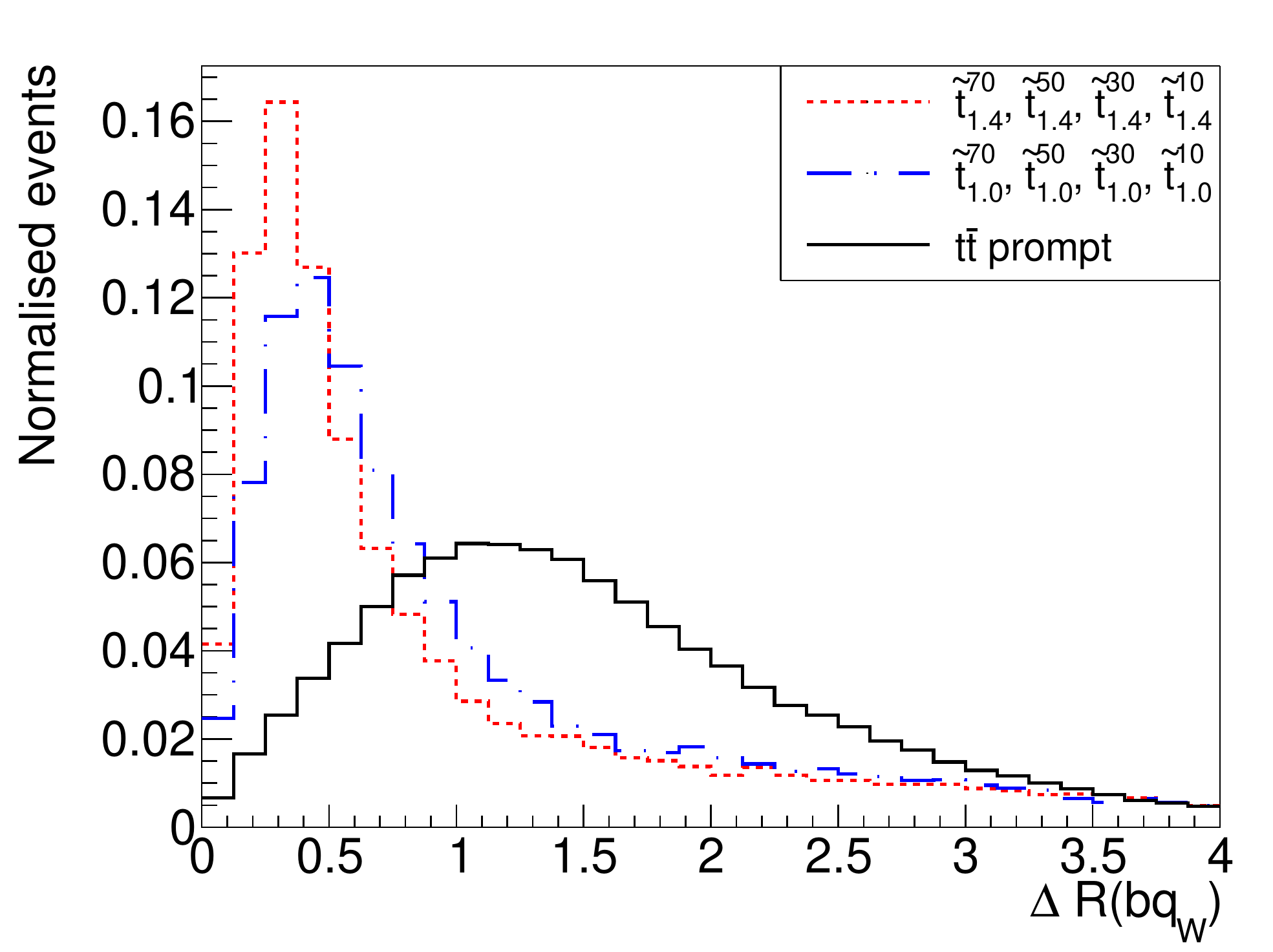}
    \caption{$\Delta R$ distributions in the stop squark $\tilde{t}$ case.}
    \label{fig:stop_deltaR}
\end{minipage}
\hspace*{1.4cm}
\begin{minipage}[t]{0.45\linewidth}
    \centering
    \includegraphics[scale=0.4]{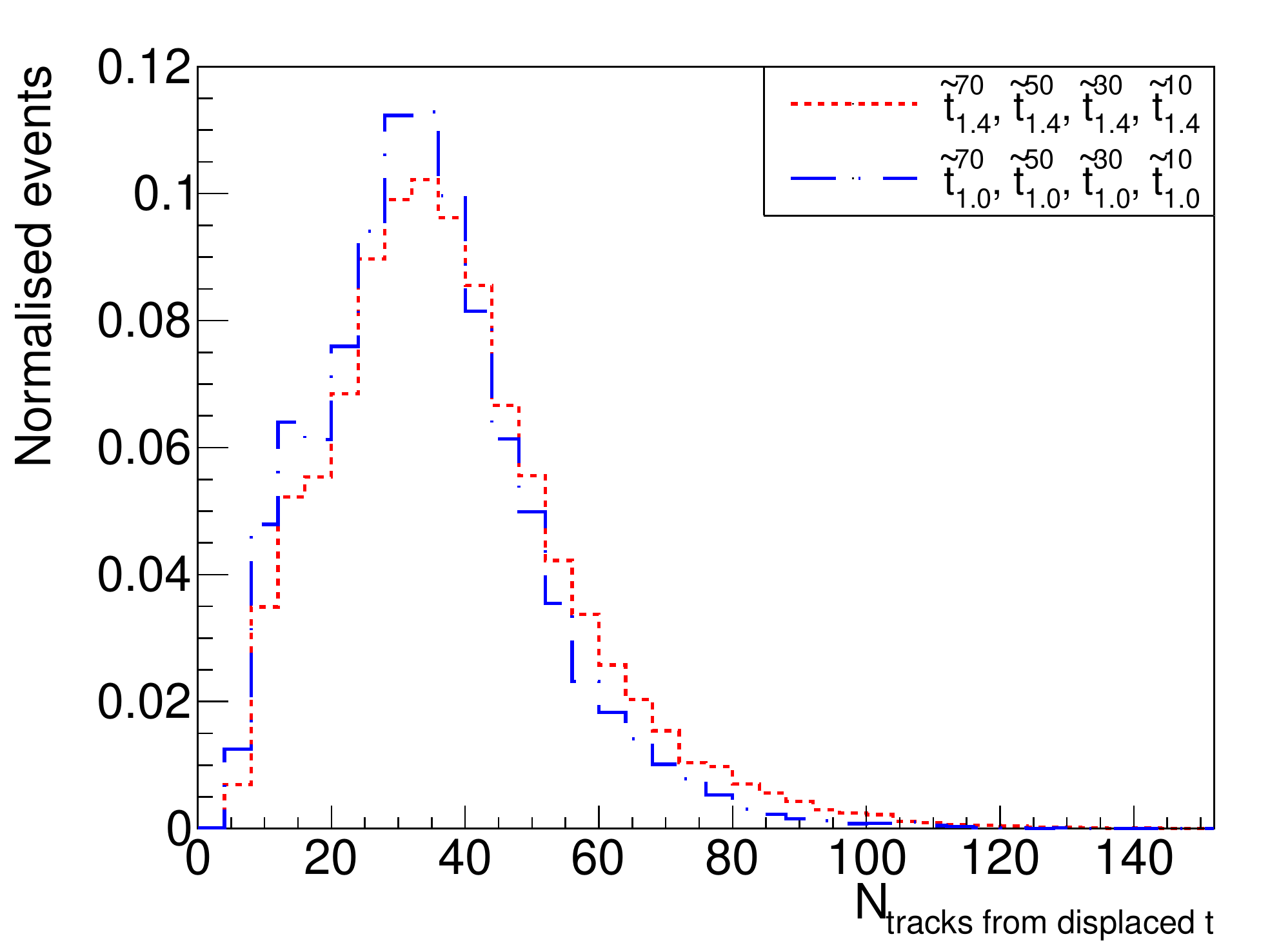}
    \caption{Multiplicity of tracks generated by the long-lived particle for the eight benchmarks in the stop squark $\tilde{t}$ case.}
    \label{fig:stop_ntracks}
\end{minipage}
\begin{minipage}[t]{0.45\linewidth}
    \centering
    \includegraphics[scale=0.4]{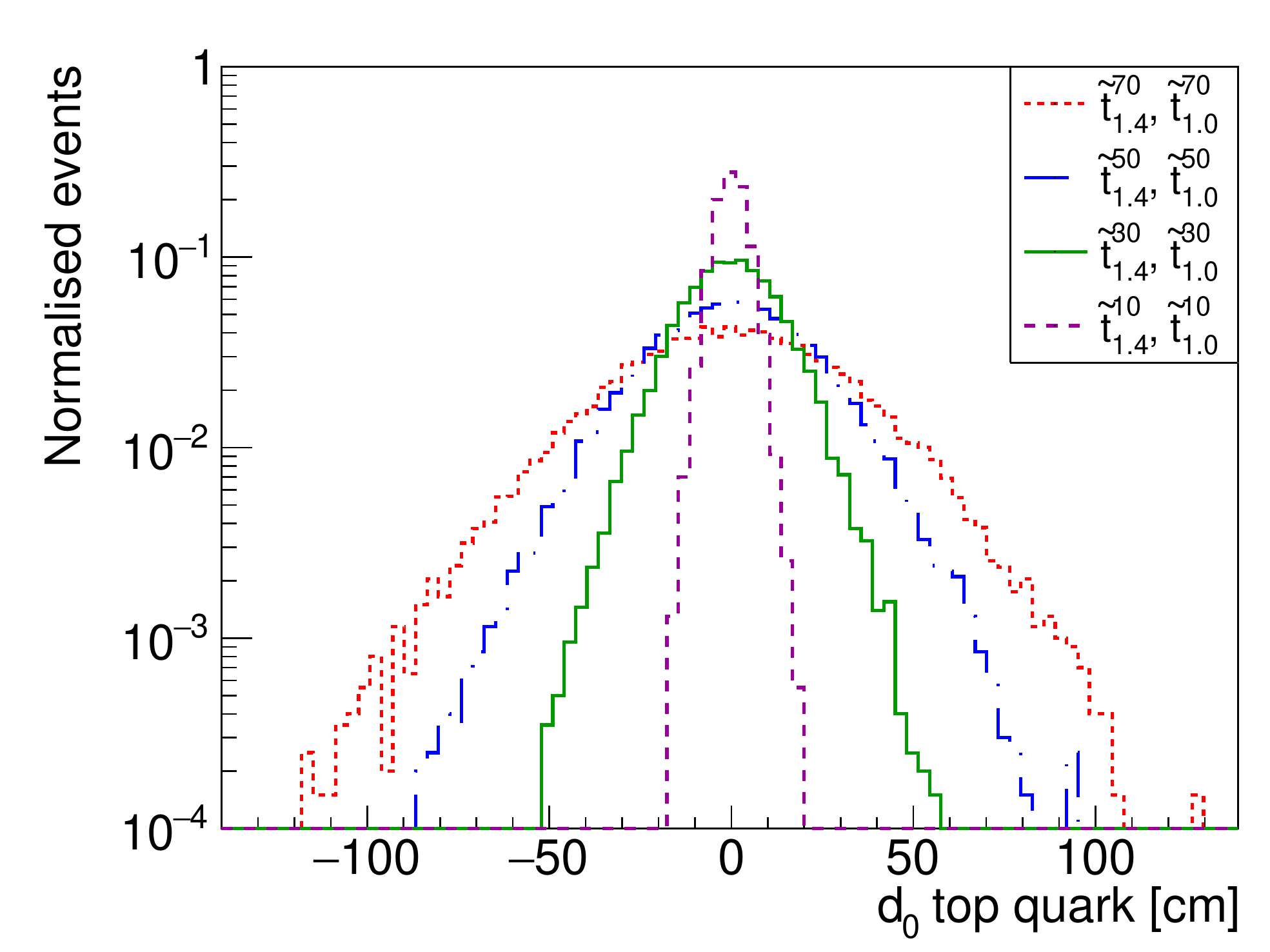}
    \caption{Transverse impact parameter $d_0$ distributions for the eight benchmarks for the long-lived stop squark with a magnetic field of $B=3.8\ \text{T}$.}
    \label{fig:stop_d0}
\end{minipage}
\hspace*{1cm}
\begin{minipage}[t]{0.45\linewidth}
    \centering
    \includegraphics[scale=0.4]{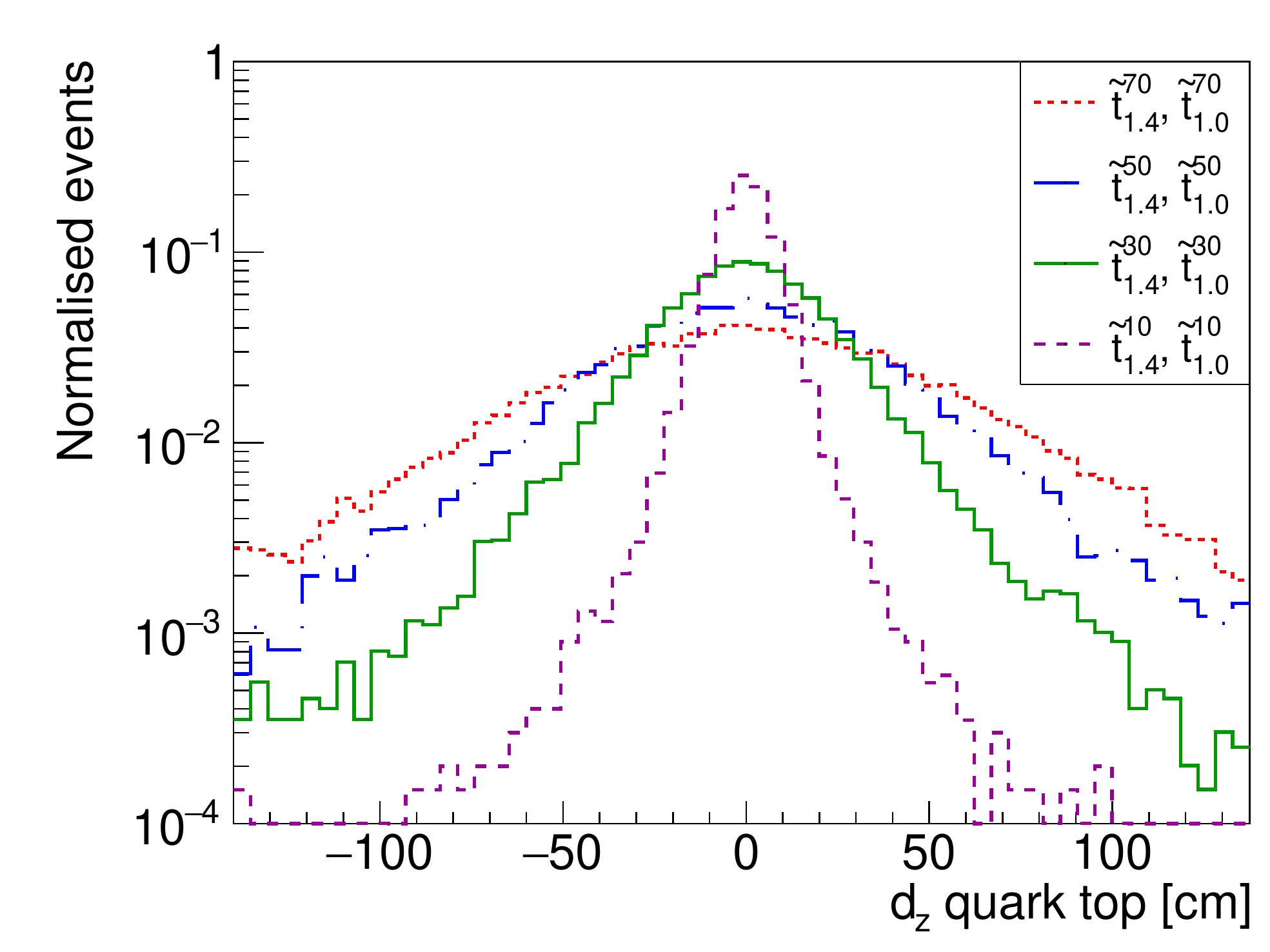}
    \caption{$d_z$ distributions for the eight benchmarks for the long-lived stop squark with a magnetic field of $B=3.8\ \text{T}$.}
    \label{fig:stop_dz}
\end{minipage}
\end{figure*}

\begin{figure*}[htbp]
\begin{minipage}[t]{0.45\linewidth}
    \centering
    \includegraphics[scale=0.4]{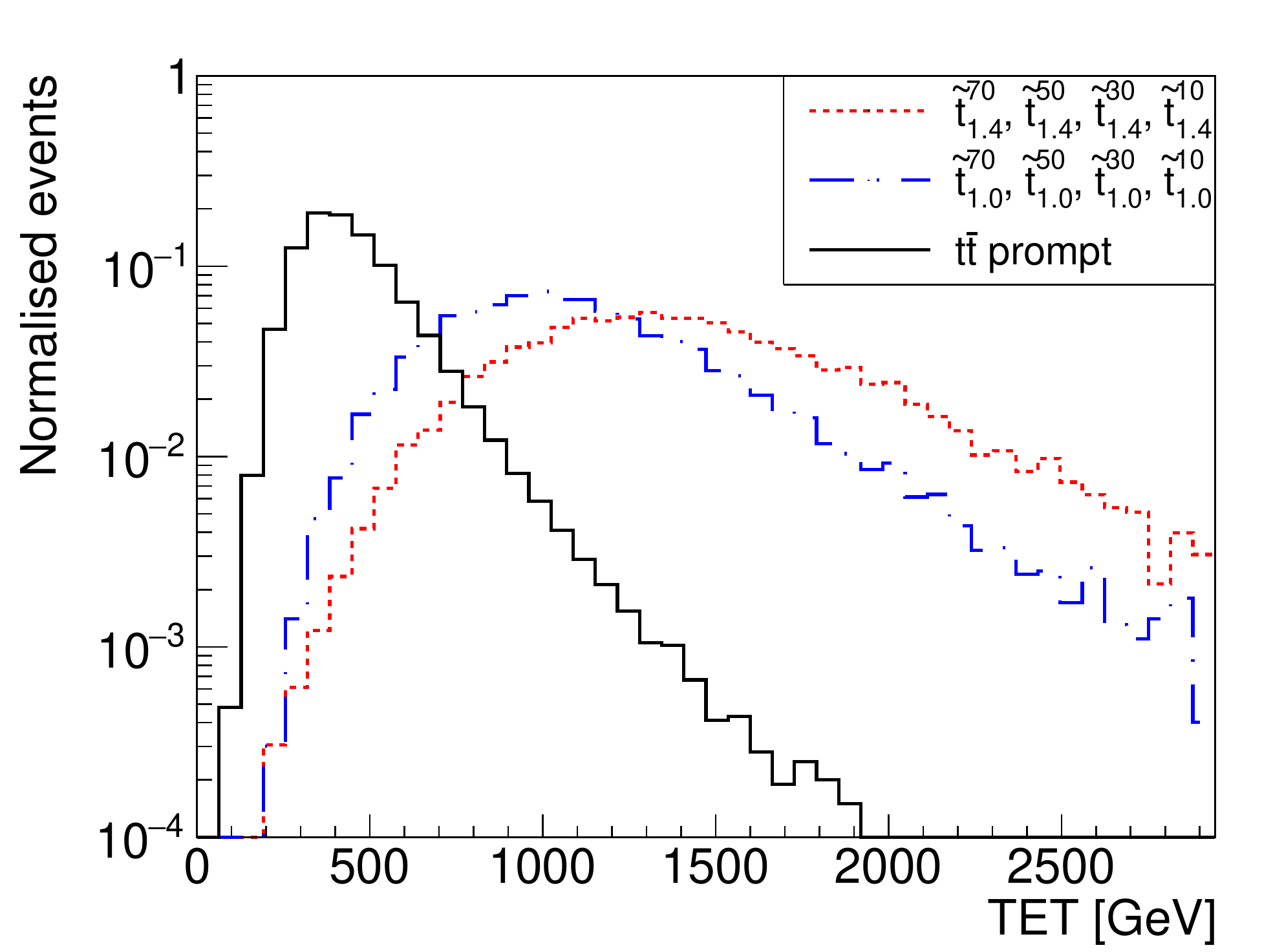}
    \caption{Total transverse energy TET distributions in the stop squark $\tilde{t}$ case.}
    \label{fig:stop_TET}
\end{minipage}
\hspace*{1cm}
\begin{minipage}[t]{0.45\linewidth}
    \centering
    \includegraphics[scale=0.4]{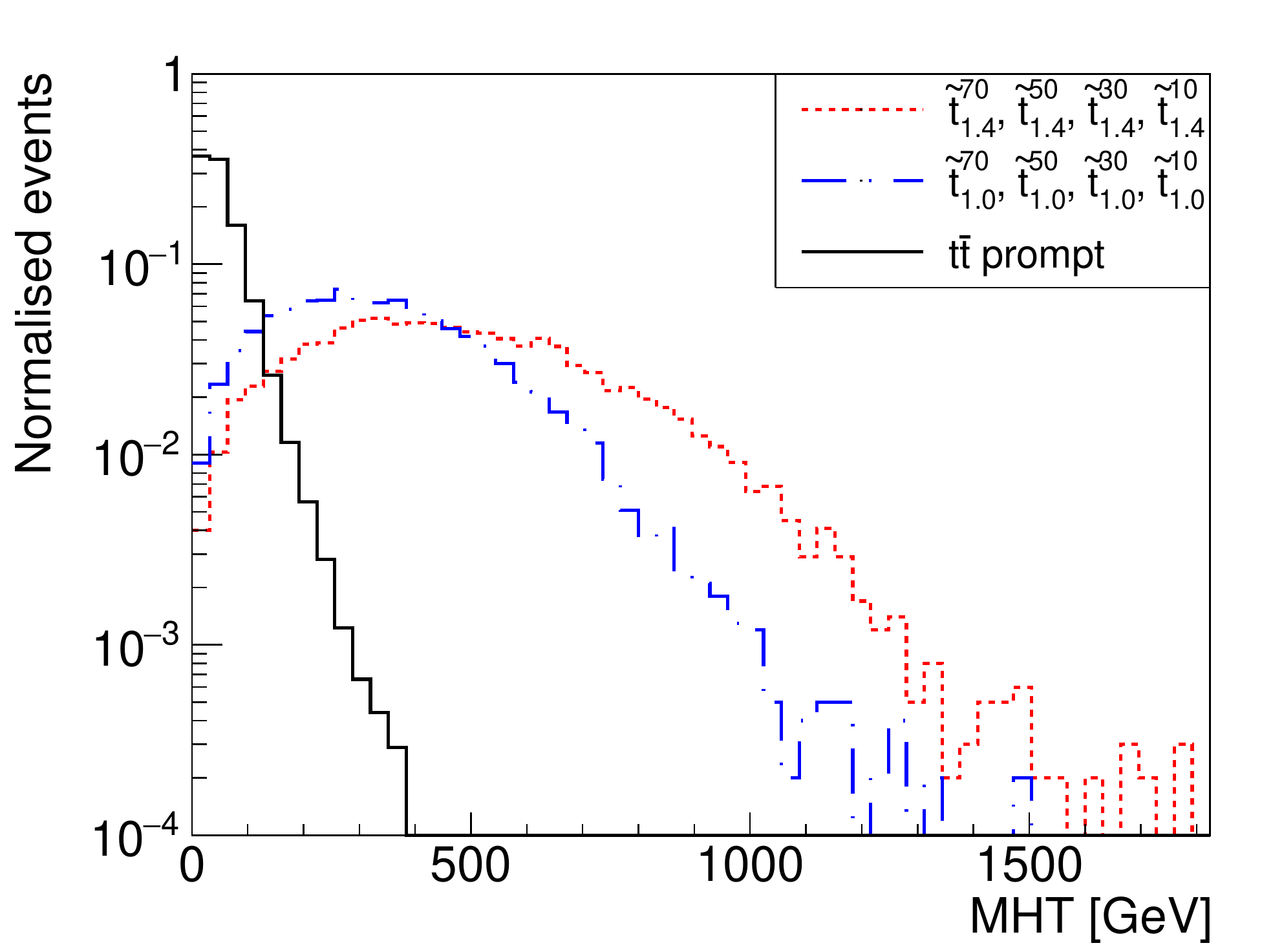}
    \caption{Missing transverse hadronic energy MHT in the stop squark $\tilde{t}$ case.}
    \label{fig:stop_mht}
\end{minipage}
\begin{minipage}[t]{0.45\linewidth}
    \centering
    \includegraphics[scale=0.4]{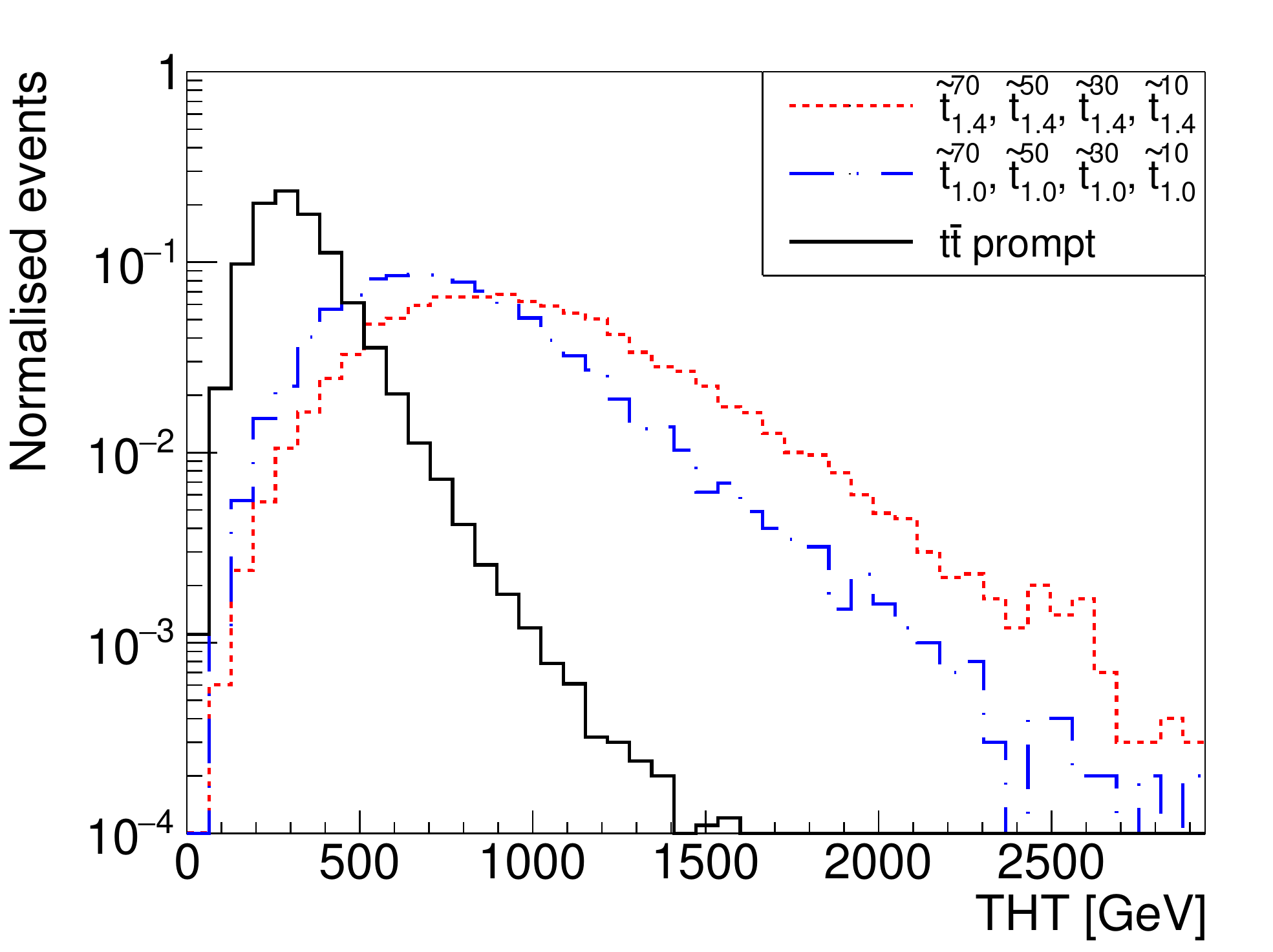}
    \caption{Total transverse hadronic energy in the stop squark $\tilde{t}$ case.}
    \label{fig:stop_THT}
\end{minipage}
\hspace*{1cm}
\begin{minipage}[t]{0.45\linewidth}
    \centering
    \includegraphics[scale=0.4]{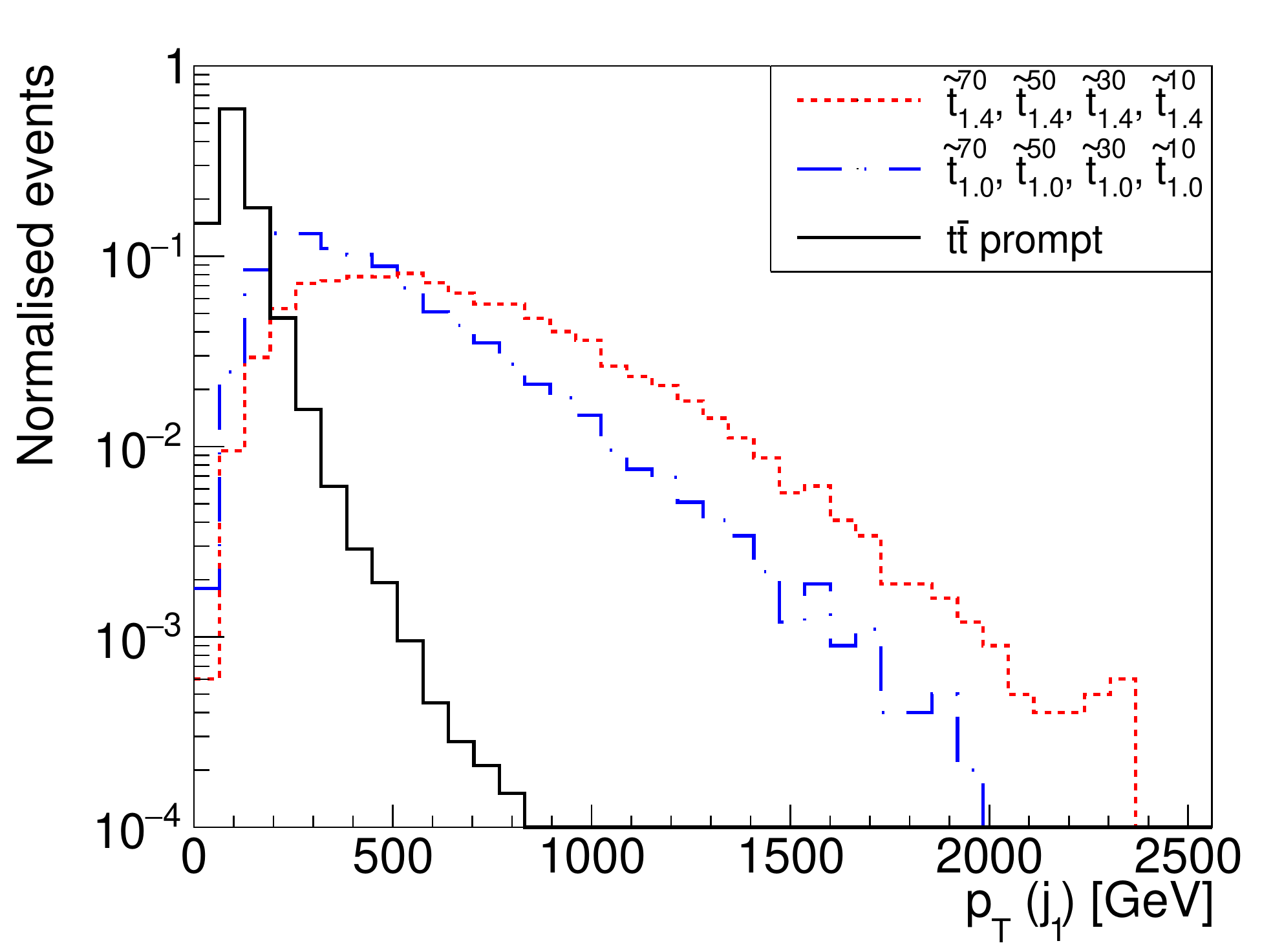}
    \caption{Transverse momentum $p_T$ of the leading jet in the stop squark $\tilde{t}$ case.}
    \label{fig:stop_ptj1}
\end{minipage}
\begin{minipage}[t]{0.45\linewidth}
    \centering
    \includegraphics[scale=0.4]{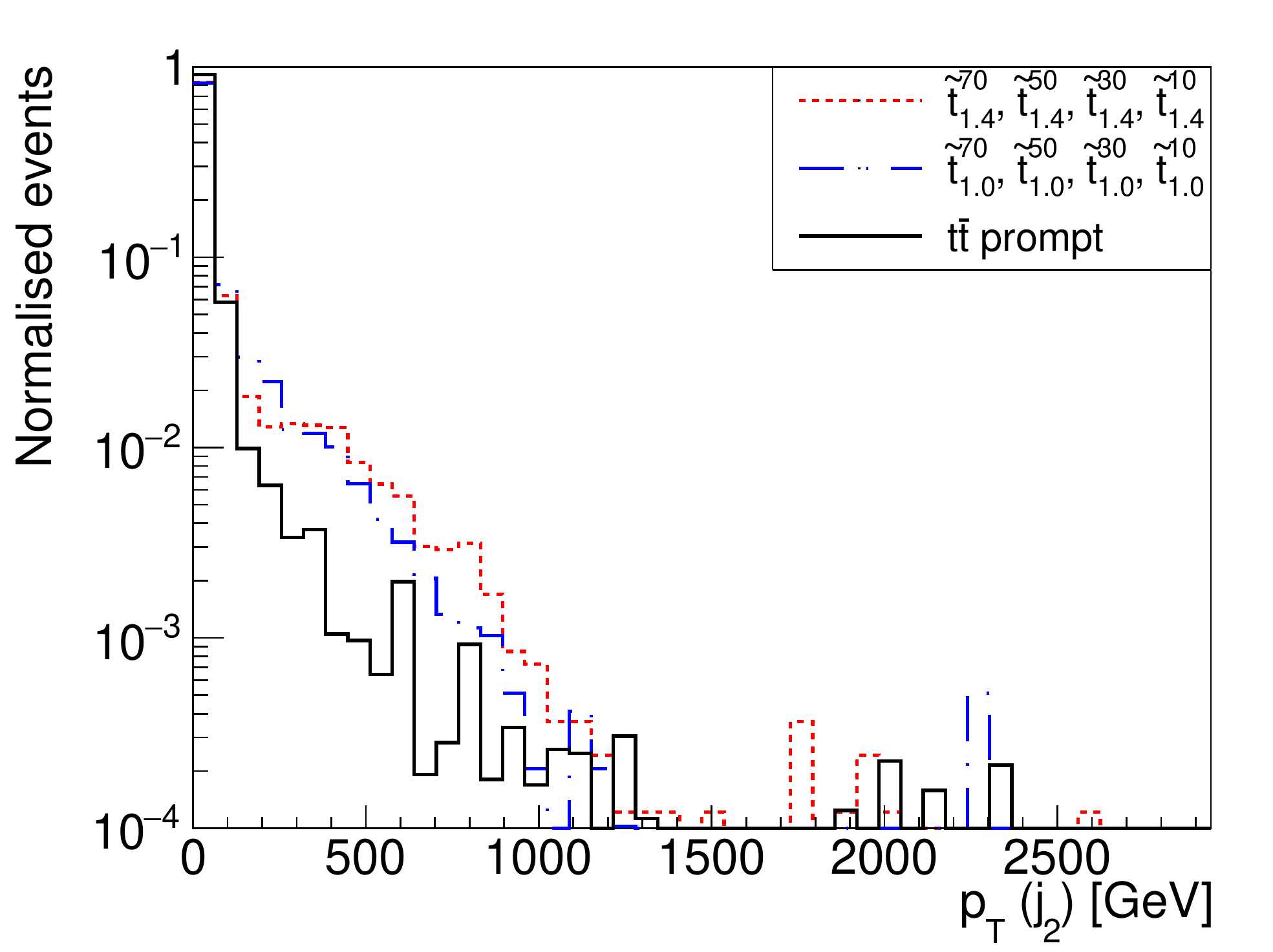}
    \caption{Transverse momentum $p_T$ of the second leading jet in the stop squark $\tilde{t}$ case.}
    \label{fig:stop_ptj2}
\end{minipage}
\hspace*{1cm}
\begin{minipage}[t]{0.45\linewidth}
    \centering
    \includegraphics[scale=0.4]{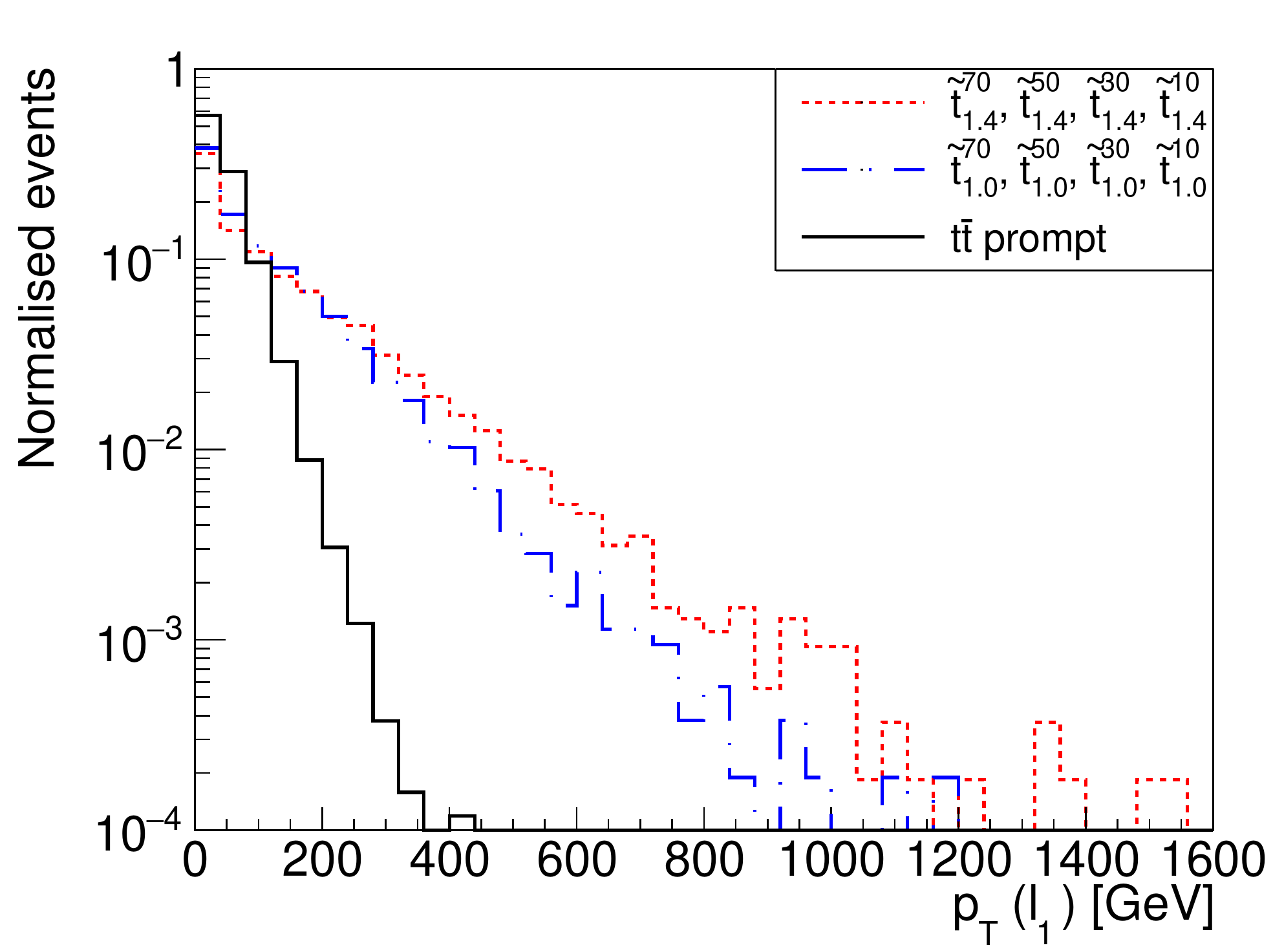}
    \caption{Transverse momentum $p_T$ of the leading lepton in the stop squark $\tilde{t}$ case.}
    \label{fig:stop_ptl1}
\end{minipage}
\end{figure*}

\begin{figure*}[htbp]
\begin{minipage}[t]{0.45\linewidth}
    \centering
    \includegraphics[scale=0.4]{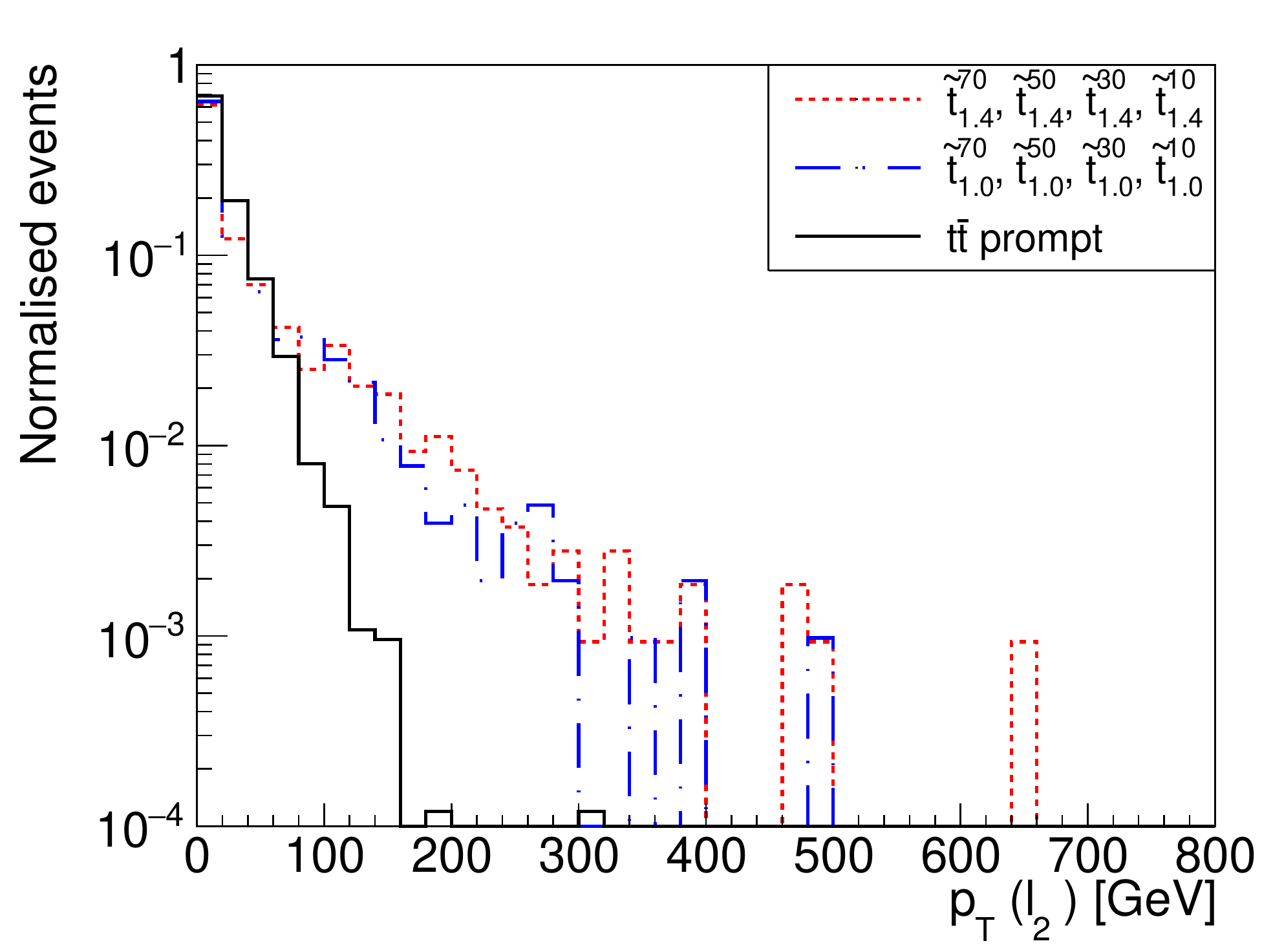}
    \caption{Transverse momentum $p_T$ of the second leading lepton in the stop squark $\tilde{t}$ case.}
    \label{fig:stop_ptl2}
\end{minipage}
\hspace*{1cm}
\begin{minipage}[t]{0.45\linewidth}
    \centering
    \includegraphics[scale=0.4]{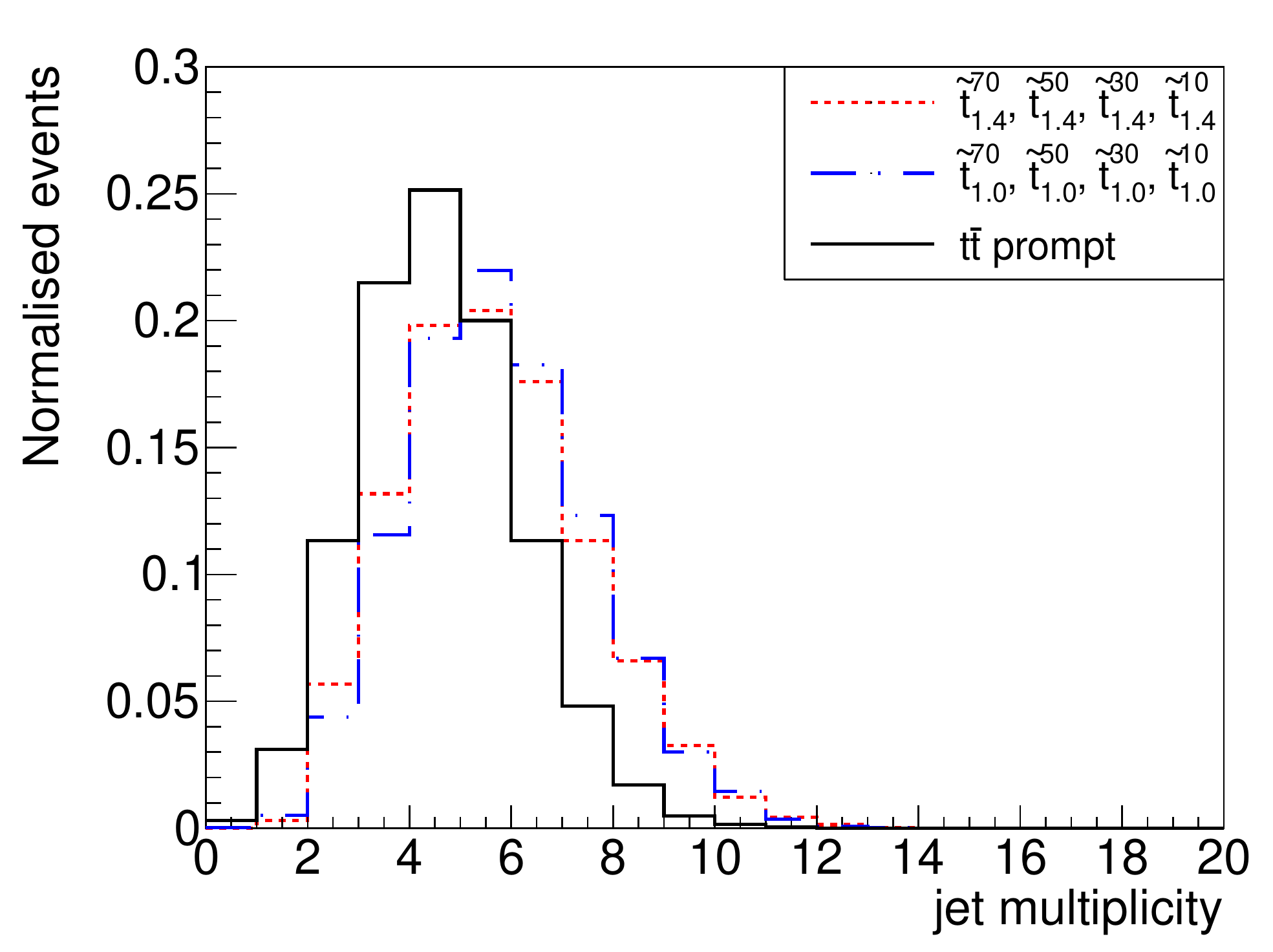}
    \caption{Jet multiplicity $N_j$ in the stop squark $\tilde{t}$ case.}
    \label{fig:stop_Nj}
\end{minipage}
\end{figure*}

\noindent Classical {geometric and kinematic} observables have been studied for each benchmark. Figs.~\ref{fig:stop_MET} to 
\ref{fig:stop_Nj} are a selection of some relevant observables for this model. A summary of the results can be found in Table \ref{table:eventninvol1}.\medskip

The first point to take into account is that the flight distance of the stop squark has no impact on several observables (see Table \ref{table:eventninvol1}) because the difference in gravitino masses in the benchmarks is negligible. The kinematics study will be mainly based on only two sets of distributions: a stop mass of $1.0\ \textrm{TeV}$ and  $1.4\ \textrm{TeV}$ regardless of the mass of the gravitino.\footnote{This simplification is valid for benchmarks with $\langle d^{\text{flight}}_{\tilde{t}} \rangle$ not {too} close to the upper limit of the tracker volume, \textit{i.e.}, 100 cm. If we assume a benchmark with $\langle  d^{\text{flight}}_{\tilde{t}} \rangle = 90\ \text{cm}$ or more, a non-negligible amount of stops will decay after the tracker volume which will impact the global observables reconstruction.} Distributions of {prompt top quark pair-production} ($pp\rightarrow t\bar{t}$) have also been added, which is one of the possible background sources. Moreover, the multiplicities of displaced tracks corresponding to one displaced signature are close for the two stop squark mass {scenarios}, with a mean value between 34 and 38.    \medskip

Nevertheless, the benchmarks differ for a fixed stop mass in terms of impact parameter distributions (see Figs.~\ref{fig:stop_d0} and \ref{fig:stop_dz} where the $d_0$ and $d_z$ distributions get wider for larger flight distance values).  The choice of the flight distance parameter also {leads} to a different multiplicity of {displaced top quark} signatures in {the same} event: two, one or {zero}. The fraction of events corresponding to each configuration is displayed in Table \ref{table:eventninvol1}. Except for the $\langle d^{\text{flight}}_{\tilde{t}}\rangle =10\ \text{cm}$ case, more than $90\%$ of the events have at least one stop squark decaying inside the tracker volume.    


\begin{table*}[htbp]
\begin{tabular*}{\textwidth}{@{\extracolsep{\fill}}lcccccccc|c}
\hline
 & $\tilde{t}_{1.0}^{10}$ & $\tilde{t}_{1.0}^{30}$ & $\tilde{t}_{1.0}^{50}$ & $\tilde{t}_{1.0}^{70}$ & $\tilde{t}_{1.4}^{10}$ & $\tilde{t}_{1.4}^{30}$ & $\tilde{t}_{1.4}^{50}$ & $\tilde{t}_{1.4}^{70}$ & $t\bar{t}$ \\
\hline
\hline
$\sigma_{LO}(pp\rightarrow \tilde{t}\bar{\tilde{t}})~[\text{fb}]$ & \multicolumn{4}{c|}{4.7} & \multicolumn{4}{c|}{0.4} &  - \\
\hline
$R_{\ =\ 2\ \textrm{displaced t}}$~[\%] & 60.0 & 92.5 & 96.3 & 93.9 & 67.8 & 94.7 & 97.3 & 94.7 & -  \\ 
$R_{\ \geq\ 1\ \textrm{displaced t}}$~[\%] & 71.8 & 97.4 & 99.0 & 96.6 & 77.2 & 98.1 & 99.0 & 96.7 & -  \\ 
\hline
$\langle \text{MET} \rangle$ [GeV]  & 495 & 494  & 494 & 493  & 696 & 692 & 698 & 698  & 43 \\ 
$\langle \text{TET} \rangle$ [GeV] & 1144  & 1138  & 1142  & 1141 & 1454 & 1447 & 1453 & 1451  & 465 \\ 
$\langle \text{THT} \rangle$ [GeV]   & 785 & 778 & 781 & 781 & 1000 & 993 & 1001 & 1003 & 327  \\ 
$\langle \text{MHT}\rangle $ [GeV]  & 351 & 348 & 349 & 348 & 491 & 486 & 491 & 488 & 52 \\
$\langle \alpha_T \rangle$& 0.55 & 0.55 & 0.55 & 0.55 & 0.53 & 0.53 & 0.54 & 0.54 & 0.48  \\  
\hline
$\langle p_T (j_1)\rangle $ [GeV]  & 459 & 459 & 460 & 460 & 668 & 664 & 667 & 663 & 118 \\ 
$\langle p_T (j_2)\rangle $ [GeV]  & 30 & 30 & 56 & 47 & 38 & 55 & 70 & 55 & 23 \\ 
$\langle p_T (\ell_1)\rangle $ [GeV]  & 105 & 105 & 106 & 105 & 133 & 134 & 136 & 136 & 32 \\
$\langle p_T (\ell_2)\rangle $ [GeV]  & 31 & 32 & 32 & 32 & 40 & 40 & 35 & 37 & 31 \\
$\langle \Delta R (bq_W)\rangle $   & 1.0 & 1.0 & 1.0 & 1.0 & 0.9 & 0.9 & 0.9 & 0.9 & 1.5  \\ 
\hline
$\langle |d_0(t)|\rangle $ [cm] & 3.2 & 9.6 & 15.9 & 22.3 & 3.6 & 10.7 & 17.9 & 25.0 & -  \\
$\langle |d_z(t)|\rangle $ [cm] & 7.2 & 21.1 & 34.0 & 46.4 & 7.1 & 20.8 & 33.9 & 46.2 & -  \\
\hline
$N_{\text{jets}}$ & 5.3 & 5.3 & 5.3 & 5.3 & 5.1 & 5.2 & 5.2 & 5.2 & 4.3 \\
$N_{\text{tracks from displaced t}}$ & 34.3 & 34.0 & 34.0 & 34.1 & 37.6 & 37.4 & 37.6 & 37.6 & -  \\
\hline
\end{tabular*}
\caption{Mean values of some observables and number of events with top quarks decaying in the tracker volume for the eight benchmarks of the stop squark pair production. We remind that $R_{\geq \textrm{1 displaced t}}$ corresponds to the number of events with at least one top quark decaying in the tracker volume. }
\label{table:eventninvol1}
\end{table*}

As expected, the heavier stop squark events  induce a larger mean value on the leading jet $p_T$, the THT, and the MET (around $460$, $780$ and $500\ \text{GeV}$, respectively, for $m_{\tilde{t}}=1.0\ \text{TeV}$ and $660$, $1000$ and $690\ \text{GeV}$, respectively, for $m_{\tilde{t}}=1.4\ \text{TeV}$). Two particles contribute to the MET in our events: the LSP gravitino and the neutrinos coming from the leptonic decay of the W bosons (this latter contribution is, however, negligible in comparison to the gravitino). The MET in the prompt-top {background} (which only comes from the neutrinos) is relatively small {(around 40 GeV, see Table~\ref{table:eventninvol1})}. The leptons coming from the decay of the W bosons {also get} a high transverse momentum with mean value of $105$ and $135\ \text{GeV}$ for $m_{\tilde{t}}=1.0$ and $1.4\ \text{TeV}$, respectively. 

Figure \ref{fig:stop_alphaT} presents the distributions of $\alpha_T$ (as defined in Section \ref{subsec:obs}). The two distributions of signals are symmetric with a central value of about $0.5$, {which is the typical value from which the QCD contribution decreases fiercely and vanishes.} This can be seen on the top-prompt distribution where the majority of the events correspond to $\alpha_T< 0.5$. An offline selection on this observable may {therefore be} useful to increase the signal over background ratio.\medskip 

From the large mean values of the transverse momentum of the hardest jet $\langle p_T(j_1)\rangle$ in Table \ref{table:eventninvol1}, we may identify boosted topologies in our events. It may impact the jet reconstruction in our signal. Indeed, as the b quark and the quarks from the W boson decay (coming from the decay of the top quark) get closer, they appear as one fat jet instead of several jets. To investigate this effect, the distributions of $\Delta R(bq)=\sqrt{\Delta\eta(bq)^2+ \Delta\phi(bq)^2}$ have been displayed (see Fig.~\ref{fig:stop_deltaR}). This variable characterises the angular distance between b quarks and the closest light quark coming from the W boson decay, which also arises from $t\rightarrow bW$. Through the anti-$k_T$ algorithm with $\Delta R =0.4$ and $p_T^{min.}=30\ \text{GeV}$, a non-negligible portion of jets merge into one single jet. 
Therefore, an analysis of the jets substructure may be relevant to identify the signal. {In this case, a larger cone $\Delta R$ must be used, leading to the reconstruction of a \textsl{fat jet} allowing us to study its substructure.} 


\subsection{Mediation by a coloured long-lived particle (R-hadron based on a gluino)} 

The review of {scenarios} involving \textsl{displaced top quarks} signatures cannot be complete without taking into account the production of long-lived gluinos. Although this event topology is already studied at the LHC, a devoted simplified model has been designed based on the split supersymmetry framework.

\subsubsection{Model description}

\begin{figure}[htbp]
    \centering
    \includegraphics[scale=0.15]{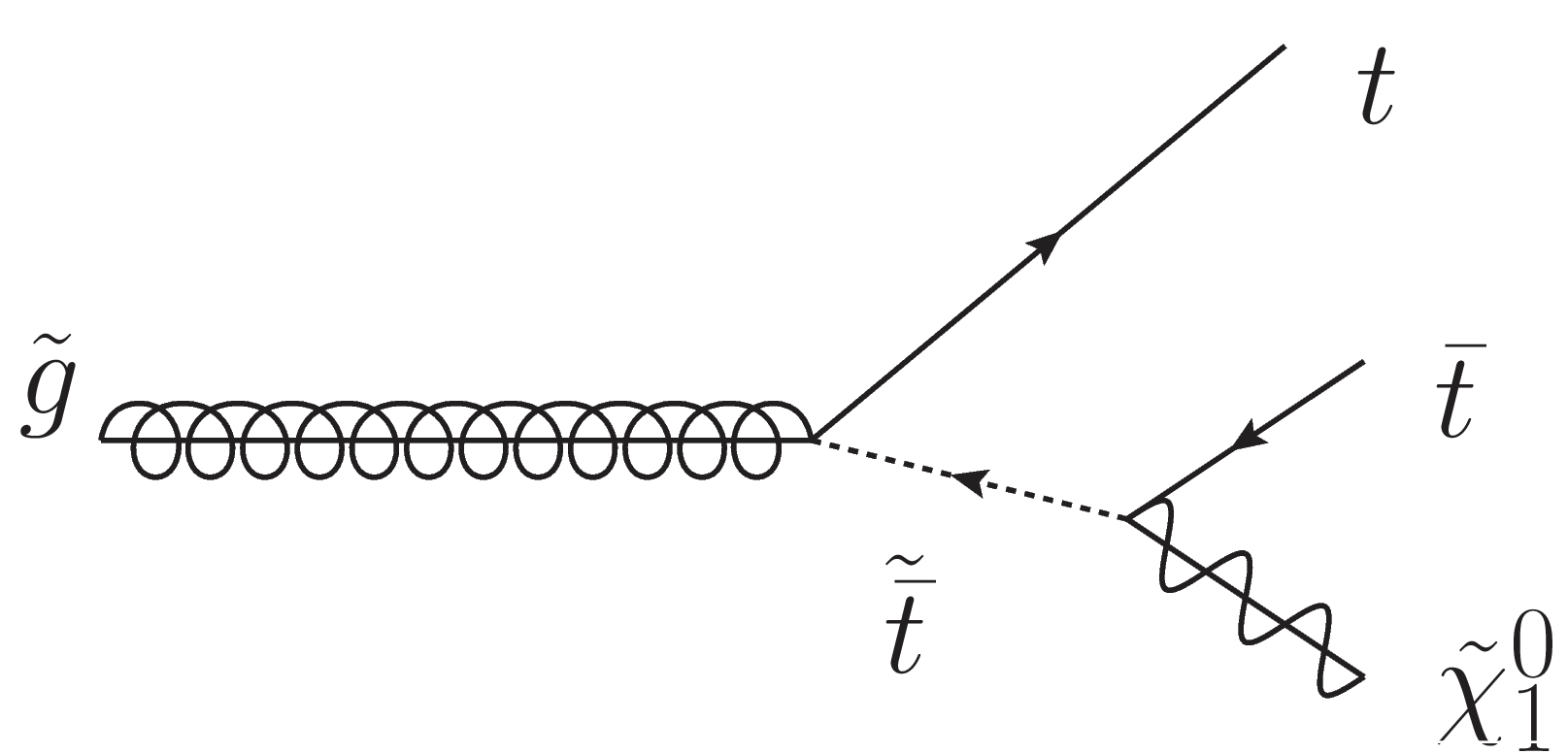}
    \caption{Feynman diagram describing the gluino decay in a top quark pair and a neutralino}
    \label{fig:gluino_lodecay}
\end{figure}

The split supersymmetry model~\cite{Giudice_2005,Arkani_Hamed_2005} offers a superpartner particle spectrum split into two categories. From one side, the scalar particles, \textsl{i.e.}, squarks and sleptons, acquire a mass close to the energy scale of supersymmetry breaking which can be as large as the GUT scale. From the other side, the fermions, the neutralinos and charginos have a mass near the electroweak scale and can potentially solve the gauge-coupling unification and dark matter puzzles of the Standard Model. In this context, assuming that the lightest neutralino is the LSP and the gluino is the NLSP, the allowed decay channels for the gluino infer a tiny decay width because they involve a heavy-massive mediator as shown in Fig.~\ref{fig:gluino_lodecay}. Therefore the gluino (or more precisely the R-hadron containing the gluino) is expected to be long-lived.\\

This simplified model is based on the R-parity conserved MSSM where the spectrum of masses satisfies the two typical scales of the split supersymmetry framework. As in the previous models,  we define the stop squark as an equal combination of left-handed and right-handed states. Concerning the neutralino, we only study the gauginos and the higgsinos, without any combinations.

\subsubsection{Cross sections and decays}

\begin{figure}[htbp]
    \centering
   \includegraphics[scale=0.30]{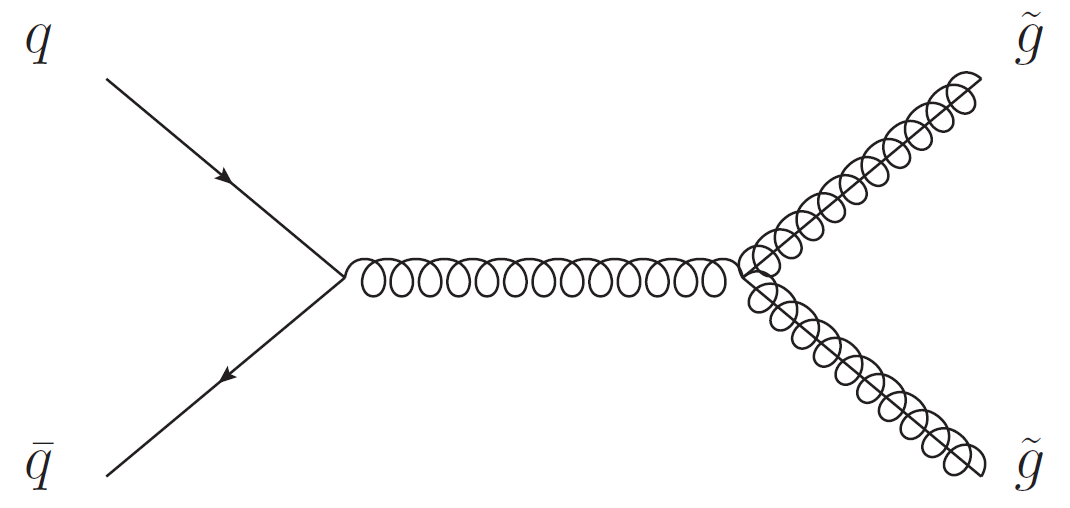}
    \caption{Feynman diagram of gluino pair production by quark annihilation}
    \label{fig:gluino-prod1}
\end{figure}

\begin{figure}[htbp]
    \centering
    \includegraphics[scale=0.2]{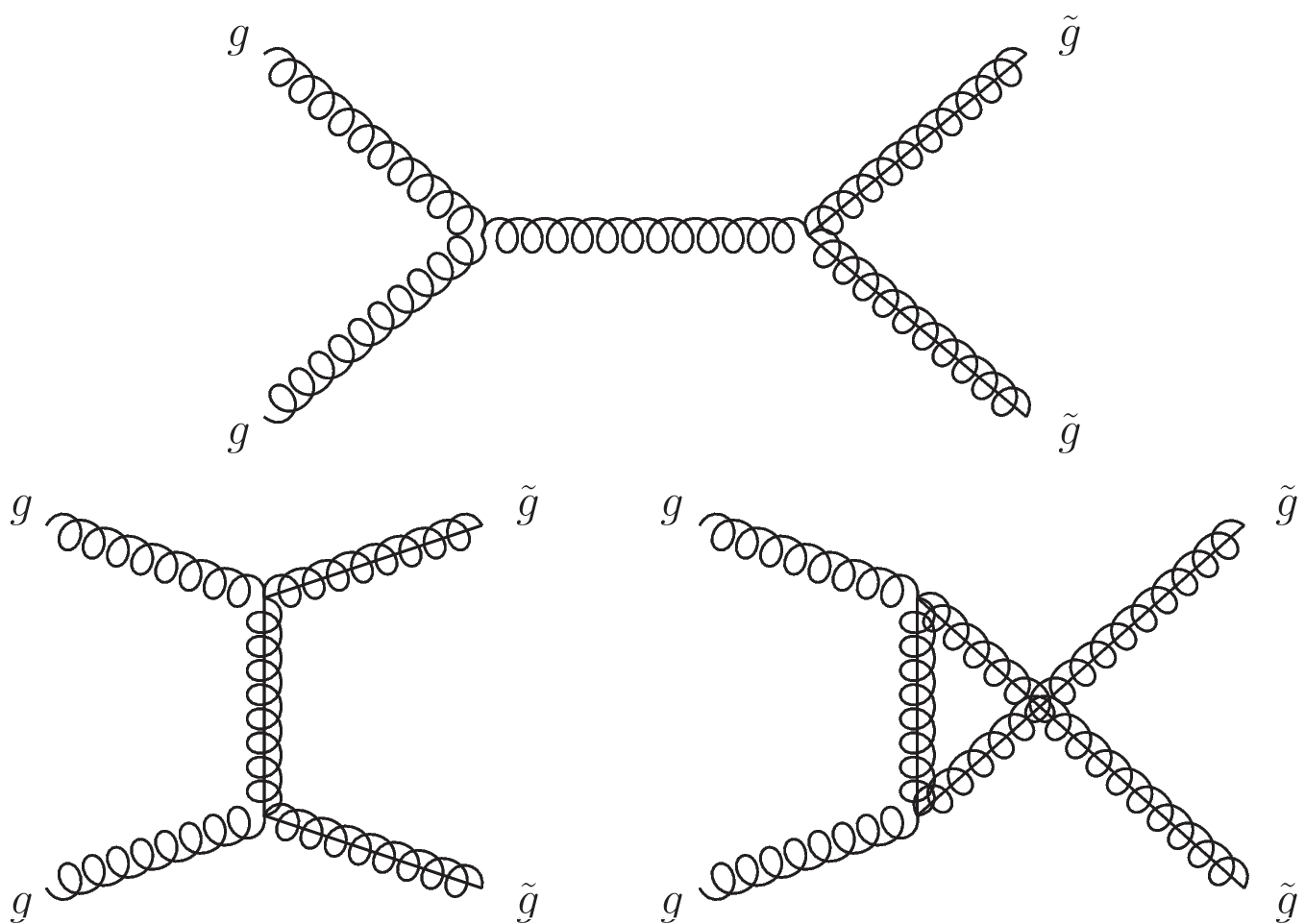}
    \caption{Feynman diagrams of gluino pair production by gluon fusion}
    \label{fig:gluino-prod2}
\end{figure}

Gluino production at the LHC is very similar to the stop production described in the previous section. At LO QCD, the gluinos can be produced by quark annihilation (Fig.~\ref{fig:gluino-prod1}) or by gluon fusion (Fig.~\ref{fig:gluino-prod2}). Diagrams involving a squark particle mediator can be discarded by requiring squark masses greater than 1 TeV, which is consistent with the split supersymmetry assumption. Such diagrams should give a negligible contribution to the gluino production. Therefore, the corresponding cross section depends only on the gluino mass.\\

\begin{figure}[htbp]
\centering
    {\centering
    \includegraphics[scale=0.43]{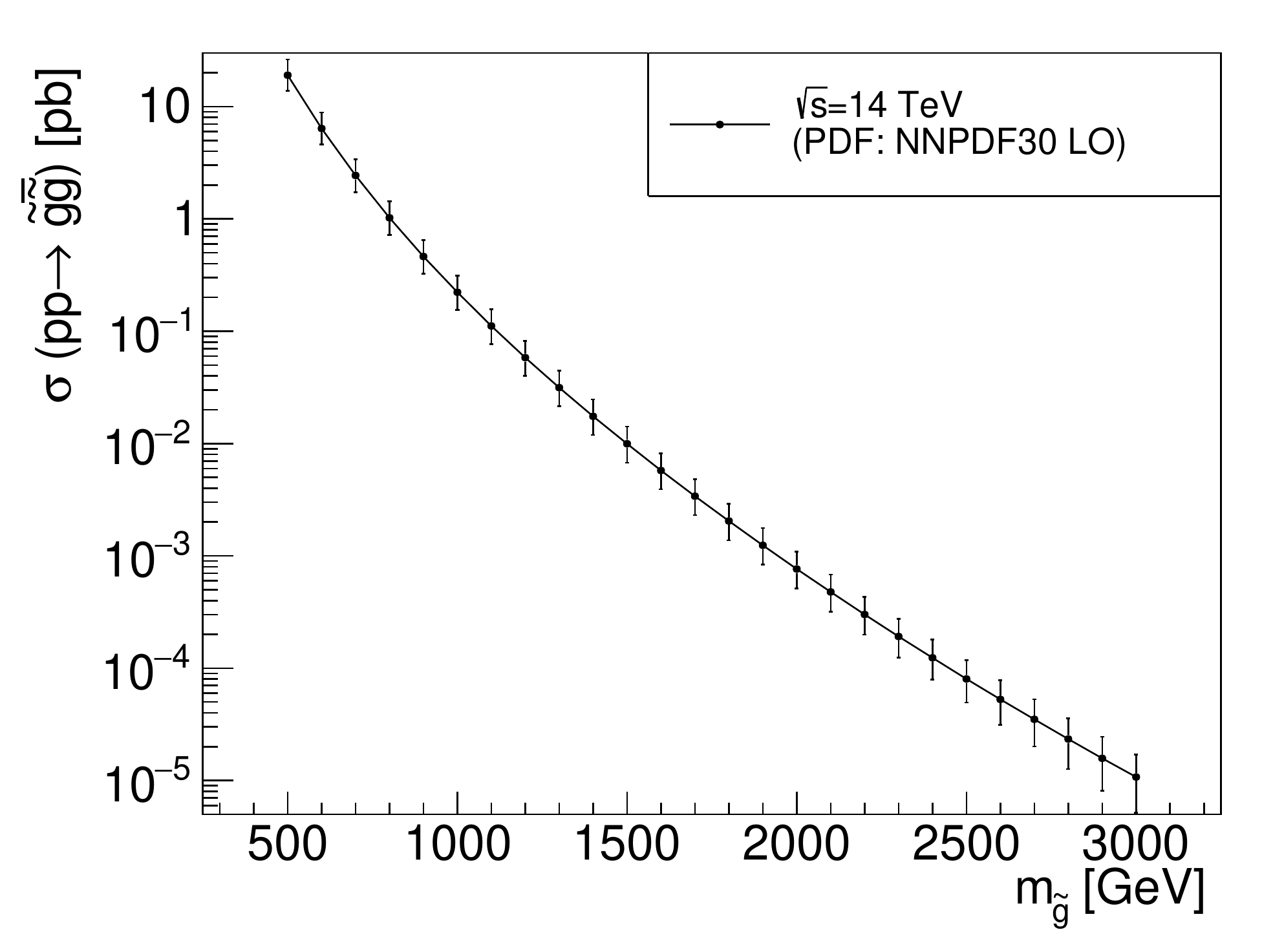}
    \caption{Leading order cross section $\sigma (pp\rightarrow \tilde{g}\tilde{g})$ at $\sqrt{s}=14\ \text{TeV}$ as a function of the gluino mass $m_{\tilde{g}}$. The uncertainties are related to PDF variations and the factorisation and renormalisation scales.}
    \label{fig:gluino-xsection}}
\end{figure}

\begin{figure}[htbp]
    {\centering
    \includegraphics[scale=0.43]{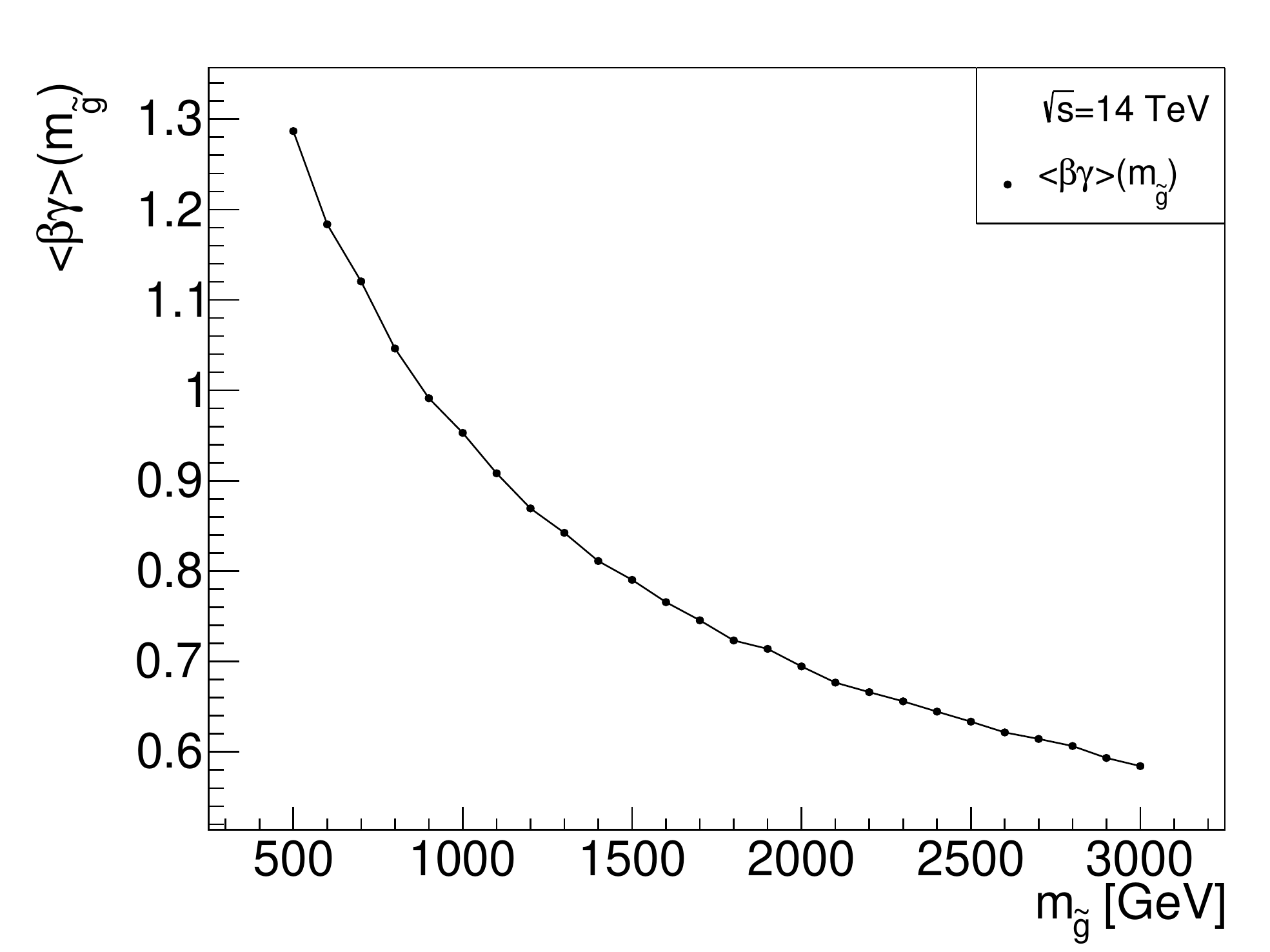}
    \caption{Average value of the Lorentz factor $\beta\gamma$ of gluinos as a function of the gluino mass $m_{\tilde{g}}$.}
    \label{fig:gluino-betagamma}}
\end{figure}

Figures~\ref{fig:gluino-xsection} and \ref{fig:gluino-betagamma} show the evolution of the cross section production and the mean Lorentz factor $\langle \beta\gamma \rangle$ of the gluino as a function of the gluino mass $m_{\tilde{g}}$. The calculations have been achieved at the LO QCD and more accurate values of the cross section can be found in the literature~\cite{Beenakker:1996ch,Kauth:2011vg}. With a gluino mass of $1\ \text{TeV}$, and assuming some theoretical assumptions, the K-factor $K=\sigma_{NLO}/\sigma_{LO}$ for the gluino production is higher than 2. We consider here only uncertainties on the PDF and the renormalisation and factorisation scales. As expected, both observables decrease with the gluino mass. The evaluation of the cross section production will help us to fix the gluino mass in the benchmarks for a reasonable number of events at the Run~3. The mean Lorentz factor $\langle \beta\gamma \rangle$ (see Fig.~\ref{fig:gluino-betagamma}) varies between 0.6 and 1.3 for the chosen masses, so the flight distance of the gluino is not enhanced, as it was the case for the stop production.\medskip


\begin{figure}[htbp]
    {\centering
    \includegraphics[scale=0.43]{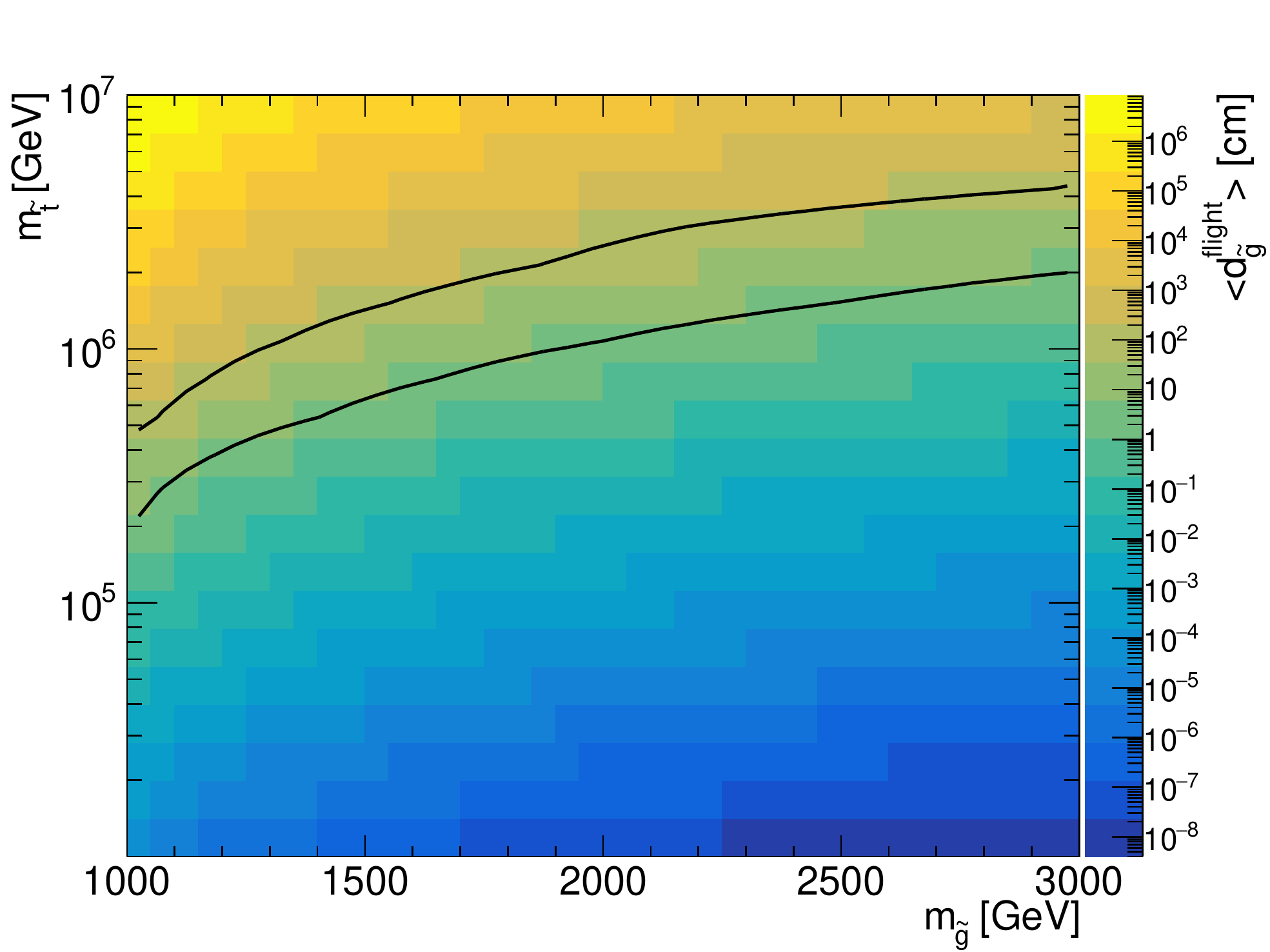}
    \caption{Average flight distance  in the laboratory frame of the gluino $\langle d_{\tilde{g}}^{\textrm{flight}}\rangle$ as a function of the gluino mass $m_{\tilde{g}}$ and the stop mass $m_{\tilde{t}}$ for a fixed value of the neutralino mass $m_{\tilde{\chi}^0_1}=250~\textrm{GeV}$ and $\tilde{\chi}^0_1=\tilde{B}$.  The black lines represent the geometrical limits of the tracker volume.}
    \label{fig:gluino-bgctau1}}
\end{figure}

\begin{figure}[htbp]
    {\centering
    \includegraphics[scale=0.43]{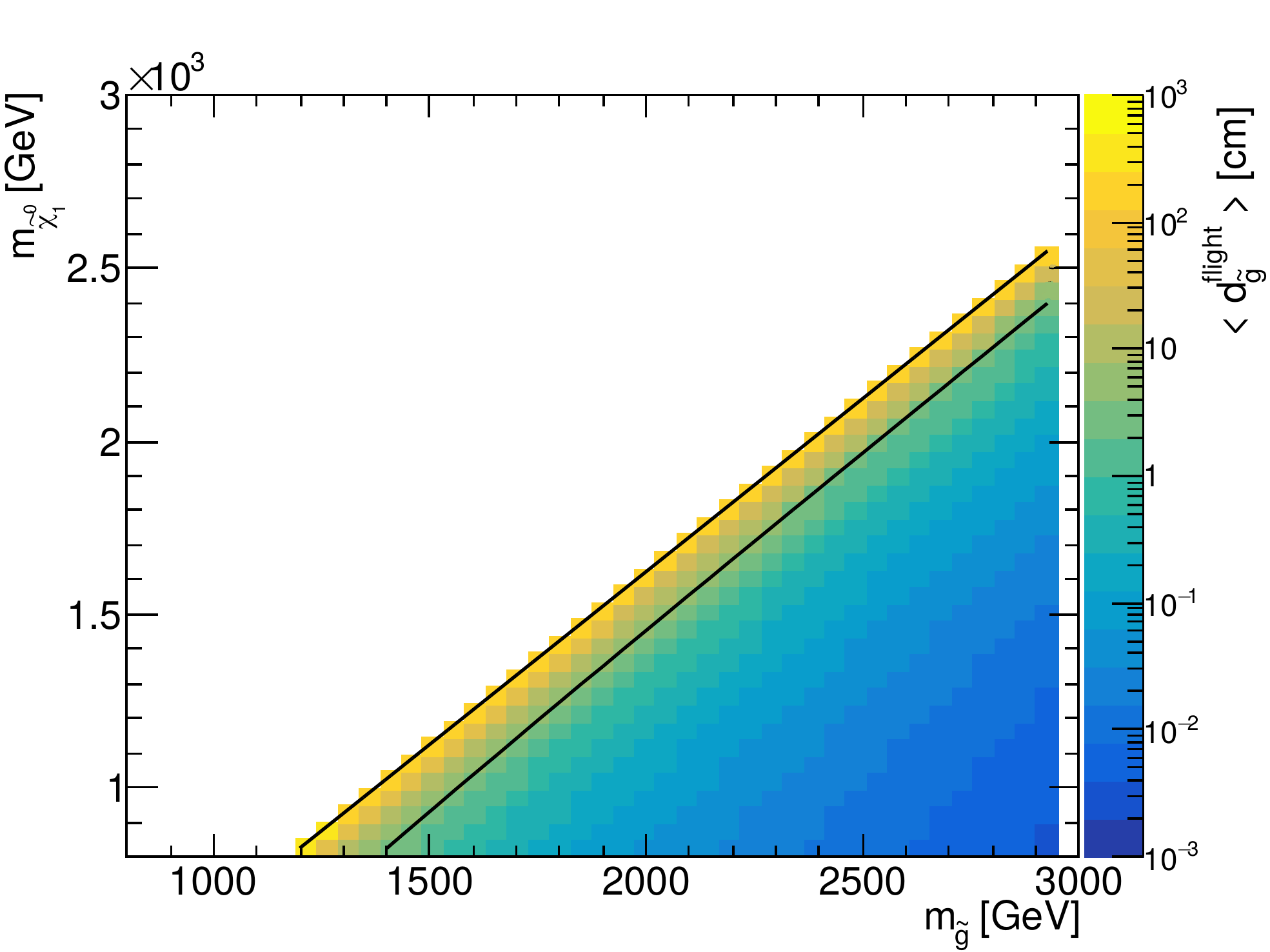}
    \caption{Average flight distance  in the laboratory frame of the gluino $\langle d_{\tilde{g}}^{\textrm{flight}}\rangle$ as a function of the gluino mass $m_{\tilde{g}}$ and the neutralino mass $m_{\tilde{\chi}^0_1}$ for a fixed value of the stop mass $m_{\tilde{t}}=500~\textrm{TeV}$ and $\tilde{\chi}^0_1=\tilde{B}$. The black lines represent the geometrical limits of the tracker volume. The white region is forbidden from the constraint $m_{\tilde{\chi}^0_1}+2m_{t} < m_{\tilde{g}}$.
    \label{fig:gluino-bgctau2}}}
\end{figure}

At LO QCD, the only possible decays have the following topology: $\tilde{g}\rightarrow\tilde{\chi}^0_1 q \bar{q}$ where $q$ is a quark, and potentially a top quark if the neutralino mass is sufficiently smaller than the gluino one (see Fig.~\ref{fig:gluino_lodecay}). {For simplification purposes, we} assume in the following that the branching fraction of this decay channel involving top quarks is 100\%. It is worth to note that other channels can appear at the NLO QCD but they are not taken into account in this study. Therefore, the gluino decay width (and the gluino flight distance) depends on three masses (the gluino mass, the stop mass, the neutralino mass) and the nature of the neutralino, \textsl{i.e.}, bino, wino or (up-like only) higgsino. Figures~\ref{fig:gluino-bgctau1} and \ref{fig:gluino-bgctau2} illustrate this dependence. The first figure shows the average flight distance of the gluino in the gluino~mass~-~stop~mass plane for a fixed value of neutralino mass equal to 250~GeV and a neutralino defined as a bino state. The second one shows the variation of the same quantity in the gluino mass - neutralino {mass} plane for a fixed value of stop mass equal to 500~TeV and a neutralino defined as a bino state. The nature of the neutralino has a negligible impact on the order of magnitude of the gluino flight distance. 
We conclude that a region of the parameter space allows us to observe a \textsl{displaced top quark} signature, especially for a stop mass between 100 TeV and 1000 TeV when the gluino mass is about 1 - 3 TeV.



\subsubsection{Benchmarks definition}

 \begin{table*}[htbp!]
\begin{tabular*}{\textwidth}{@{\extracolsep{\fill}}lcccccccc@{}}
\hline
Name 
& $\tilde{g}_{1.8\cdot1.3}^{10}$ 
& $\tilde{g}_{1.8\cdot1.3}^{30}$ 
& $\tilde{g}_{1.8\cdot1.3}^{50}$ 
& $\tilde{g}_{1.8\cdot1.3}^{70}$
& $\tilde{g}_{2.2\cdot1.3}^{10}$
& $\tilde{g}_{2.2\cdot1.3}^{30}$
& $\tilde{g}_{2.2\cdot1.3}^{50}$ 
& $\tilde{g}_{2.2\cdot1.3}^{70}$ \\
\hline
$\langle d^\textrm{flight}_{\tilde{g}}\rangle~[\textrm{cm}]$ & 10 & 30 & 50 & 70 & 10 & 30 & 50 & 70 \\ 
$m_{\tilde{g}}~[\textrm{TeV}]$ & 1.8 & 1.8 & 1.8 & 1.8 & 2.2 & 2.2 & 2.2 & 2.2 \\
$m_{\tilde{\chi}_1^0}~[\textrm{TeV}]$ & 1.3 & 1.3 & 1.3 & 1.3 & 1.3 & 1.3 & 1.3 & 1.3\\
\hline
$m_{\tilde{t}}$~[TeV] & 489 & 643 & 731 & 795 & 1347 & 1770 & 2013 & 2189 \\
$\tilde{\chi}^0_1$ type & $\tilde{B}$ & $\tilde{B}$ & $\tilde{B}$ & $\tilde{B}$ & $\tilde{B}$ & $\tilde{B}$ & $\tilde{B}$ & $\tilde{B}$ \\ 
\hline
\end{tabular*}
\caption{Definition of the eight benchmarks for the long-lived gluino analysis}
\label{tab:gluino_bcmk}
\end{table*}

\noindent By using the results from Figs.~\ref{fig:gluino-xsection} to 
\ref{fig:gluino-bgctau2}, we have built eight different benchmark points and their definition is given in Table~\ref{tab:gluino_bcmk}. Two values of cross section have been chosen by fixing the gluino mass: 1.8~TeV and 2.2~TeV. Concerning the neutralino, we choose the bino state and its mass is fixed to 1.3 TeV. The mean flight distance of the gluino in the laboratory frame is a parameter of the benchmark point, and the stop mass is adjusted for this purpose. These benchmarks take into account the current experimental limits coming from the search of prompt-gluino production (for ATLAS~\cite{ATLAS:2021gld,ATLAS:2020syg,ATLAS:2019fag} and for CMS~\cite{CMS:2019zmd,CMS:2019ybf,CMS:2021beq,CMS:2020cur,CMS:2020cpy,CMS:2017qxu,CMS:2017okm,CMS:2017abv}). The best experimental constraints forbid a gluino with a mass lower than 2.2 or 2.3 TeV. The limits on the neutralino mass depend on the gluino mass and can reach 1.2 or 1.3 TeV. Searches for long-lived gluinos have been also scrutinised at LHC Run 1 and 2 by using the displaced jets signature~\cite{CMS:2020iwv,CMS:2018ncl} or the displaced vertices signature~\cite{PhysRevD.104.052011,ATLAS:2017tny}. \\

 We {note} that the long-lived particle is not the gluino but the R-hadron made up of the gluino. Three kinds of R-hadrons can be produced: the gluinoball $R(\tilde{g}g)$, the R-meson $R(\tilde{g}q\bar{q})$ and the R-baryon $R(\tilde{g}qqq)$. The gluinoball is always electrically neutral whereas R-mesons and R-baryons can have a negative, neutral or positive electric charge according to the flavour of the quarks. When it is
electrically charged, the R-hadron has a trajectory bent by the magnetic field and can induce a reconstructed track in the tracker.  For instance, charged R-hadrons based on gluino and decaying outside the tracker (and so corresponding to HSCP signatures) have been also studied by ATLAS and CMS collaborations on the basis of the \textsl{dE/dx} energy deposit~\cite{ATLAS:2022pib,CMS:2016ybj}. In the following, we do not take into account this potential property and, by misuse of language, we refer {to} the R-hadron as gluino.

\subsubsection{{Event } kinematics}

\begin{figure*}[htbp]
\begin{minipage}[t]{0.45\linewidth}
    \centering
    \includegraphics[scale=0.45]{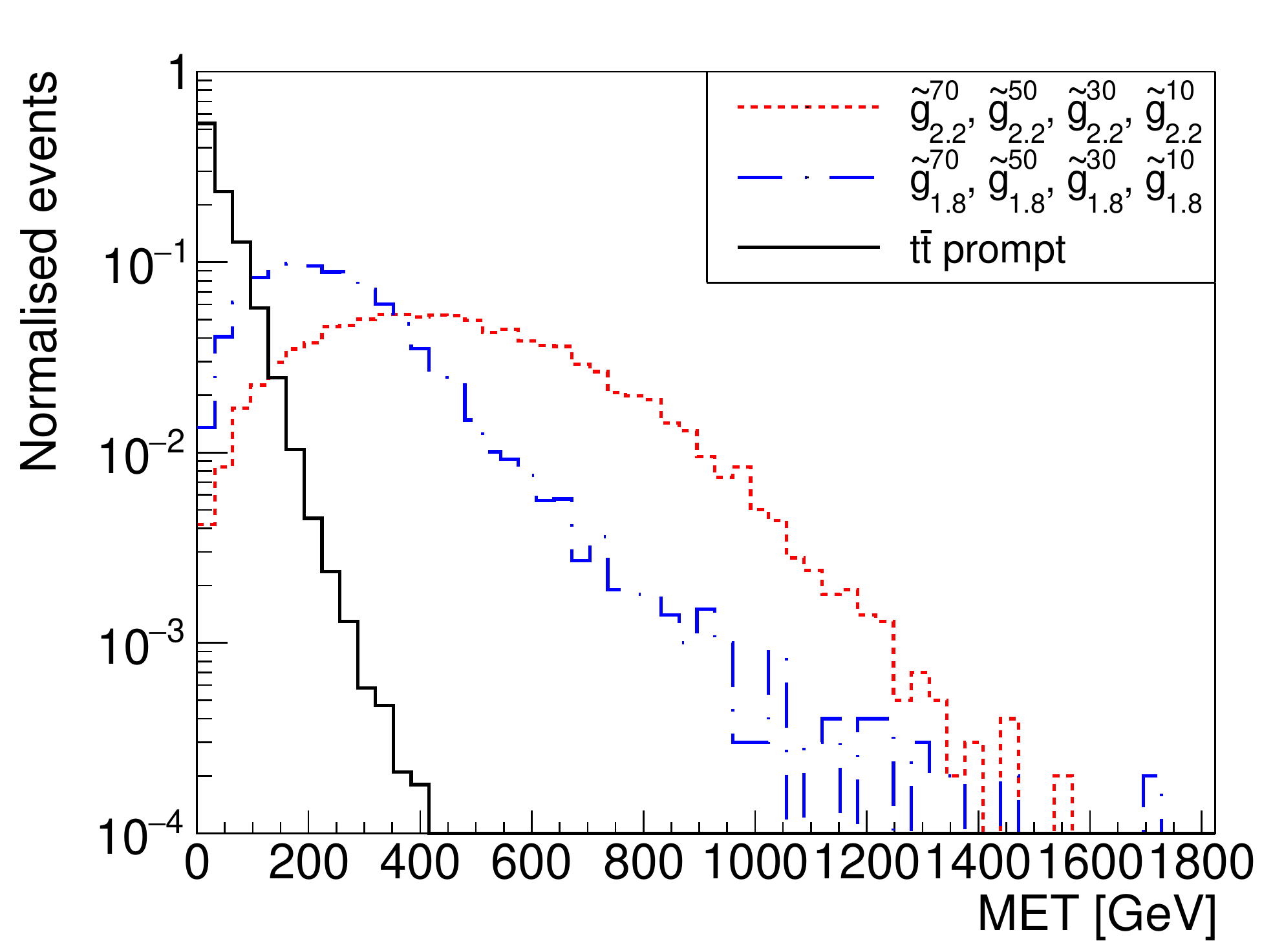}
    \caption{Missing transverse energy MET in the gluino $\tilde{g}$ case. The main contribution is coming from the invisible neutralino.}
    \label{fig:gluinoMET}
\end{minipage}
\hspace*{1cm}
\begin{minipage}[t]{0.45\linewidth}
    \centering
    \includegraphics[scale=0.45]{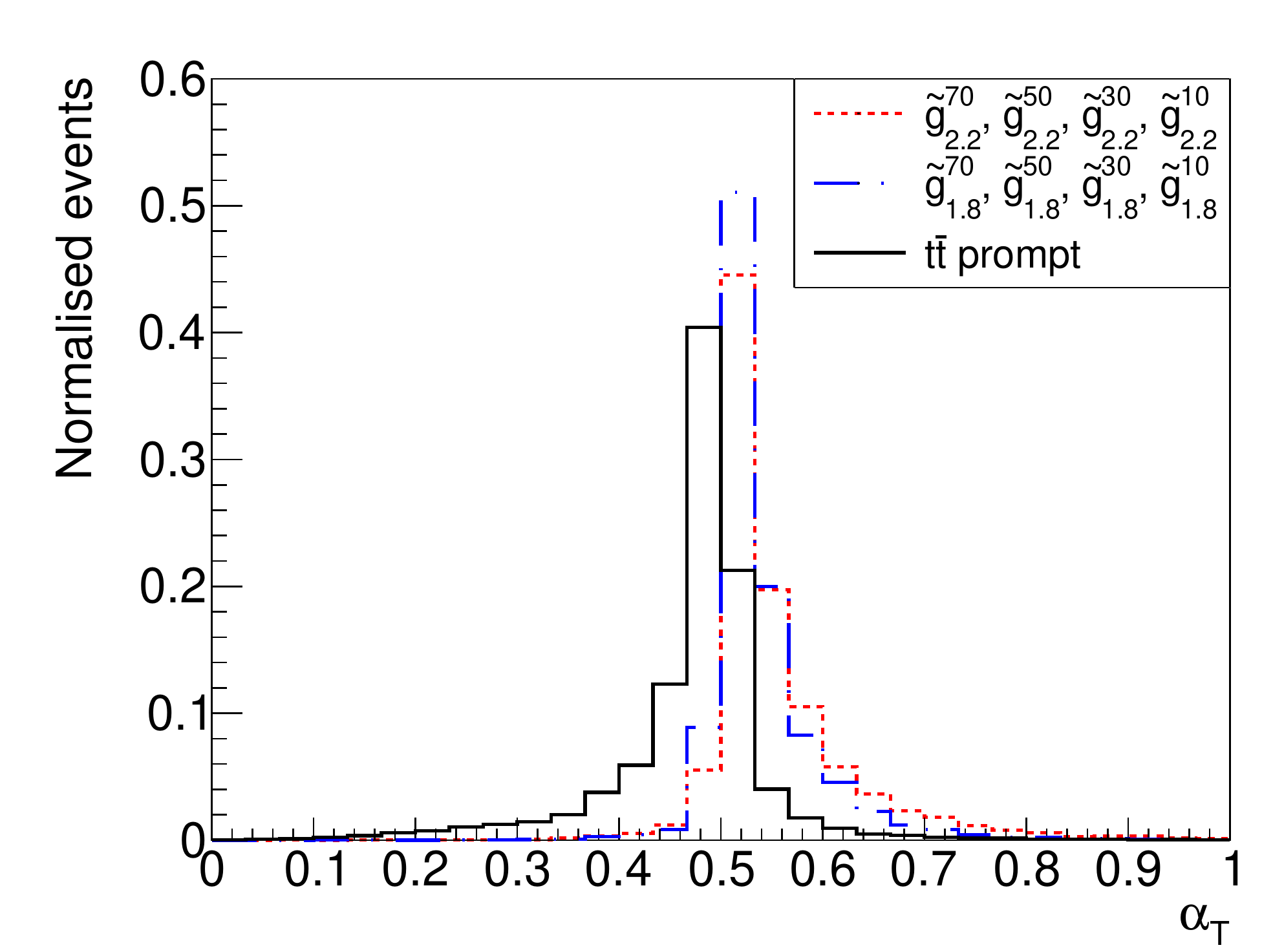}
    \caption{$\alpha_T$ distributions in the gluino $\tilde{g}$ case.}
    \label{fig:gluinodeltaR}
\end{minipage}
\begin{minipage}[t]{0.45\linewidth}
    \centering
    \includegraphics[scale=0.45]{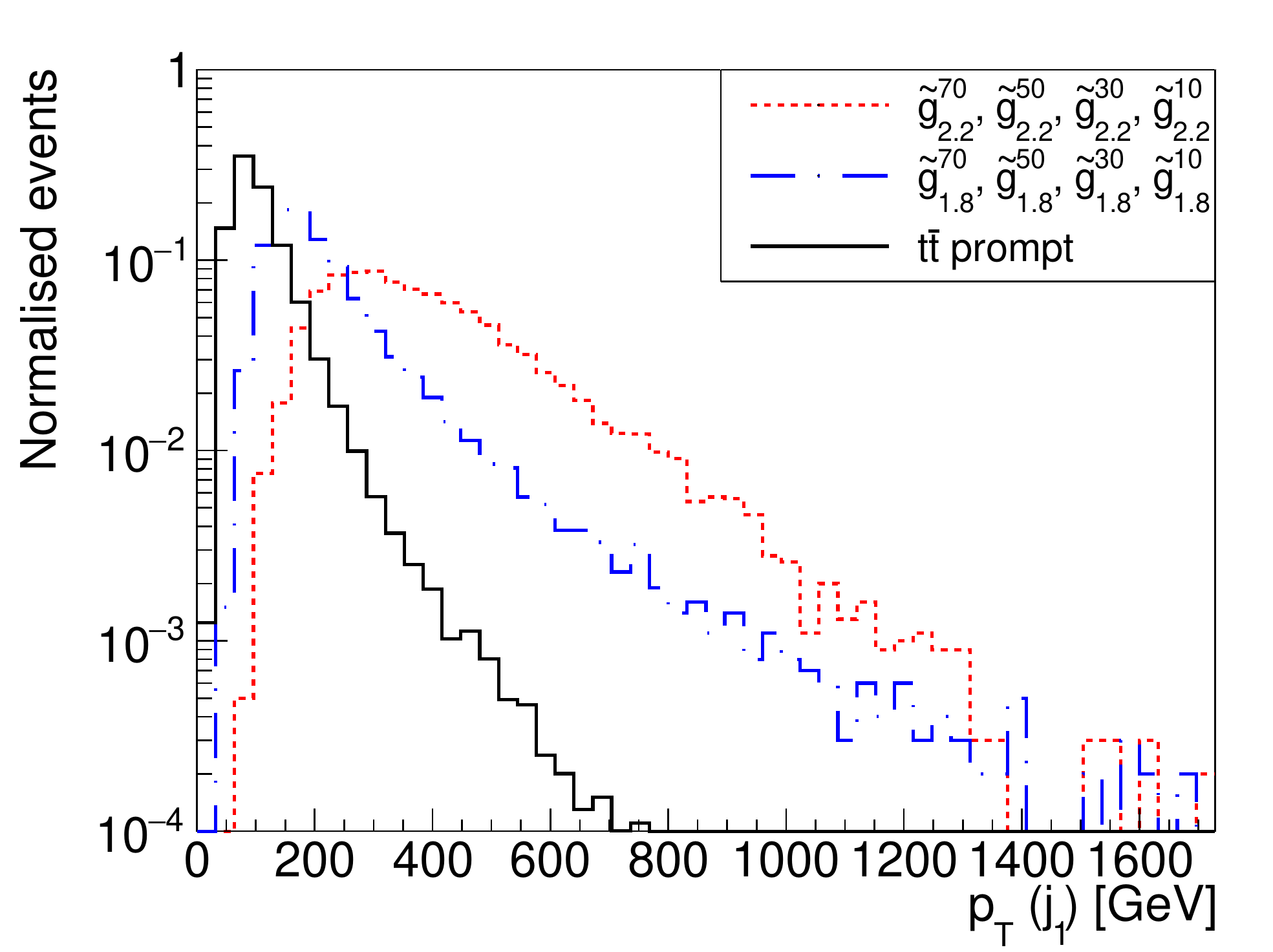}
    \caption{Transverse momentum of the leading jet in the gluino $\tilde{g}$ case.}
    \label{fig:gluinoPTJ}
\end{minipage}
\hspace*{1cm}
\begin{minipage}[t]{0.45\linewidth}
    \centering
    \includegraphics[scale=0.45]{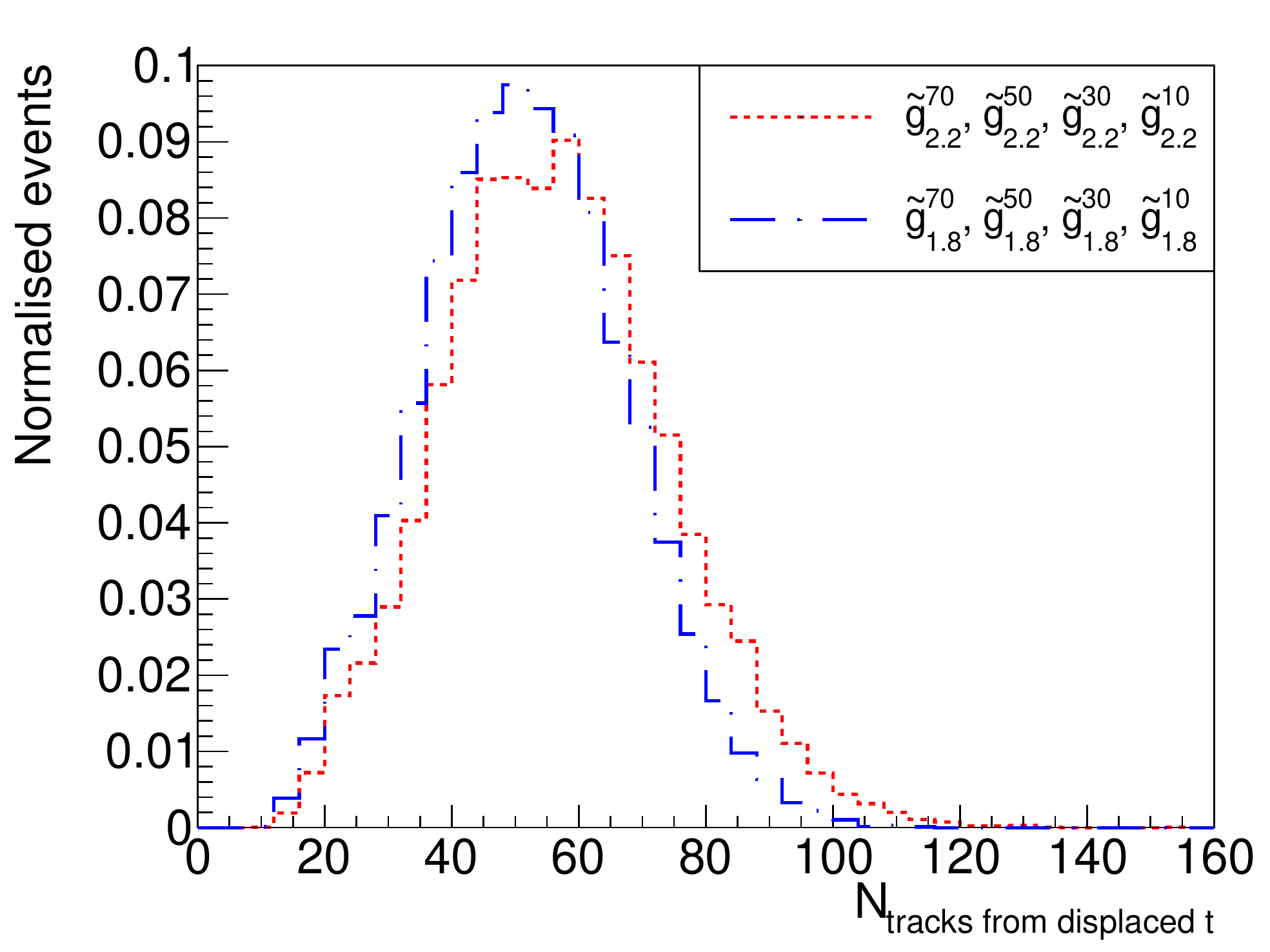}
    \caption{Multiplicity of displaced tracks for each displaced top quark decay vertex for a gluino mass of $1.8\ \mathrm{TeV}$ and $2.2\ \mathrm{TeV}$.o}
    \label{fig:gluinontrack}
\end{minipage}
\begin{minipage}[t]{0.45\linewidth}
    \centering
    \includegraphics[scale=0.45]{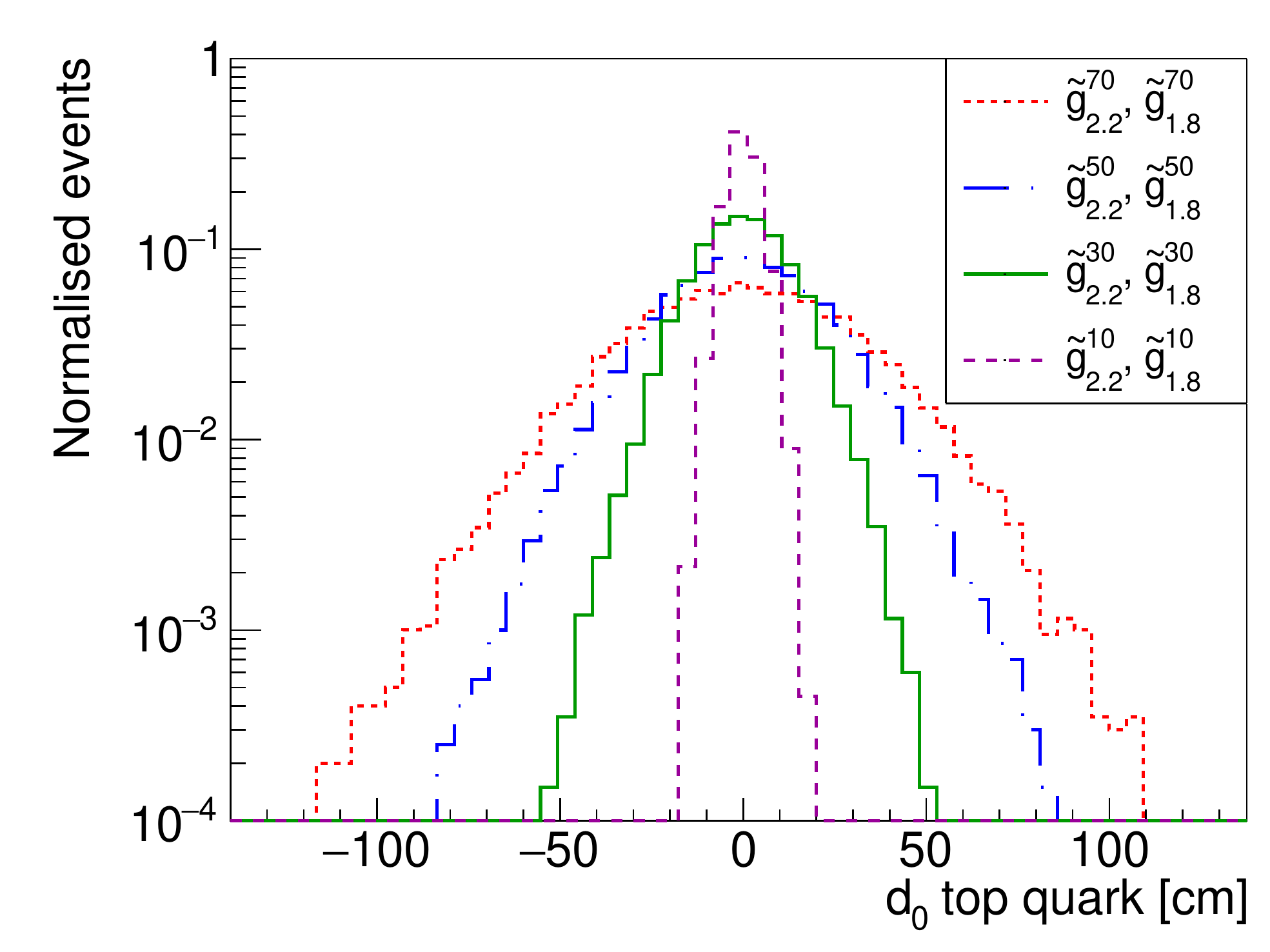}
    \caption{Transverse impact parameter $d_0$ distributions for the eight benchmarks with a magnetic field of $B=3.8\ \text{T}$.}
    \label{fig:gluinod0}
\end{minipage}
\hspace*{1cm}
\begin{minipage}[t]{0.45\linewidth}
    \centering
    \includegraphics[scale=0.45]{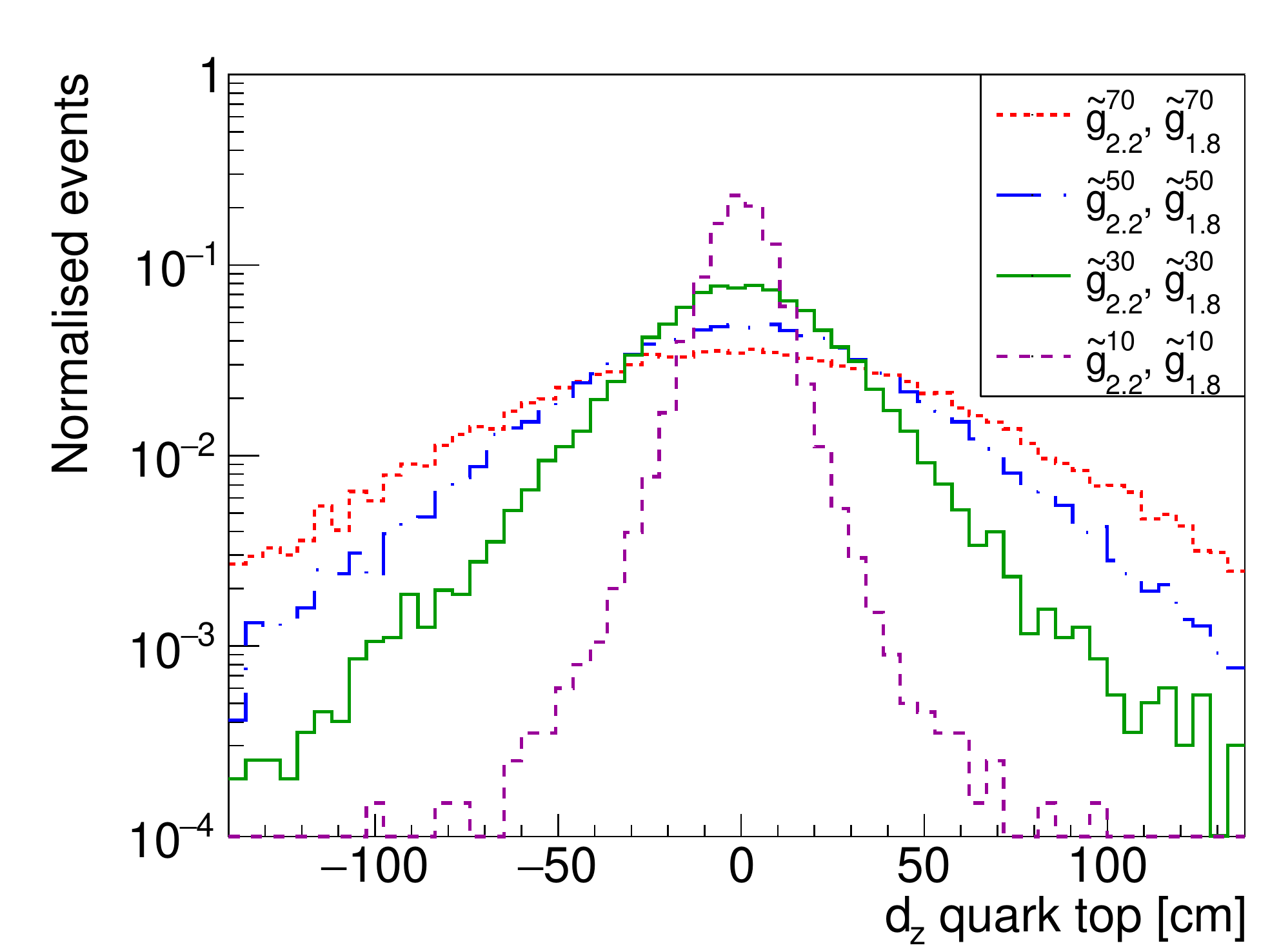}
    \caption{$d_z$ distributions for the eight benchmarks with a magnetic field of $B=3.8\ \text{T}$.}
    \label{fig:gluinodz}
\end{minipage}
\end{figure*}

\begin{figure*}[htbp]
\begin{minipage}[t]{0.45\linewidth}
    \centering
    \includegraphics[scale=0.45]{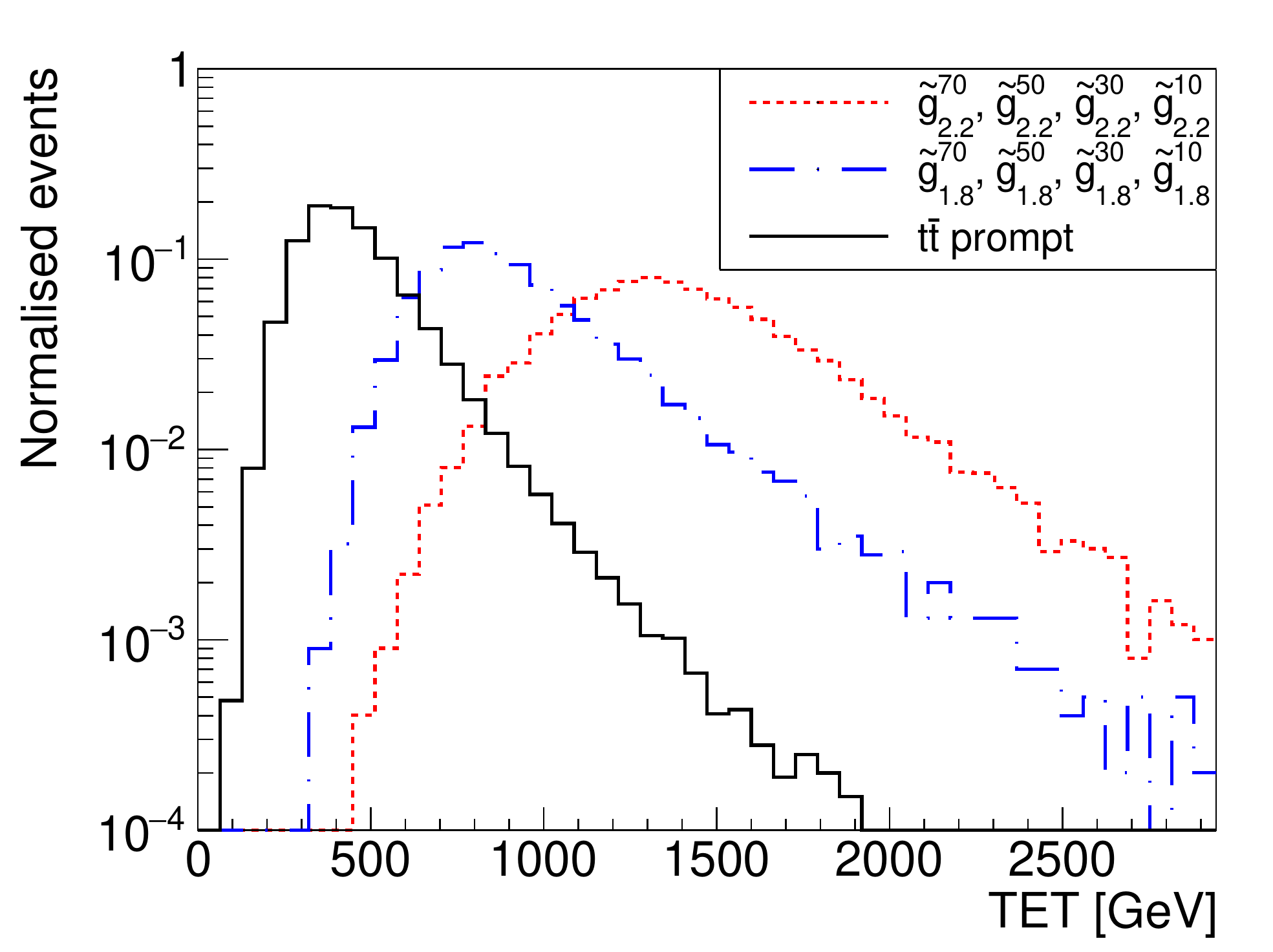}
    \caption{Total transverse energy TET in the gluino $\tilde{g}$ case}
    \label{fig:gluinoTET}
\end{minipage}
\hspace*{1cm}
\begin{minipage}[t]{0.45\linewidth}
    \centering
    \includegraphics[scale=0.45]{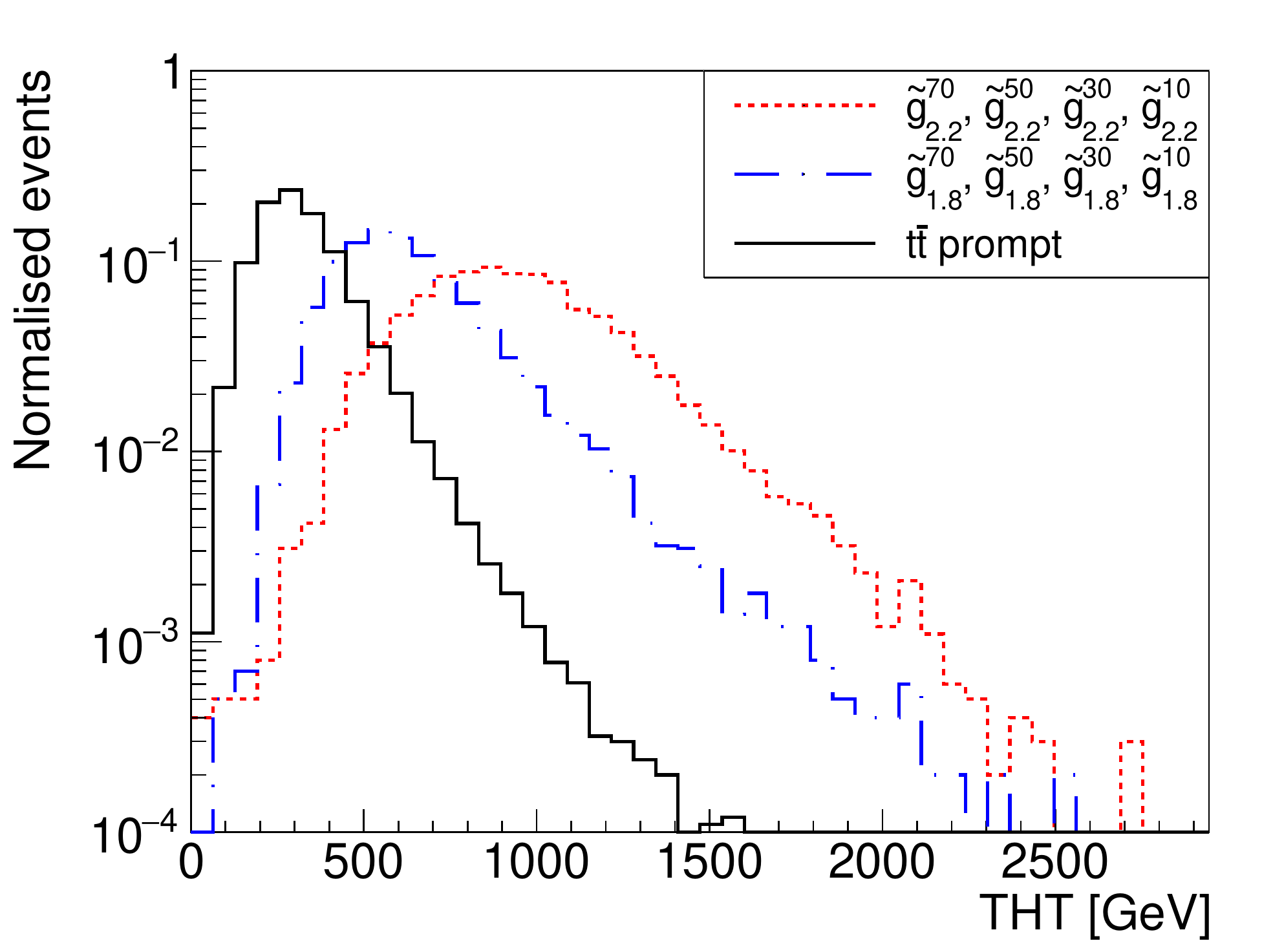}
    \caption{Total transverse hadrnoic energy THT in the gluino $\tilde{g}$ case}
    \label{fig:gluinoTHT}
\end{minipage}
\begin{minipage}[t]{0.45\linewidth}
    \centering
    \includegraphics[scale=0.45]{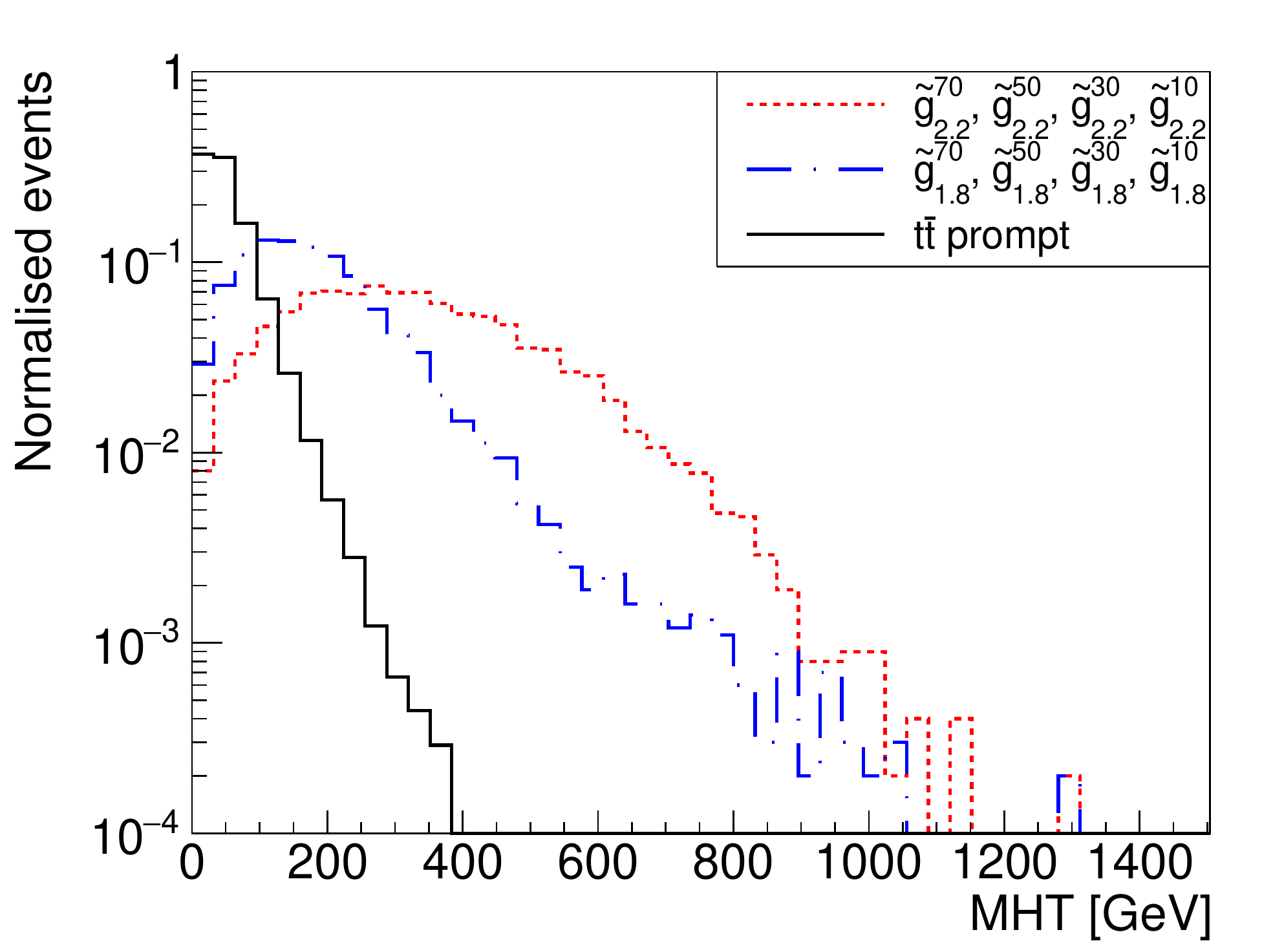}
    \caption{Missing transverse hadronic energy MHT in the gluino $\tilde{g}$ case.}
    \label{fig:gluinoMHT}
\end{minipage}
\hspace*{1cm}
\begin{minipage}[t]{0.45\linewidth}
    \centering
    \includegraphics[scale=0.45]{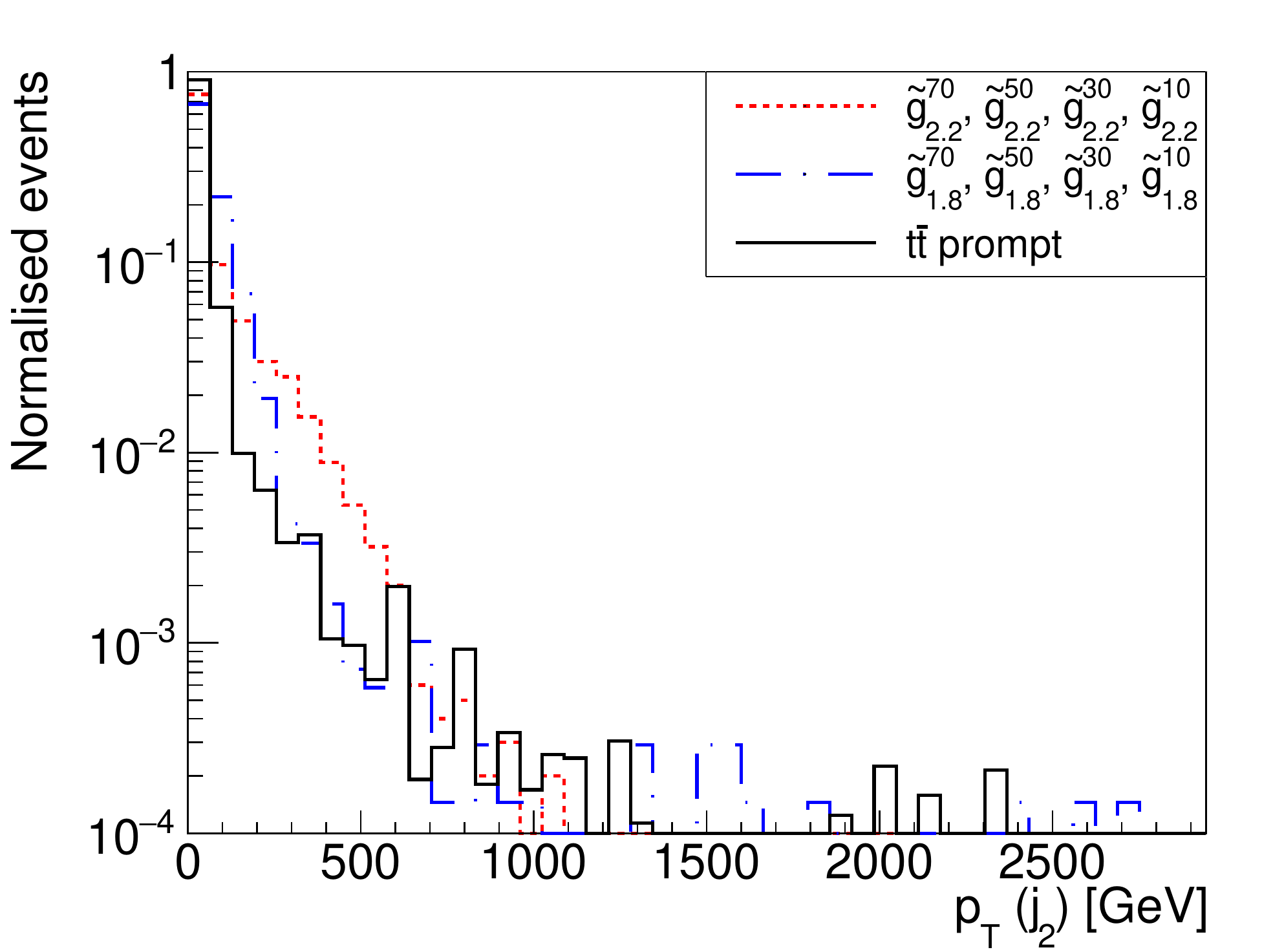}
    \caption{Transverse momentum $p_T$ of the second leading jet in the gluino $\tilde{g}$ case.}
    \label{fig:gluinoptj2}
\end{minipage}
\begin{minipage}[t]{0.45\linewidth}
    \centering
    \includegraphics[scale=0.45]{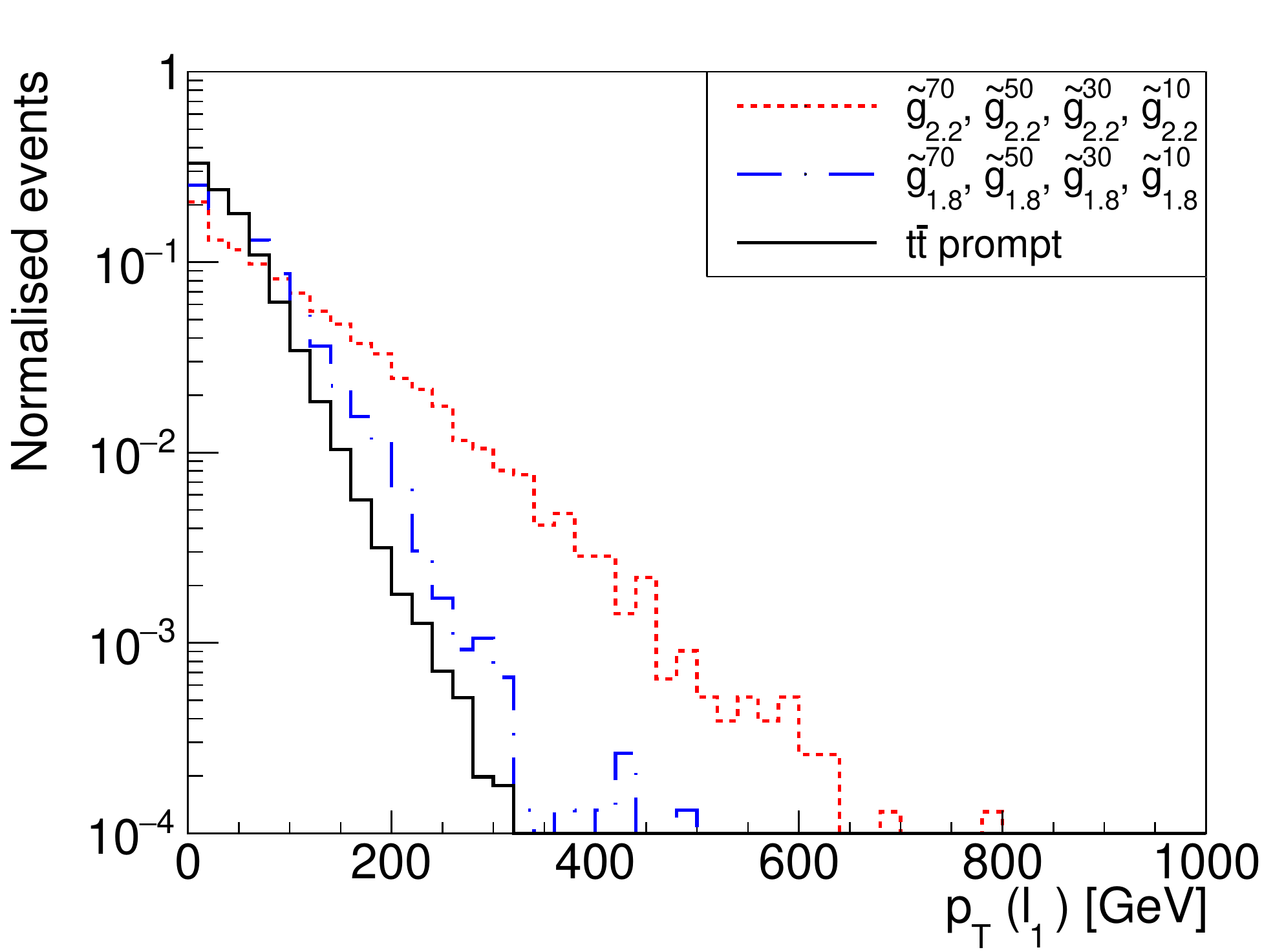}
    \caption{Transverse momentum $p_T$ of the leading lepton in the gluino $\tilde{g}$ case.}
    \label{fig:gluinoptl1}
\end{minipage}
\hspace*{1cm}
\begin{minipage}[t]{0.45\linewidth}
    \centering
    \includegraphics[scale=0.45]{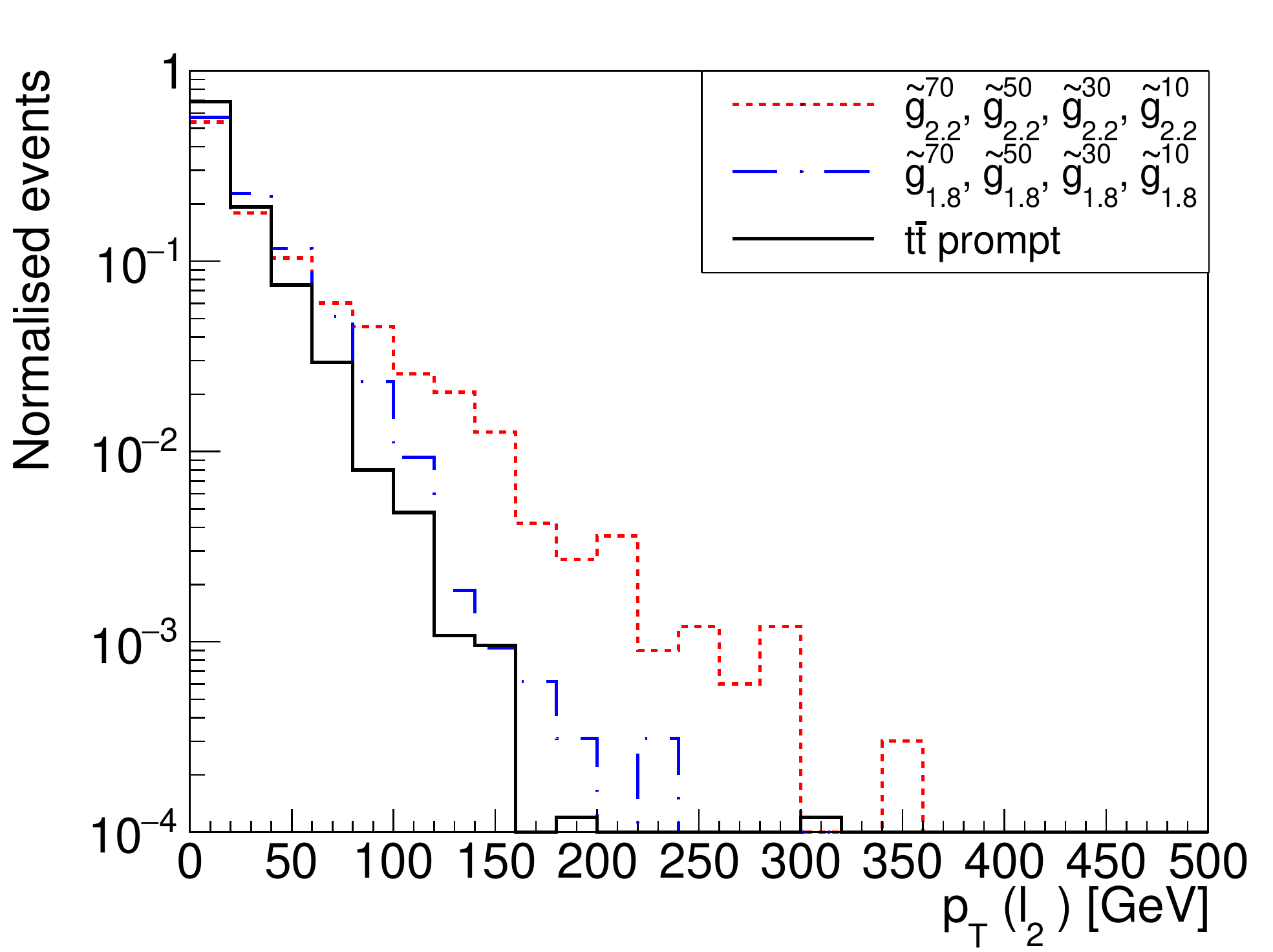}
    \caption{Transverse momentum $p_T$ of the second leading lepton in the gluino $\tilde{g}$ case.}
    \label{fig:gluinoptl2}
\end{minipage}
\end{figure*}

\begin{figure*}[htbp]
\begin{minipage}[t]{0.45\linewidth}
    \centering
    \includegraphics[scale=0.45]{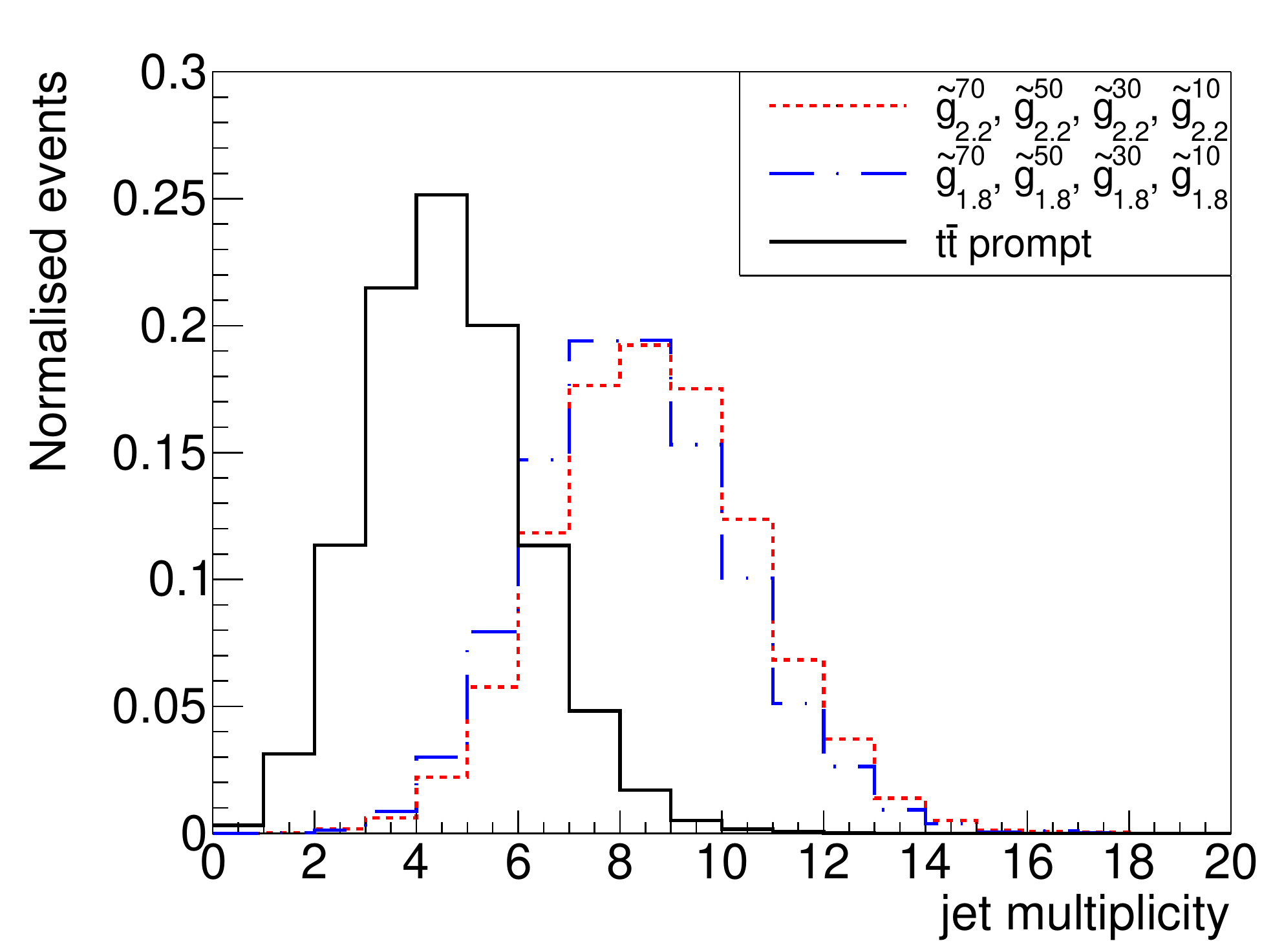}
    \caption{Jet multiplicity $N_j$ in the gluino $\tilde{g}$ case.}
    \label{fig:gluinoNj}
\end{minipage}
\hspace*{1cm}
\begin{minipage}[t]{0.45\linewidth}
    \centering
    \includegraphics[scale=0.45]{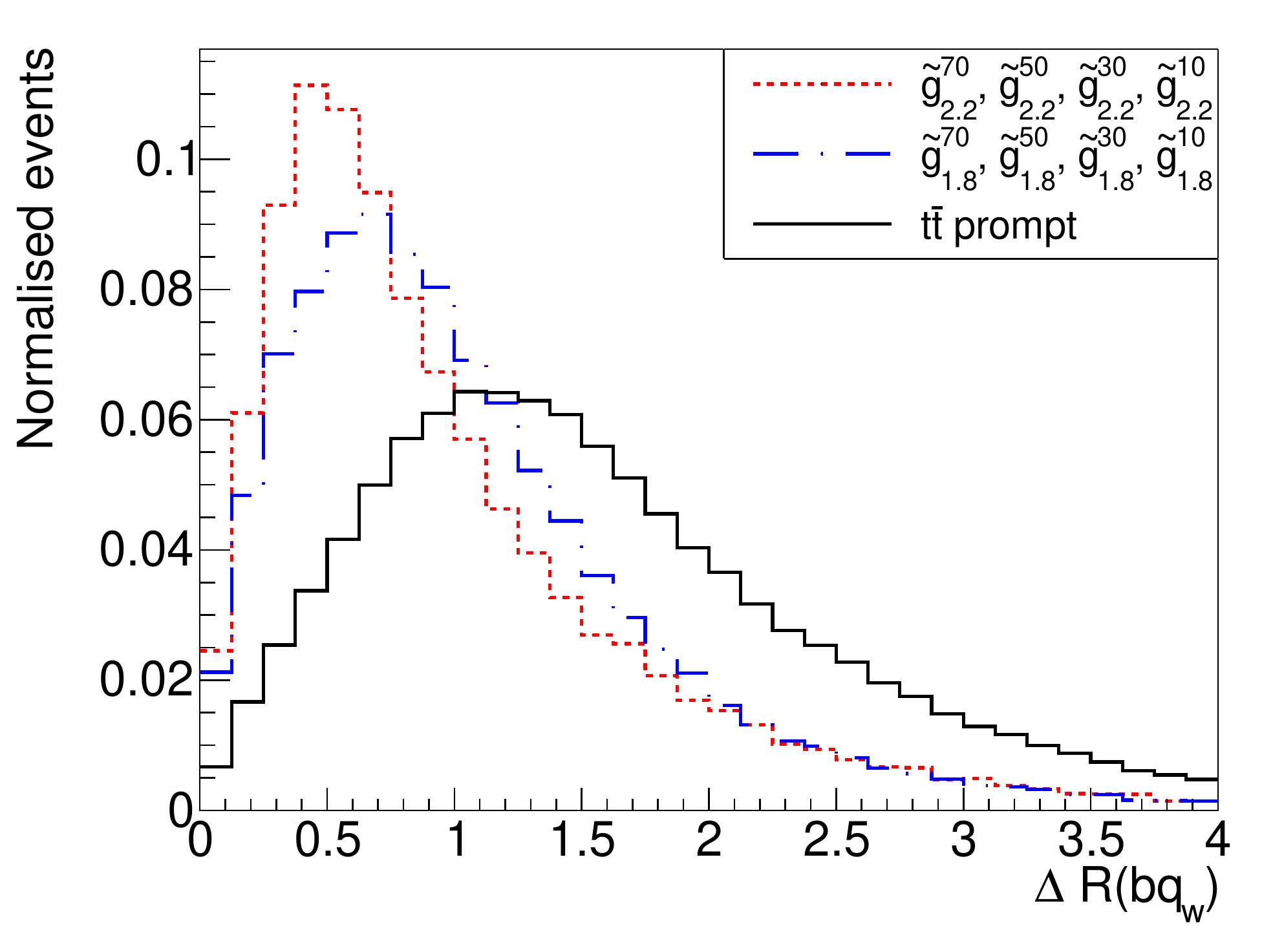}
    \caption{$\Delta R$ distributions in the gluino $\tilde{g}$ case.}
    \label{fig:gluinoDR}
\end{minipage}
\end{figure*}

Classical geometrical and kinematics observables have been studied for each benchmark.
Figures~\ref{fig:gluinoMET} to 
\ref{fig:gluinoDR} are a selection of some relevant observables for this model. A
summary of the results can be found in Table~\ref{table:gluino_properties}.\medskip

First, the main property of this signal topology is to offer 4 produced top quarks per event instead of 2 in the previous cases. This multiplicity of top quarks coming from the decay of {a} heavy R-hadron leads to a high multiplicity of reconstructed jets (about 8 on average) with a large transverse momentum (the hardest jet has a mean transverse momentum of about 230~GeV and 410~GeV respectively for a gluino mass of 1.8~TeV and 2.2 TeV according to Fig.~\ref{fig:gluinoPTJ}) and a large hadronic activity in the calorimeter (the mean total hadronic energy, THT, is about 640 GeV and 950 GeV for both gluino masses). \medskip

The probability to have a four \textsl{displaced top quark} signature is about 63\% for a mean flight distance of the R-hadron of 10~cm and it can increase to more than 97\% for a mean flight distance of 50~cm. As the Figs.~\ref{fig:gluinod0} and \ref{fig:gluinodz} show, the {distributions} of impact parameter $d_0$ and $d_z$ of the top quarks get wider when the mean flight distance of the gluino increases and their mean values are similar to the previous cases of stop benchmarks. \medskip

Another key observable is the MET distribution (see Fig.~\ref{fig:gluinoMET}) since this observable is expected to be larger for the gluino events than for the $t\bar{t}$ events. In the signal sample, the MET is mainly dominated by the neutralino contribution coming from the decays of the two gluinos. The other contributions are the neutrinos coming from the top quark decays and the rare cases of gluino decaying outside the calorimeter. The MET mean value is about 260~GeV and 480~GeV for gluino with a mass of 1.8 TeV and 2.2 TeV, respectively. \medskip

Similarly to the stop production case, Figure~\ref{fig:gluinodeltaR} shows that the indicator $\alpha_T$, usually used in {searches} for supersymmetric signatures, can be also used in this case to discriminate the signal from Standard Model background sources like the $t\bar{t}$ one. This indicator can be also relevant for discriminating the signal from the standard four top quarks production if the selection strategy requires a such multiplicity of quark tops with a small displacement. Contrary to the stop production case, the b quark and the W boson coming from the top decay are separated enough in the transverse plane in order to reconstruct non-merged jets, see Figure~\ref{fig:gluinoDR}. Obviously, the flight distance of the gluino, driven by the stop mass, has a negligible impact on the angle between the b quark and the W decay products.

\begin{table*}[htbp]
\begin{tabular*}{\textwidth}{@{\extracolsep{\fill}}lcccccccc|c}
\hline
Name 
& $\tilde{g}_{1.8\cdot1.3}^{10}$ 
& $\tilde{g}_{1.8\cdot1.3}^{30}$ 
& $\tilde{g}_{1.8\cdot1.3}^{50}$ 
& $\tilde{g}_{1.8\cdot1.3}^{70}$
& $\tilde{g}_{2.2\cdot1.3}^{10}$
& $\tilde{g}_{2.2\cdot1.3}^{30}$
& $\tilde{g}_{2.2\cdot1.3}^{50}$ 
& $\tilde{g}_{2.2\cdot1.3}^{70}$ & $t\bar{t}$ \\
\hline
$\sigma_{LO}(pp\rightarrow \tilde{g}\tilde{g})~[\text{fb}]$ & \multicolumn{4}{c|}{2.0} & \multicolumn{4}{c|}{0.3} & - \\
\hline
$R_{\ =\ 4\ \textrm{displaced t}}$~[\%] & 63.2 & 94.3 & 97.5 & 96.4 & 65.5 & 94.3 &  97.3 & 96.3 & - \\ 
$R_{\ \geq\ 2\ \textrm{displaced t}}$~[\%] & 72.1 & 97.2 & 99.0 & 98.1 & 72.4 & 97.1 & 98.7 & 97.8 & - \\ 
\hline
$\langle MET \rangle$ [GeV]  & 254 &  253 & 253 & 257 & 473 & 470 & 470 & 478 & 43 \\ 
$\langle TET \rangle$ [GeV] &  940 & 937  & 939  & 941 & 1422  & 1417 & 1417 & 1424 & 465 \\ 
$\langle THT \rangle$ [GeV]   & 642 & 639 & 641 & 641 & 962 & 953 & 955 & 955 & 327 \\ 
$\langle MHT\rangle $ [GeV]  & 189 & 188 & 188 & 190 & 340 & 338 & 339 & 335 & 52 \\
$\langle \alpha_T \rangle$&  0.54 &  0.54 & 0.54 & 0.54 & 0.56 & 0.56 & 0.56 & 0.56 & 0.48 \\  
\hline
$\langle p_T (j_1)\rangle $ [GeV]  & 234 & 230 & 231 & 235 & 410 & 409 & 408 & 409 & 118 \\ 
$\langle p_T (j_2)\rangle $ [GeV]  & 40 & 33 & 51 & 57 & 50 & 53 & 80 & 52 & 23 \\ 
$\langle p_T (\ell_1)\rangle $ [GeV]  & 56 & 57 & 56 & 56 & 92 & 94 & 93 & 95 & 32 \\
$\langle p_T (\ell_2)\rangle $ [GeV]  & 23 & 23 & 22 & 23 & 32 & 33 & 32 & 33 & 31 \\
$\langle \Delta R (bq_W)\rangle $   &  1.0 & 1.0 & 1.0 & 1.0 & 1.0 & 1.0 & 1.0 & 1.0 & 1.5 \\ 
\hline
$\langle |d_0(t)|\rangle $ [cm] & 3.0 & 9.3 & 15.3 & 21.3 & 3.4 & 10.2 & 17.2 & 24.0 & - \\
$\langle |d_z(t)|\rangle $ [cm] & 6.9 & 20.3 & 32.3 & 44.3 & 7.5 & 22.1 & 36.0 & 49.2 & - \\
\hline
$ N_{\text{jets}} $ & 7.8 & 7.8 & 7.8 & 7.8 & 8.2 & 8.2 & 8.2 & 8.2 & 4.3 \\
$N_{\text{tracks from displaced t}}$ & 51 & 51 & 51 & 51 & 56 & 56 & 56 & 56 & - \\
\hline
\end{tabular*}
\caption{Mean values of some observables and number of events with top quarks decaying in the tracker volume for the eight benchmarks of the long-lived gluino analysis. We remind that $R_{\geq \textrm{2 displaced t}}$ corresponds to the number of events with at least two top quarks decaying in the tracker volume.}
\label{table:gluino_properties}
\end{table*}

\section{Discussion on the design of experimental analyses devoted to displaced top quark signature}\label{sec:general-guidelines}

Using the various properties and distributions of the studied benchmarks, we discuss possible strategies which can be implemented in future experimental analyses for extracting the signal from the background. The discussion will be split in two parts: one will be devoted to the online selection by the trigger system of the ATLAS and CMS multipurpose detectors, the other will be devoted to the offline selection achieved on the data. 

\subsection{Selecting online the signal events with the detector trigger system}

Due to the high frequency of proton-proton collisions at the LHC ($40\ \text{MHz}$), the CMS and ATLAS collaborations use (hardware and software) trigger systems to extract events of {interest} for physics studies and detector commissioning. It is crucial for experimentalists to define a strategy for triggering on the signal events. Traditional observables used for prompt physics can be relevant to the signal topologies studied in this paper.
\begin{itemize}
    \item [$\bullet$] The main interest of the production of displaced top quarks by a pair of sleptons remains in the presence of two prompt leptons in the final state. The transverse momentum of these muons for a small mass of neutralino (mean $p_T$ of 90~GeV for a neutralino mass of 200~GeV) is ideal for using lepton triggers. This kind of trigger can be also used, to a lesser extent, for firing events where the top quarks decay leptonically and with a small displacement. For instance, leptons coming from top quarks produced by a neutralino pair have an average transverse momentum of $\sim 45-50\ \text{GeV}$.
    \item[$\bullet$] Displaced top quark events are very energetic. For this reason, experimentalists have the opportunity to use the total hadronic energy (THT) in the event or the presence of jets with a large $p_T$. For instance, in the case where the long-lived particle is a R-hadron, THT can reach more than 1~TeV and the hardest jets have a mean $p_T$ between 400 and 700~GeV.
    \item[$\bullet$] A trigger associated with MET may be relevant to suppress the QCD background and highlight the signal from the SM $t\bar{t}$ production. This MET is the sum of the contributions of the LSP (except the case where it is the long-lived particle), the neutrinos coming from the top quark decay and the small number of events where the long-lived particle decayed outside the detector. Despite the extremely small LSP mass, the gravitino contributes to most of the MET in the stop benchmarks (between 500 and 700~GeV, depending on the stop mass). This allows us to use a trigger based on a MET threshold. 
\end{itemize}
Undeniably, the trigger strategy can be based on only one of the previous discriminating observables or on the combination of several observables with lower thresholds (if they are not fully correlated). As an illustration, some HSCP analyses have a trigger based on the criteria $\text{MET}>170\ \text{GeV}$ and a reconstructed muon with $p_T>50\ \text{GeV}$ \cite{Khachatryan_2016}. {Thus, such selection thresholds on MET can be applied in the case of the benchmarks containing large MET sources (stop squark $\tilde{t}$ and gluino $\tilde{g}$ models). Moreover, all studied signal sources present a large value of THT and large transverse momenta for jets. A criterion such as $\text{THT}>430\ \text{GeV}$ is relevant, as well as $p_{T}(j_{1})>150\ \text{GeV}$. For the bino case, a selection rule on the leading muon may help to trigger the production of a smuon pair with $p_T(\mu_{1})>24\ \text{GeV}$. Assuming such criteria, the $t\bar{t}$ background is decreased by 90\% and 99\%, without and with a cut on MET, respectively. According to the considered benchmark, the signal is reduced by a ratio varying between 60 and 30\% for the bino $\tilde{B}$-like case, 70\% and 30\% for the higgsino $\tilde{h}$-like case, 20\% and 10\% for the stop squark $\tilde{t}$ case and between 30\% and 10\% for the gluino $\tilde{g}$ case.} \smallskip

This discussion cannot be complete without mentioning the alternative to design and develop a specific trigger path for the displaced top quark signatures. The main difficulty comes from the difficulty to reconstruct displaced vertices or tracks in the first level of the trigger system. The design of a sophisticated discriminator taking into account the displacement requires the information coming from the tracker and the calorimeters while satisfying the constraint of processing time budget. A review of ongoing or scheduled {trigger development} dedicated to long-lived {particles} for the LHC Run~3 can be consulted~\cite{Alimena:2021mdu}.

\subsection{Selecting the signal after the trigger system}

For designing a complete experimental {analysis}, the main background sources must be considered. The most important one for signal events with a low displacement is composed of top quark pairs $t\bar{t}$ and QCD processes in general with an associated MET coming from an incorrect reconstruction, $tW$ processes or $t\bar{t}+V$ (with $V=Z$ or $W$). Drell-Yan events with associated jets are also relevant in the case of top quarks produced indirectly by sleptons, due to the presence of leptons in the final state. This list of background sources must be completed by taking under consideration all diboson cases, or even tribosons (associated to jets). For extracting the signal from the Standard Model contributions, experimentalists must keep a close eye on several points.

\begin{itemize}
    \item[$\bullet$] The offline selection should refine the criteria imposed by the trigger system, \textsl{i.e.} the {transverse momenta of the leptons and jets}, the total hadronic energy THT and the missing transverse energy MET. For the slepton benchmarks, Drell-Yann background can be reduced by requiring a specific range of values for the dilepton invariant mass or, more safely, by applying thresholds on hadronic variables.
    \item [$\bullet$] As an offline selection devoted to the benchmarks introducing a long-lived R-hadron, cuts on various observables such as $\alpha_T$ with $\alpha_T>0.5$ (see Fig.~\ref{fig:stop_alphaT}) may help to suppress standard background events. 
    \item [$\bullet$] A special care must be taken with the reconstruction of jets in the context of stop-based R-hadrons. Indeed, the study of the {kinematic} properties reveals that quarks coming, directly or indirectly, from the top quark decay are very close for a large fraction of events. Collimated jets can be reconstructed and the jet substructure can be investigated for identifying properly the displaced top quark signature.
\end{itemize}
The majority of the displaced vertex background is identified as effects that are not related to the pp collision, but to secondary interactions and instrumental noises. For instance, nuclear interactions within the detector matter can also induce secondary vertices, as well as reconstruction {algorithm-induced} fakes (see \cite{Alimena:2019zri} for more information).\\

Moreover, assuming a full simulation for the background process, a lot of $t\bar{t}$ events will induce secondary interaction with layers of detection which can be confused with displaced processes. Since the production cross section of $t\bar{t}$ is much larger than for the studied signals, the multiplicity of displaced tracks is then not enough to highlight the signal over the background. Moreover, applying a reconstruction efficiency on the reconstructed displaced tracks will decrease such multiplicity. Note that the precise experimental analyse based on the signal-background ratio is out of the scope of this paper.

\subsection{Towards a specific algorithm devoted to displaced top quark signature}

Currently, there are  no specific algorithms devoted to the identification of displaced top quarks. Nonetheless, such algorithms could be designed by gathering information from displaced jets, b-tagging, and displaced vertices (taking advantage of the multiplicity of displaced tracks coming from the top quark decay). Note also that each method cited here can also be used on its own. Thus, including such reconstruction in the analysis, it would be possible to require at least one displaced top quark in the events, and thus increase the signal/background ratio. The amount of events offering several displaced quark top signatures raises the question to identify and to reconstruct several displaced candidates. Defining signal regions sorted according to the multiplicity of displaced candidates can increase the sensitivity of the analysis. Moreover, 
specific Multi-Variate Analysis can also be set up using techniques such as Boosted Decision Trees or neutral networks in order to improve the performance of the selection.

\section{Summary}
In this paper, we have addressed  the \textsl{displaced top quark} signature which extends the list of displaced objects (displaced jets, displaced vertex, displaced leptons, ...) already studied by the CMS and the ATLAS collaborations and which should lead to a complementary approach to the inclusive research of exotic long-lived particles at the LHC. We have presented a comprehensive review of event topologies that offer such a signature. For each topology, a simplified model has been designed, the relevant region of the parameter space has been highlighted and some promising benchmark points have been identified. Some of these physics processes are known and studied with other displaced objects like the long-lived gluino, and others are still unexplored like the displaced neutralino stemming from slepton pair production. The simplified models studied here cover also various theoretical scenarios: the presence of {a} dark matter candidate with the GMSB mechanism, the non-conservation of R-parity in the RPV scenario and the non-naturalness in the Split-SUSY model. The properties of the benchmarks have been illustrated by distributions of key observables and their relevance has been shown by comparing the obtained distributions with the distributions of Standard Model top quark pair production.\\

This work is the first concrete step toward the exploration of TeV physics with the \textsl{displaced top quark} signature. It provides practical tools for both experimentalist and phenomenologist communities for LHC Run~3. First, the benchmarks can be used for evaluating the performance of reconstruction and identification algorithms of the ATLAS and CMS detector in extreme and challenging conditions. Devoted algorithms to \textsl{displaced top quark} can be built in order to optimise the sensitivity to this signature and to probe data with it. The benchmarks can also interest phenomenologists working in the reinterpretation of the LHC experimental limits by using some framework like \textsc{Rivet}~\cite{Buckley_2013}, \textsc{CheckMate}~\cite{Dercks_2017} or \textsc{MadAnalysis~5}~\cite{Conte_2014}.

\section{Acknowledgments}
The authors are grateful to the CMS and ATLAS collaborations for the interesting discussions about this prospective work. We would like to thank also Gilbert Moultaka and Michel Rausch de Traubenberg for their precious help on the GMSB and the supergravity models. We cannot conclude without thanking Benjamin Fuks and Jack Araz for their valuable developments within the \textsc{MadAnalysis~5} framework.

\bibliographystyle{spphys}
\bibliography{displaced_tops}

\end{document}